\documentclass[ 
	fontsize=12pt,          
	numbers=noenddot,    	
    parskip=half,        	
    listof=totoc,        	
    bibliography=totoc,  	
	headsepline=true,       
	footsepline=false, 		
    DIV=12,                	
    toc=sectionentrywithdots
]{scrartcl}
\usepackage{amsthm}
\usepackage{amsmath, amsfonts, amssymb, bbm, bm}

\usepackage[super, comma]{natbib}
\usepackage{placeins}
\usepackage{array}
\usepackage{prodint}

\usepackage[english]{babel}

\usepackage[a4paper, left=2cm, right=2cm, top=3cm, bottom=3cm]{geometry}

\usepackage[hidelinks]{hyperref} 

\usepackage{todonotes} 

\definecolor{color1}{RGB}{122,0,63}

\usepackage{floatrow}
\DeclareFloatFont{notsotiny}{\fontsize{7.2}{7.2}\selectfont}
\usepackage[morefloats = 100]{morefloats} 

\theoremstyle{plain}
\newtheorem{theorem}{Theorem}
\newtheorem{ex}{Example}
\newtheorem{lemma}{Lemma}
\newtheorem{proposition}{Proposition}

\usepackage{apptools}
\AtAppendix{
\setcounter{table}{0}
\setcounter{equation}{0}
\setcounter{lemma}{0}
\setcounter{theorem}{0}
\setcounter{figure}{0}

}

\newcommand{\R}{\mathbb{R}}
\newcommand{\T}{\mathcal{T}}
\newcommand{\E}{\mathbb{E}}
\renewcommand{\P}{P}
\newcommand{\Var}{\mathbb{V}ar}
\newcommand{\Cov}{\mathbb{C}ov}

\let\OLDthebibliography\thebibliography
\renewcommand\thebibliography[1]{
  \OLDthebibliography{#1}
  \setlength{\parskip}{0pt}
  \setlength{\itemsep}{0pt plus 0.3ex}
}

\newenvironment{myproof}[1][\proofname]{%
  \proof[\bfseries\scshape #1]%
}{\endproof}

\renewcommand{\title}[1]{ \noindent{\centering \Large \textbf{ #1 } \\} }
\newcommand{\inst}[1]{\textsuperscript{#1}}
\newcommand{\institute}[1]{{\centering \footnotesize{#1}} \vspace{2ex}}

\begin{document}

\title { RMST-Based Multiple Contrast Tests in General Factorial Designs }
\begin{center}
		Merle Munko\inst{1,*}\let\thefootnote\relax\footnote{*Corresponding author. Email address: \url{merle.munko@ovgu.de}},
		Marc Ditzhaus\inst{1},
        Dennis Dobler\inst{2}
        and
        Jon Genuneit\inst{3}
\end{center}

	\institute{
		\inst{1} Otto-von-Guericke University Magdeburg; Magdeburg (Germany)\newline
        \inst{2} Vrije Universiteit Amsterdam; Amsterdam (Netherlands)\newline
        \inst{3} Leipzig University; Leipzig (Germany)
	}

\hrule

 
\paragraph{Abstract}  
Several methods in survival analysis are based on the proportional hazards assumption. However, this assumption is very restrictive and often not justifiable in practice. Therefore, effect estimands that do not rely on the proportional hazards assumption
 are highly desirable in practical applications.
 One popular example for this is the restricted mean survival time (RMST).
It is defined as the area under the survival curve up to a prespecified time point and, thus, summarizes the survival curve into a meaningful estimand. For two-sample comparisons based on the RMST, 
previous research found the
inflation of the type I error of the asymptotic test for small samples and, therefore, a two-sample permutation test has already been developed. 
The first goal of the present paper is to further extend the permutation test for general factorial designs and general contrast hypotheses by considering a Wald-type test statistic and its asymptotic behavior. Additionally, a groupwise bootstrap approach is considered.
Moreover, when a global test detects a significant difference by comparing the RMSTs of more than two groups, it is of interest which specific RMST differences cause the result. However, global tests do not provide this information. Therefore, multiple tests for the RMST are developed in a second step to infer several null hypotheses simultaneously. Hereby, the asymptotically exact dependence structure between the local test statistics is incorporated to gain more power. Finally, the small sample performance of the proposed global and multiple testing procedures is analyzed in simulations and illustrated in a real data example.

\textit{Keywords:}
{	Factorial design; Multiple testing; 
 Resampling; Restricted mean survival time; Survival analysis. }\\

\hrule

\section{Introduction} \label{sec:Intro}

In survival analysis, various methods for comparing different groups rely on the proportional hazards assumption, e.g., the famous Cox proportional hazards model.\cite{cox} 
However, verifying this assumption can be challenging, and its fulfillment is not always guaranteed. Hence, alternatives that do not require the proportional hazards assumption are of great interest. Furthermore, easy-to-interpret effect estimands, which summarize treatment and interaction effects in factorial designs, are desired.
Beyond the average hazard ratio\cite{averageHR1,averageHR}, concordance and Mann-Whitney effect\cite{concordance2009,concordance2017,concordance2020},  
and the median survival time\cite{7Brookmeyer1982Median,8Median,SMMR}, the restricted mean survival time (RMST) is becoming increasingly popular and does not rely on the proportional hazards assumption.\cite{alternative}

The RMST is defined as the area under the survival curve up to a prespecified time point and it has an intuitive interpretation as the expected minimum of the survival time and the specified time point. Thus, the RMST reduces the whole survival curve to a meaningful estimand.
The asymptotic behaviour and statistical inference of the RMST have already been considered in the literature. Horiguchi and Uno\cite{Perm} detected an inflation of the type I error of the asymptotic test for small samples in the two-sample case and, therefore, proposed an unstudentized permutation approach under exchangeability. Ditzhaus et al.\cite{RMST} extended this approach by developing a studentized permutation test, such that different censoring distributions in the two groups can be handled.
A similar approach has been further analyzed in the context of cure models, in both non- and semiparametric models.\cite{MST2023}

Such studentized permutation tests could be of interest for more complex factorial designs or more general linear hypotheses in practice, e.g., when more than two different treatments are to be compared in a clinical study. 
Thus, our first aim is the extension of the studentized permutation test of Ditzhaus et al.\cite{RMST} for general factorial designs and general linear hypotheses by employing a Wald-type test statistic. Furthermore, other resampling methods as the groupwise and, in the supplement, the wild bootstrap are considered for this general setup.

On the other hand, when a global test detects a significant result by comparing the RMSTs of more than two groups, it is of interest which particular RMSTs differ significantly. Unfortunately, global tests do not yield this information. Therefore, multiple linear hypothesis testing (MLHT) procedures are desired. They offer the information which of the local hypotheses are rejected in addition to the global one.
Moreover, their power is not necessarily lower than the power of a global testing procedure.\cite{konietschke2013}
To the best of our knowledge, the MLHT problem for the RMST in general factorial designs has not been tackled in the literature up until now. 
The present paper shall fill this gap.
For gaining more power, we aim to take the exact asymptotic dependency structure between the different test statistics into account. In order to improve the small sample performance, we propose a groupwise bootstrap procedure for approximating the limiting null distribution and we show its validity.

The remainder of this paper is organized as follows. Section~\ref{sec:Inference} includes four subsections. In Section~\ref{ssec:Setup}, the factorial survival setup and a general contrast testing problem is presented. For this testing problem, a suitable test statistic is defined and studied in Section~\ref{ssec:Wald}. The studentized permutation approach of Ditzhaus et al.\cite{RMST} is extended for more general factorial designs in Section~\ref{ssec:Perm}. Furthermore, a groupwise bootstrap procedure is investigated in Section~\ref{ssec:gw}. In Section~\ref{sec:Multiple}, multiple contrast tests are constructed and the consistency of the groupwise bootstrap in this setup is shown. Additionally, several subtopics of practical interest are covered in Section~\ref{sec:Multiple}. This includes the calculation of adjusted $p$-values, the construction of simultaneous confidence regions, a stepwise extension of the multiple testing procedure and simultaneous non-inferiority and equivalence tests.
The small sample performance of the proposed tests is analyzed in extensive simulation studies in Section~\ref{sec:Simu}. In Section~\ref{sec:Data}, we illustrate the proposed methodologies by analyzing a real data example. Finally, the results are discussed in Section~\ref{sec:Discussion}. Detailed simulation results, a wild bootstrap approach and all technical proofs can be found in the supplement.

\section{Statistical Inference}\label{sec:Inference}
In this section, the factorial survival setup and statistical methodologies for the global testing problem are presented.

\subsection{Factorial Survival Setup}\label{ssec:Setup}

We consider the following factorial design as in Ditzhaus et al.\cite{SMMR}, i.e., as $k$-sample setup, $k\in\mathbb{N}$.
We suppose that the survival and censoring times
\begin{align*}
    T_{ij} \sim S_i, \quad C_{ij} \sim G_i, \quad j\in\{1,...,n_i\}, i\in\{1,...,k\},
\end{align*}
respectively, are mutually independent.
Here, $S_i$ and $G_i$ denote the survival functions of the survival and censoring times, respectively, 
and
$n_i \in\mathbb{N}$ represent the numbers of individuals in group $i$ for all $i\in\{1,...,k\}$.
Of note, 
we do not assume the continuity of the survival functions. Consequently, ties in the data are explicitly allowed.
However, we assume that the $S_i$ do not have jumps of size 1, i.e., the survival times are not deterministic. 
Moreover, we define the right-censored observable event times $X_{ij} := \min\{T_{ij}, C_{ij}\}$ and the censoring status \mbox{$\delta_{ij} := \mathbbm{1}\{X_{ij} = T_{ij}\}$} for all $j\in\{1,...,n_i\}, i\in\{1,...,k\}.$

Furthermore, we assume that the group sizes do not vanish asymptotically, i.e., 
\begin{align}\label{eq:kappa}
    \frac{n_i}{n} \to \kappa_i \in (0,1)
\end{align} as $n\to\infty$ for all $i\in\{1,...,k\},$ where $n := \sum_{i=1}^k n_i$ represents the total sample size.

The restricted mean survival time (RMST) of group $i$ is defined as
$$ \mu_i := \int\limits_0^{\tau} S_i(t) \;\mathrm{d}t  = \E[\min\{T_{i1},\tau\}]$$
for all $i\in\{1,...,k\}.$ Here, $\tau > 0$ should be a pre-specified constant such that $\P({X}_{i1} \geq \tau)  = \P(T_{i1} \geq \tau)\P(C_{i1} \geq \tau) > 0$ and $\P(T_{i1} < \tau) > 0$ 
holds for all $i\in\{1,...,k\}$. 
By replacing $S_i$ through the Kaplan-Meier estimator $\widehat{S}_i$, a natural estimator for the RMST of group $i$ is
\begin{align*}
    \widehat{\mu}_i := \int\limits_0^{\tau} \widehat{S}_i(t) \;\mathrm{d}t 
\end{align*}
for all $i\in\{1,...,k\}.$
Let
$\boldsymbol{\mu} := (\mu_1,...,\mu_k)^{\prime}$ be the vector of the RMSTs and $\widehat{\boldsymbol{\mu}} := (\widehat{\mu}_1,...,\widehat{\mu}_k)^{\prime}$ be the vector of their estimators. In addition, let $r\in\mathbb{N}$, $\mathbf{c}\in\R^r$ be a fixed vector and $\mathbf{H}\in \R^{r \times k}$ be a contrast matrix
, i.e., $\mathbf{H}\mathbf{1}_k = \mathbf{0}_r,$ where and throughout $\mathbf{1}_k\in\R^k$ and $\mathbf{0}_r\in\R^r$ denote the vectors of ones and zeros, respectively. Moreover, we assume that $\text{rank}(\mathbf{H}) > 0$. Then, we consider the null and alternative hypothesis
\begin{align}\label{eq:NullHyp}
    \mathcal{H}_0: \mathbf{H}\boldsymbol{\mu} = \mathbf{c} \quad \text{vs.} \quad \mathcal{H}_1: \mathbf{H}\boldsymbol{\mu} \neq \mathbf{c}.
\end{align}
The formulation of this testing framework is very general. In particular, it includes the null hypothesis of equal RMSTs in all groups by choosing, for example, $\mathbf{c}=\mathbf{0}_k$ and the \textit{Grand-mean-type} contrast matrix\cite{djira_hothorn_2009} \mbox{$\mathbf{H} := \mathbf{P}_k := \mathbf{I}_k - \mathbf{J}_k/k$.}
Here, $\mathbf{I}_k \in\R^{k\times k}$ represents the unit matrix and \mbox{$\mathbf{J}_k := \mathbf{1}_k\mathbf{1}_k^{\prime} \in\R^{k\times k}$} represents the matrix of ones.
Moreover, by splitting up indices, different kinds of factorial structures can be covered.
For example, in a two-way design with factors A ($a$~levels) and B ($b$ levels), we set $k := a b$ and split up the group index $i$ in two subindices \mbox{$(i_1, i_2)\in\{1,...,a\}\times\{1,...,b\}$.}
Then, hypotheses about no main or interaction effect can be formulated by choosing $\mathbf{c}$ as the zero vector and one of the following contrast matrices, respectively:
\begin{itemize}
    \item  $\mathbf{H}_A := \mathbf{P}_a \otimes (\mathbf{1}_b^{\prime}/b)$ \quad (no main effect of factor A),
    \item $\mathbf{H}_B := (\mathbf{1}_a^{\prime}/a)  \otimes \mathbf{P}_b$ \quad (no main effect of factor B),
    \item $\mathbf{H}_{AB} := \mathbf{P}_a  \otimes \mathbf{P}_b$ \quad (no interaction effect).
\end{itemize}
Here, $\otimes$ represents the Kronecker product. Higher-way designs or hierarchically nested layouts can be incorporated similarly as in Pauly et al.\cite{paulyETAL2015}

\subsection{The Wald-Type Test Statistic and its Asymptotic Behaviour}
\label{ssec:Wald}

In this section, a suitable test statistic for the testing problem (\ref{eq:NullHyp}) is constructed and its asymptotic behaviour is studied.
First of all, let us introduce some notation.
In the following, let $Y_i(x) := \sum\limits_{j=1}^{n_i} \mathbbm{1}\{X_{ij} \geq x\}$ represent the number of individuals at risk just before time $x\geq 0$ and $N_i(x) := \sum_{j=1}^{n_i} \delta_{ij} \mathbbm{1}\{X_{ij} \leq x\}$ denote the number of observed individuals with an event before or at time $x\geq 0$ in group $i$ with $i\in\{1,...,k\}.$
Furthermore, $\widehat{A}_i(x) := \int_{[0,x]} \frac{1}{Y_i} \;\mathrm{d}N_i$ denotes the Nelson-Aalen estimator of the cumulative hazard function $A_i(x) := \int_{[0,x]} \frac{1}{S_{i-}} \;\mathrm{d}F_i = \int_{[0,x]} \frac{1}{y_i} \;\mathrm{d}\nu_i$  at time $x$ with $\nu_i(x):= \int_{[0,x]} G_{i-}\;\mathrm{d}F_i $, $y_i(x) :=  S_{i-}(x) G_{i-}(x)$ and $F_i(x) := 1-S_i(x)$ for all $x\geq 0, i\in\{1,...,k\}$.
Here and throughout, $\Delta M = M - M_-$ denotes the increment and 
$M_-$ denotes the left-continuous version of 
a monotone function $M$.\\
Then, we define the Wald-type test statistic for the testing problem (\ref{eq:NullHyp}) as
\begin{align*}
    W_n(\mathbf{H},\mathbf{c}) := n (\mathbf{H}\widehat{\boldsymbol\mu}-\mathbf{c})^{\prime} (\mathbf{H} \widehat{\boldsymbol{\Sigma}} \mathbf{H}^{\prime})^+ (\mathbf{H}\widehat{\boldsymbol\mu}-\mathbf{c}),
\end{align*}
where $\widehat{\boldsymbol\Sigma} := \text{diag}(\widehat{\sigma}_1^2,...,\widehat{\sigma}_k^2) $ with 
\begin{align}\label{eq:varest}
     \widehat{\sigma}_i^2 := n \int\limits_0^{\tau} \left(\int\limits_x^{\tau} \widehat{S}_i(t) \;\mathrm{d}t \right)^2    \frac{1}{(1-\Delta\widehat{A}_i(x))Y_{i}(x)}  \;\mathrm{d}\widehat{A}_i(x) 
\end{align} being an estimator regarding the asymptotic variance of $\sqrt{n}(\widehat{\mu}_i - \mu_i)$ for all $i\in\{1,...,k\}$.\cite{RMST} Here {and throughout}, we use the convention $0/0 := 0$.
   
The following theorem provides the asymptotic distribution of the Wald-type test statistic.

\begin{theorem}\label{Asy}
    Under the null hypothesis in (\ref{eq:NullHyp}), we have
    \begin{align*}
    W_n(\mathbf{H},\mathbf{c}) \xrightarrow{d} \chi^2_{\text{rank}(\mathbf{H})}
\end{align*}
as $n\to\infty$.
\end{theorem}

Thus, we obtain an asymptotically valid level-$\alpha$-test
\begin{align}\label{eq:test}
    \varphi_n := \mathbbm{1}\{W_n(\mathbf{H},\mathbf{c}) > q_{\text{rank}(\mathbf{H}), 1-\alpha}\},
\end{align}
where $q_{\text{rank}(\mathbf{H}), 1-\alpha}$ denotes the $(1-\alpha)$-quantile of the $\chi^2_{\text{rank}(\mathbf{H})}$ distribution for $\alpha\in (0,1)$.

\subsection{Studentized Permutation Test}\label{ssec:Perm}
For two-sample comparisons, Horiguchi and Uno\cite{Perm} pointed out that RMST-based tests derived from asymptotic methods have an increased type I error.
Hence, we aim to improve the type I error control by extending the studentized permutation approach of Ditzhaus et al.\cite{RMST} to the present general factorial design setting.
In the already treated two-sample case, the approach has the advantage that it also works asymptotically without the assumption of exchangeable data. 
In this section, we will transfer these good properties to general factorial designs to construct a resampling-based test that serves as an alternative for (\ref{eq:test}).

For this purpose, let $(\mathbf{X}, \boldsymbol\delta) := (X_{ij},\delta_{ij})_{j\in\{1,...,n_i\}, i\in\{1,...,k\}}$ denote the observed data and $(\mathbf{X}^{\pi}, \boldsymbol\delta^{\pi}) := (X_{ij}^{\pi},\delta_{ij}^{\pi})_{j\in\{1,...,n_i\}, i\in\{1,...,k\}}$ be the permuted version.
That is, the groups of the original data are randomly shuffled in the sense that the data pairs $(X_{ij},\delta_{ij})$ are permuted. 
In the following, we denote the permutation counterparts of the statistics $\widehat{\boldsymbol\mu}$ and $\widehat{\boldsymbol\Sigma}$ defined in the previous sections with a superscript $\pi$: $\widehat{\boldsymbol\mu}^{\pi}$ and $\widehat{\boldsymbol\Sigma}^{\pi}.$ 
Then, we define the permutation counterpart of the Wald-type test statistic as
\begin{align*}
    W_n^{\pi}(\mathbf{H}) := n (\mathbf{H}\widehat{\boldsymbol\mu}^{\pi})^{\prime} (\mathbf{H} \widehat{\boldsymbol{\Sigma}}^{\pi} \mathbf{H}^{\prime})^+ \mathbf{H}\widehat{\boldsymbol\mu}^{\pi} . 
\end{align*}
{Since we do not have convergence in distribution of this statistic for all observations in the conditional space, let $\xrightarrow{d^{{*}}}$ denote conditional convergence in distribution in probability given the data $(\mathbf{X}, \boldsymbol\delta)$. This means that the conditional distribution converges in probability. Another possibility to explain this convergence is to use another way to state the convergence in distribution via uniform convergence of the conditional distribution function as in the following theorem.
To this end, let $\xrightarrow{P}$ denote convergence in probability.}

\begin{theorem}\label{Perm}
 Under both hypotheses $\mathcal{H}_0$ and $\mathcal{H}_1$, we have
\begin{align}\label{eq:Perm}
    W_n^{\pi}(\mathbf{H}) \xrightarrow{d^{{*}}} \chi^2_{\text{rank}(\mathbf{H})}
\end{align}
as $n\to\infty$. 
Mathematically, (\ref{eq:Perm}) means
    \begin{align*}
        \sup\limits_{z \in\R} \left|\P\left(W_n^{\pi}(\mathbf{H}) \leq z  \mid (\mathbf{X}, \boldsymbol\delta)\right) - \P\left(Z \leq z \right)\right|\xrightarrow{P} 0
    \end{align*} as $n\to\infty$, where $Z\sim \chi^2_{\text{rank}(\mathbf{H})}.$
\end{theorem}

From this result, we can construct a permutation test
\begin{align*}
    \varphi_n^{\pi} := \mathbbm{1}\{W_n(\mathbf{H}, \mathbf{c}) > q_{1-\alpha}^{\pi}\},
\end{align*}
where $q_{1-\alpha}^{\pi}$ denotes the $(1-\alpha)$-quantile of the conditional distribution of $W_n^{\pi}(\mathbf{H})$ given $(\mathbf{X}, \boldsymbol\delta)$.
Lemma~1 of Janssen and Pauls\cite{janssenPauls2003} ensures that $\varphi_n^{\pi}$ is asymptotically valid.

\subsection{Groupwise Bootstrap Test}\label{ssec:gw}
Another possible solution for approximating the limiting distribution is the groupwise bootstrap. 
An advantage over the studentized permutation approach is that the groupwise bootstrap can mimic the different variance structures in the groups. This ensures that the groupwise bootstrap is also applicable for the multiple testing problem, see Section~\ref{sec:Multiple}.\\
For the groupwise bootstrap, the bootstrap observations are drawn randomly with replacement from the observations of the corresponding group, i.e., $(X_{ij}^*,\delta_{ij}^*), j\in\{1,...,n_i\},$ are drawn randomly from the $i$th sample $(X_{ij},\delta_{ij}), j\in\{1,...,n_i\},$ for all $i\in\{1,...,k\}.$ Then, we denote the groupwise bootstrap counterparts of the statistics $\widehat{\boldsymbol\mu}$ and $\widehat{\boldsymbol\Sigma}$ defined in Section~\ref{ssec:Wald} with a superscript~$*$: $\widehat{\boldsymbol\mu}^*$ and $\widehat{\boldsymbol\Sigma}^*$. The groupwise bootstrap test statistic is defined by
\begin{align*}
    W_n^*(\mathbf{H}) := n \left(\mathbf{H}(\widehat{\boldsymbol\mu}^*-\widehat{\boldsymbol\mu})\right)^{\prime} (\mathbf{H} \widehat{\boldsymbol{\Sigma}}^* \mathbf{H}^{\prime})^+ \left(\mathbf{H}(\widehat{\boldsymbol\mu}^*-\widehat{\boldsymbol\mu})\right).
\end{align*}

The following theorem provides the consistency of the groupwise bootstrap.

\begin{theorem}\label{gwBS}
     Under both hypotheses $\mathcal{H}_0$ and $\mathcal{H}_1$, we have
    \begin{align*}
        W_n^*(\mathbf{H}) \xrightarrow{d^{{*}}} \chi^2_{\text{rank}(\mathbf{H})}
    \end{align*} 
    as $n\to\infty$.
\end{theorem}

Hence, we obtain a groupwise bootstrap test
\begin{align*}
    \varphi_n^* := \mathbbm{1}\{ W_n(\mathbf{H}, \mathbf{c}) > q_{1-\alpha}^* \},
\end{align*}
where $q_{1-\alpha}^*$ denotes the $(1-\alpha)$-quantile of the conditional distribution of $W_n^*(\mathbf{H})$ given $(\mathbf{X},\boldsymbol{\delta}).$ By Lemma~1 in Janssen and Pauls\cite{janssenPauls2003}, $\varphi_n^*$ is an asymptotically valid level-$\alpha$ test.

Note that we do not need the property that $\mathbf H$ is a contrast matrix in the proofs of Theorems~\ref{Asy} and \ref{gwBS}. Hence, the groupwise bootstrap test is also valid for general matrices  $\mathbf H \in \R^{r\times k}$ with $\mathrm{rank}(\mathbf{H})>0.$

\section{Multiple Tests}\label{sec:Multiple}
Let us now interpret the contrast matrix $\mathbf{H}$ as a partitionized matrix $\mathbf{H} = [\mathbf{H}_1^{\prime}, ...,  \mathbf{H}_L^{\prime} ]^{\prime}$ with $\mathbf{H}_{\ell} \in \R^{r_{\ell}\times k}$ for all $\ell\in\{1,...,L\}$ such that $\sum_{\ell=1}^L r_{\ell} = r$ and, analogously, $\mathbf{c} = (\mathbf{c}_1^{\prime},...,\mathbf{c}_L^{\prime})^{\prime}$ with $\mathbf{c}_{\ell} \in \R^{r_{\ell}}$ for all $\ell\in\{1,...,L\}$. 
Moreover, we assume $\mathrm{rank}(\mathbf{H}_{\ell}) >0$ for all $\ell\in\{1,...,L\}$. 
In this section,
we aim to construct a testing procedure for the multiple testing problem with null and alternative hypotheses
\begin{align}\label{eq:MultipleHypos}
    \mathcal{H}_{0,\ell}: \mathbf{H}_{\ell} \boldsymbol\mu = \mathbf{c}_{\ell} \quad \text{vs.} \quad \mathcal{H}_{1,\ell}: \mathbf{H}_{\ell} \boldsymbol\mu \neq \mathbf{c}_{\ell}, \qquad \text{for } \ell\in\{1,...,L\}.
\end{align}
Thereby, we aim to incorporate the asymptotically exact dependence structure between the test statistics of the $L$ local tests to gain more power than, for example, by using a Bonferroni-correction.

\begin{ex}\label{example}
A global null hypothesis which is of interest in many applications is the equality of the RMSTs, i.e., $\mathcal{H}_0: \mu_1 = ... = \mu_k$ versus the alternative \mbox{$\mathcal{H}_1: \mu_{i_1} \neq   \mu_{i_2}$} for some $i_1,i_2 \in\{1,...,k\}$. 
However, there are different possible choices of the contrast matrix $\mathbf{H}$ which lead to this global null hypothesis.\cite{konietschke2013}
A popular choice is the Grand-mean-type contrast matrix as introduced in Section~\ref{ssec:Setup}, where the RMSTs of the different groups are compared with the overall mean of the RMSTs $\overline{\mu} := \frac{1}{k}\sum_{i=1}^k \mu_i$ for the different contrasts, respectively.
 Many-to-one comparisons can be considered by choosing the \textit{Dunnett-type} contrast matrix\cite{dunnett_1955}
\begin{align*}
    \mathbf{H} = [-\mathbf{1}_{k-1}, \mathbf{I}_{k-1}]   = \begin{bmatrix}
-1 & 1 & 0 & \cdots & 0 \\
-1 & 0 & 1 & \cdots & 0 \\
\vdots  & \vdots  & \ddots & \vdots  \\
-1 & 0 & 0 & \cdots  & 1
\end{bmatrix} \in \R^{(k-1) \times k}
\end{align*} 
 and $\mathbf{c} = \mathbf{0}_{k-1}$, where the RMSTs $\mu_2, ..., \mu_k$ are compared to the RMST $\mu_1$ of the first group regarding the different contrasts.
In order to compare all pairs of RMSTs $\mu_{i_1},\mu_{i_2}, i_1,i_2 \in\{1,...,k\}$ with $i_1 \neq i_2$, the \textit{Tukey-type} contrast matrix\cite{tukey}
\begin{align*}
    \mathbf{H} = \begin{bmatrix}
-1 & 1 & 0 & 0 & \cdots & \cdots & 0 \\
-1 & 0 & 1 & 0 &\cdots & \cdots & 0 \\
\vdots  & \vdots &\vdots & \vdots & \ddots & \vdots & \vdots  \\
-1 & 0 & 0 & 0& \cdots & \cdots & 1\\
0 & -1 & 1 & 0& \cdots & \cdots & 0 \\
0 & -1 & 0 & 1& \cdots & \cdots & 0 \\
\vdots  & \vdots & \vdots  & \vdots & \ddots & \vdots & \vdots \\
0 & 0 & 0 & 0 & \cdots & -1 & 1
\end{bmatrix} \in \R^{k(k-1)/2 \times k}
\end{align*} 
 and $\mathbf{c} = \mathbf{0}_{k(k-1)/2}$ can be used.
An overview of different contrast tests can be found in Bretz et al.\cite{bretz2001}\\
Furthermore, the choice of the considered partition of the matrix $\mathbf{H}= [\mathbf{H}_1^{\prime}, ..., \mathbf{H}_R^{\prime}]^{\prime}$ and, therefore, the resulting local hypotheses depend on the question of interest.
This general formulation of the multiple testing problem covers the post-hoc testing problem and includes, for example, the local null hypotheses 
$\mathcal H_{0,\ell} : \: {\mu}_{\ell} = \overline{{\mu}}$, for $\ell \in \{1,\ldots, k\},$ by choosing \mbox{$\mathbf{H}_{\ell} = \mathbf{e}_{\ell}^{\prime} - \frac{1}{k}\mathbf{1}_k^{\prime}$} for all $\ell\in\{1,...,k\}$, where $\mathbf{e}_{\ell}\in\R^k$ denotes the $\ell$th unit vector. Analogously, we can perform many-to-one comparisons and all-pair comparisons of the mean functions simultaneously by considering the $r$ rows of the Dunnett-type and Tukey-type contrast matrix, respectively, as blocks $\mathbf{H}_1,...,\mathbf{H}_r$.\\
Furthermore, the formulation of this testing problem allows to perform multiple tests with more than one contrast matrix simultanously. In a two-way design, we may choose $\mathbf{H}_1 = \mathbf{H}_A, \mathbf{H}_2 = \mathbf{H}_B$ and $\mathbf{H}_3 = \mathbf{H}_{AB}$ as introduced in Section~\ref{ssec:Setup}, for example. 
This allows for simultaneous testing of the factors A and B and their interaction.
\end{ex}

For all local hypotheses in (\ref{eq:MultipleHypos}), we can calculate the Wald-type test statistics $W_n(\mathbf{H}_{\ell}, \mathbf{c}_{\ell}), \ell\in\{1,...,L\}$. Since we aim to use the asymptotically exact dependence structure of the test statistics, we have to investigate the joint asymptotic behavior.

Therefore, let $\mathbf{Z} \sim \mathcal{N}_k(\mathbf{0}_k, \boldsymbol\Sigma)$ with $\boldsymbol\Sigma := \text{diag}(\sigma_1^2,...,\sigma_k^2)$ in the following, where here and throughout
\begin{align*}
    \sigma_i^2 := \frac{1}{\kappa_i} \int\limits_0^{\tau} \left(\int\limits_x^{\tau} {S}_i(t) \;\mathrm{d}t \right)^2
    \frac{1}{(1-\Delta{A}_i(x))y_i(x)}  \;\mathrm{d}{A}_i(x), \quad i\in\{1,...,k\}.
\end{align*} 
In Section~S.5 of the supplement of Ditzhaus et al.\cite{RMST}, it is shown that $\sigma_i^2$ is the almost sure limit of (\ref{eq:varest}) for all $i\in\{1,...,k\}$.

\begin{theorem}\label{MultiAsy}
Under the null hypotheses (\ref{eq:MultipleHypos}), we have
    \begin{align}\label{eq:MultiAsymptotic}
\begin{split}
    (W_n(\mathbf{H}_{\ell},\mathbf{c}_{\ell}))_{\ell\in\{1,...,L\}}
    &=  \left( n \left(\mathbf{H}_{\ell} (\widehat{\boldsymbol\mu} - \boldsymbol\mu )\right)^{\prime} ( \mathbf{H}_{\ell}\widehat{\boldsymbol{\Sigma}} \mathbf{H}_{\ell}^{\prime} )^+ \mathbf{H}_{\ell}(\widehat{\boldsymbol\mu} - \boldsymbol\mu )    \right)_{\ell\in\{1,...,L\}}
    \\&\xrightarrow{d}
      \left( (\mathbf{H}_{\ell} \mathbf{Z})^{\prime} ( \mathbf{H}_{\ell}\boldsymbol{\Sigma} \mathbf{H}_{\ell}^{\prime}  )^+ \mathbf{H}_{\ell}\mathbf{Z}    \right)_{\ell\in\{1,...,L\}}
      \end{split}
\end{align}
as $n\to\infty$.
\end{theorem}

Note that $\boldsymbol\Sigma$ is generally unknown such that we do not know the exact asymptotic joint limiting distribution of $(W_n(\mathbf{H}_1,\mathbf{c}_1),...,W_n(\mathbf{H}_L,\mathbf{c}_L))$. 

Unfortunately, we cannot use the studentized permutation approach for approximating the joint limiting distribution. That is because 
\begin{align}\label{eq:PermLimitingDistr}
    {(W_n^{\pi}(\mathbf{H}_{\ell}))_{\ell\in\{1,...,L\}} \xrightarrow{d^*}} \left( (\mathbf{H}_{\ell} \mathbf{Z}^{\pi})^{\prime} ( \mathbf{H}_{\ell}\boldsymbol{\Sigma}^{\pi} \mathbf{H}_{\ell}^{\prime}  )^+ \mathbf{H}_{\ell}\mathbf{Z}^{\pi}    \right)_{\ell\in\{1,...,L\}}
\end{align}
as $n\to\infty$ {holds} 
similarly as in the proof of Theorem~\ref{Perm}, where $\mathbf{Z}^{\pi}\sim\mathcal{N}_k(\mathbf{0}_k, \boldsymbol{\Sigma}^{\pi})$. Since {the limiting distributions in} (\ref{eq:PermLimitingDistr}) and 
(\ref{eq:MultiAsymptotic}) are generally not equal in distribution, the studentized permutation approach is not consistent for the multiple testing problem.
\\
However, we can approximate the critical values via the groupwise bootstrap as introduced above. The difference here is that the covariance structures of the groups are not altered since the bootstrap observations are drawn within each group.
The asymptotic validity is guaranteed by the following theorem.

\begin{theorem}\label{Multigw}
    Under all hypotheses $\mathcal{H}_{0,\ell}$ and $\mathcal{H}_{1,\ell}$, we have
    \begin{align*}
       (W_n^*(\mathbf{H}_{\ell}))_{\ell\in\{1,...,L\}} \xrightarrow{d^{{*}}}  \left( (\mathbf{H}_{\ell} \mathbf{Z})^{\prime} ( \mathbf{H}_{\ell}\boldsymbol{\Sigma} \mathbf{H}_{\ell}^{\prime}  )^+ \mathbf{H}_{\ell}\mathbf{Z}    \right)_{\ell\in\{1,...,L\}}
    \end{align*} 
    as $n\to\infty$. 
\end{theorem}

 A naive approach for compatible local and global test decisions would be to calculate a critical value for the maximum statistic $\max\limits_{\ell\in\{1,...,L\}} W_n(\mathbf{H}_{\ell},\mathbf{c}_{\ell})$ in view of the global hypothesis (\ref{eq:NullHyp}). A local hypothesis in (\ref{eq:MultipleHypos}) is rejected whenever the corresponding local test statistic exceeds the critical value. In the special case that $\text{rank}(\mathbf{H}_{\ell})$ are equal for all $\ell\in\{1,...,L\}$, the limiting distributions of the Wald-type test statistics are equal, as we have seen in Section~\ref{ssec:Wald}. Then, the maximum statistic can be used for testing the global hypothesis and every contrast is treated in the same way. \\
However, if the ranks 
are not equal, the limiting distributions of the Wald-type test statistics are not equal; cf.\ Section~\ref{ssec:Wald}. 
Hence, the contrasts are not treated in the same way by considering the maximum statistic. 
Thus, we adopt the idea for the construction of simultaneous confidence bands proposed by Bühlmann\cite{Buehlmann}.


To this end, let $(W_n^{*, b}(\mathbf{H}_1),...,W_n^{*, b}(\mathbf{H}_L)), b\in\{1,...,B\},$ denote $B$ groupwise bootstrap test statistics. For each $b\in\{1,...,B\}$, the same bootstrap samples are used for calculating the groupwise bootstrap counterparts $(W_n^{*, b}(\mathbf{H}_1),...,W_n^{*, b}(\mathbf{H}_L))$ for the different contrasts.
This reflects the real world situation that the same original samples are used for testing all local hypotheses.
Let $q_{\ell,1-\beta}^{*}$ denote the $(1-\beta)$-quantile of $W_n^{*, b}(\mathbf{H}_{\ell}), b\in\{1,...,B\},$ for all \mbox{$\ell\in\{1,...,L\}$.}
Our strategy is to adjust the local level $\beta$ such that the level $\alpha$ is controlled globally.
To this end, we let
\begin{align*}
    \mathrm{FWER}_n^*(\beta) := \frac{1}{B} \sum\limits_{b=1}^B \mathbbm{1}\left\{ \exists \ell\in\{ 1,...,L \} : W_n^{*, b}(\mathbf{H}_{\ell}) > q_{\ell,1-\beta}^{*} \right\}
\end{align*} denote the estimated family-wise type I error rate by using the $(1-\beta)$-quantiles as critical values for all $\beta\in [0,1].$ Then, we define the local level ${\beta}_n(\alpha)$ as the largest value such that the family-wise type I error rate is bounded by the level of significance $\alpha$, i.e.,
\begin{align*}
    {\beta}_n(\alpha) := \max\left\{ \beta \in \left\{ 0, \frac{1}{B}, ..., \frac{B-1}{B}  \right\} \mid  \mathrm{FWER}_n^*(\beta) \leq \alpha \right\}.
\end{align*}
Note that we only have to consider $\beta \in \left\{ 0, \frac{1}{B}, ..., \frac{B-1}{B}  \right\}$ since the quantiles can only take $B$ different values, respectively.
Additionally, we only have to search for ${\beta}_n(\alpha)$ within the interval $\left[\frac{1}{B}\left\lfloor\frac{B\alpha}{L}\right\rfloor, \frac{B-1}{B}\right]$.
The lower bound can be interpreted as Bonferroni bound and results from the following inequalities:
\begin{align*}
    \mathrm{FWER}_n^*\left(\frac{1}{B}\left\lfloor\frac{B\alpha}{L}\right\rfloor\right)
    \leq \sum\limits_{\ell=1}^L  \frac{1}{B} \sum\limits_{b=1}^B\mathbbm{1}\left\{  W_n^{*, b}(\mathbf{H}_{\ell}) > q_{\ell,1-\frac{1}{B}\left\lfloor\frac{B\alpha}{L}\right\rfloor}^{*} \right\}
    \leq L \frac{1}{B}\left\lfloor\frac{B\alpha}{L}\right\rfloor \leq \alpha.
\end{align*}
 
The decision rules are constructed as follows:
\begin{itemize}
	\item For each $\ell\in\{1,...,L\}$, we reject $\mathcal H_{0,\ell}$ in (\ref{eq:MultipleHypos}) if and only if $W_n(\mathbf{H}_{\ell},\mathbf{c}_{\ell})>q_{\ell,1-{\beta}_n(\alpha)}^{*}$ or, equivalently, ${W_n(\mathbf{H}_{\ell},\mathbf{c}_{\ell})}/{q_{\ell,1-{\beta}_n(\alpha)}^{*}}>1$. Here, we set ${0}/{0} := 0$.
	\item We reject the global null hypothesis $\mathcal H_0$ in (\ref{eq:NullHyp}) whenever at least one of the hypotheses $\mathcal H_{0,1},...,\mathcal H_{0,L}$ is rejected. Hence, we reject the global null hypothesis $\mathcal H_0$ in (\ref{eq:NullHyp}) if and only if
	\begin{align*}
		\max_{\ell \in\{ 1, \ldots,L\}} \frac{W_n(\mathbf{H}_{\ell},\mathbf{c}_{\ell})}{q_{\ell,1-{\beta}_n(\alpha)}^{*}} > 1.
	\end{align*}
\end{itemize}
Each test statistic $W_n(\mathbf{H}_{\ell},\mathbf{c}_{\ell}), \ell\in\{1,...,L\},$ is treated in the same way and has the same impact since we use the same local level of significance ${\beta}_n(\alpha)$ for each contrast. Moreover, the 
following theorem provides
that the level of significance of the global test and the family-wise type I error rate for the multiple testing problem is controlled asymptotically. 

\begin{theorem}\label{gwTest}
   Let $\T \subset \{1,...,L\}$ denote the subset of true hypotheses, i.e., let $\mathcal H_{0,\ell}, \ell\in\T,$ in (\ref{eq:MultipleHypos}) be true.
   With 
    $B=B(n) \to \infty$
    as $n\to\infty$, we have 
$$ \lim\limits_{n\to\infty} \P \left( \exists {\ell \in\T}: \: W_n(\mathbf{H}_{\ell},\mathbf{c}_{\ell}) >q_{\ell,1-{\beta}_n(\alpha)}^{*} \right) \leq \alpha .$$
    The inequality becomes an equality if $\T = \{1,...,L\}$.
\end{theorem}


\paragraph{Adjusted p-values}
   The method described above for constructing multiple test decisions is accompanied by an adjusting of p-values. To see this, let $w_n(\mathbf{H}_{\ell}, \mathbf{c}_{\ell} )$ be the realization of $W_n(\mathbf{H}_{\ell}, \mathbf{c}_{\ell} )$ for all $\ell\in\{1,...,L\}$. \\
   Firstly, we determine the local p-values by
   $$ \beta_{n,\ell} := \frac{1}{B} \sum\limits_{b=1}^B \mathbbm{1}\left\{ W_n^{*, b}(\mathbf{H}_{\ell})  \geq w_n(\mathbf{H}_{\ell}, \mathbf{c}_{\ell} ) \right\}
   $$ for all $\ell\in\{1,...,L\}$.
   Comparing the local p-values to $\beta_n(\alpha)$ yields multiple test decisions that are consistent to the method described above. 
   Translating this comparison to a comparison with the level of significance $\alpha$ is intuitive due to the definition of $\beta_n(\alpha) $.
   Hence, by plugging the local p-value in $\mathrm{FWER}_n^*$, the adjusted p-value for the $\ell$th hypothesis can be defined by
   $$p_{\ell} := \mathrm{FWER}_n^*\left(\beta_{n,\ell}\right) $$ for all $\ell\in\{1,...,L\}$ and the global p-value by $p := \min\{p_1,...,p_L\}$. 

   The following proposition ensures that the test decisions based on these p-values are unchanged.\\
   
   \begin{proposition}\label{pValues}
   \mbox{} 
       \begin{itemize}
       \item[(i)] For each $\ell\in\{1,...,L\}$, it holds $p_{\ell} \leq \alpha$ whenever $w_n(\mathbf{H}_{\ell}, \mathbf{c}_{\ell}) > q_{l,1-{\beta_n}(\alpha)}^{*}$,
       \item[(ii)] it holds $p \leq \alpha$ whenever $\max\limits_{\ell \in\{ 1, \ldots,L\}} {w_n(\mathbf{H}_{\ell},\mathbf{c}_{\ell})}/{q_{\ell,1-{\beta_n}(\alpha)}^{*}} > 1$.
   \end{itemize}
   \end{proposition}

\paragraph{Simultaneous confidence regions and intervals}
   Furthermore, we can use the constructed multiple testing procedure for defining  simultaneous confidence regions for $\mathbf{H}_{\ell}\boldsymbol{\mu}$ with asymptotic global confidence level $1-\alpha$. Therefore, we define the $\ell$th confidence region as
   $$ CR_{n,\ell} := \left\{ \boldsymbol\xi \in \R^{r_{\ell}} \mid   W_n(\mathbf{H}_{\ell},\boldsymbol\xi) \leq q_{\ell,1-{\beta_n}(\alpha)}^{*} \right\} $$ for all $\ell\in\{1,...,L\}$.
It can be easily checked that
$\P( \mathbf{H}\boldsymbol{\mu} \in \otimes_{\ell=1}^L CR_{n,\ell} ) \to 1 - \alpha$ as $n\to\infty$.\\
In the case that $\mathbf{H}_{\ell} \in \R^{1\times k}$, i.e., $r_{\ell}=1$, we can simplify the confidence regions to confidence intervals $CR_{n,\ell} := [L_{n,\ell}(\alpha/2), U_{n,\ell}(\alpha/2)]$ by solving the equation $W_n(\mathbf{H}_{\ell},\xi) \leq q_{\ell,1-{\beta_n}(\alpha)}^{*}$ for $\xi\in\mathbb{R}$. This yields
$$L_{n,\ell}(\alpha/2) := \mathbf{H}_{\ell}\widehat{\boldsymbol{\mu}} - \frac{\sqrt{\mathbf{H}_{\ell}\widehat{\boldsymbol\Sigma}\mathbf{H}_{\ell}^{\prime}}}{\sqrt{n}}\sqrt{q_{\ell,1-{\beta_n}(\alpha)}^{*}} \quad\text{ and }\quad U_{n,\ell} (\alpha/2) := \mathbf{H}_{\ell}\widehat{\boldsymbol{\mu}} + \frac{\sqrt{\mathbf{H}_{\ell}\widehat{\boldsymbol\Sigma}\mathbf{H}_{\ell}^{\prime}}}{\sqrt{n}}\sqrt{q_{\ell,1-{\beta_n}(\alpha)}^{*}}.$$

\paragraph{Simultaneous non-inferiority and equivalence tests}
Let us consider again the case \mbox{$r_{\ell}=1$} for a\textit{}ll $\ell \in \{1,...,L\}.$
In this special case, we write $c_{\ell}$ instead of $\mathbf{c}_\ell$ in non-bold type for all $\ell\in\{1,...,L\}$.
Based on the previous constructed confidence intervals, we can also define simultaneous non-inferiority and equivalence tests by using the \textit{two one-sided test procedure}:\cite{TOST} let $\epsilon_1, ..., \epsilon_L > 0$ be prespecified equivalence bounds; the hypotheses of interest are
\begin{align}\label{eq:noninf}
    \mathcal{H}_{0,\ell}^i: \mathbf{H}_{\ell}\boldsymbol{\mu} -  {c}_{\ell} \geq \epsilon_{\ell}
   \quad \text{vs.} \quad \mathcal{H}_{1,\ell}^i: \mathbf{H}_{\ell}\boldsymbol{\mu} -  {c}_{\ell} < \epsilon_{\ell}, \qquad \text{for } \ell\in\{1,...,L\}
\end{align} for the non-inferiority testing problem and
\begin{align}\label{eq:equivalence}
    \mathcal{H}_{0,\ell}^e: |\mathbf{H}_{\ell}\boldsymbol{\mu} -  {c}_{\ell}| \geq \epsilon_{\ell}
   \quad \text{vs.} \quad \mathcal{H}_{1,\ell}^e: |\mathbf{H}_{\ell}\boldsymbol{\mu} -  {c}_{\ell}| < \epsilon_{\ell}, \qquad \text{for } \ell\in\{1,...,L\}
\end{align}
for the equivalence testing problem.

For each $\ell\in\{1,...,L\}$, we reject $\mathcal{H}_{0,\ell}^i$ in (\ref{eq:noninf}) if and only if $U_{n,\ell}(\alpha) - {c}_{\ell} < \epsilon_{\ell}$. Furthermore, for each $\ell\in\{1,...,L\}$, we reject $\mathcal{H}_{0,\ell}^e$ in (\ref{eq:equivalence}) if and only if $$U_{n,\ell}(\alpha) - {c}_{\ell} < \epsilon_{\ell} \quad\text{ and }\quad L_{n,\ell}(\alpha) - {c}_{\ell} > -\epsilon_{\ell}.$$

\paragraph{Stepwise extension}
For gaining more power, our methodologies can be combined with the closed testing procedure as in Blanche et al.\cite{stepwise} if only multiple test decisions but not the construction of (simultaneous) confidence regions are of interest: for each $\ell\in\{1,...,L\}$, the hypothesis $\mathcal{H}_{0,\ell}$ in (\ref{eq:MultipleHypos}) is rejected at level $\alpha$ if and only if for each $\mathcal{J} \ni \ell$ the intersection hypothesis $\mathcal{H}_{0,\mathcal{J}} := \bigcap\limits_{j\in\mathcal{J}} \mathcal{H}_{0,j}$ is rejected at level $\alpha$. For testing an intersection hypothesis  $\mathcal{H}_{0,\mathcal{J}}$, we can use the procedure as described above. To be specific, $\mathcal{H}_{0,\mathcal{J}}$ is rejected at level $\alpha$ whenever 
\begin{align*}
		\max_{j \in\mathcal{J}} \frac{W_n(\mathbf{H}_{j},\mathbf{c}_{j})}{q_{j,1-{\beta}_{n,\mathcal{J}}(\alpha)}^{*}} > 1
\end{align*} holds, where
$$
{\beta}_{n,\mathcal{J}}(\alpha) := \max\left\{ \beta \in \left\{ 0, \frac{1}{B}, ..., \frac{B-1}{B}  \right\} \mid  \frac{1}{B} \sum\limits_{b=1}^B \mathbbm{1}\left\{ \exists j\in\mathcal{J} : W_n^{*, b}(\mathbf{H}_{j}) > q_{j,1-\beta}^{*} \right\} \leq \alpha \right\}
$$ for all $\mathcal{J} \subset \{1,...,L\}$.


\section{Simulation Study}\label{sec:Simu}
\begin{figure}[htb]
    \centering
    \includegraphics[height=6cm]{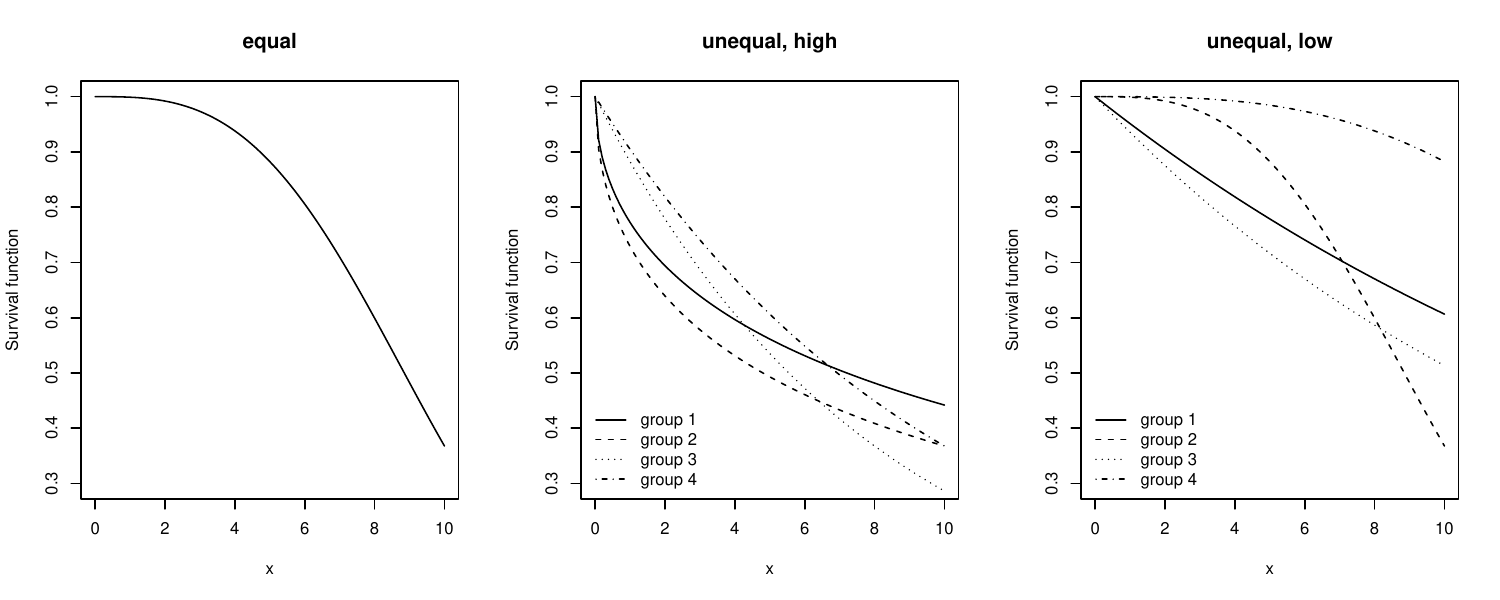}
    \vspace{-0.5cm}\caption{Survival functions of the censoring times.}
    \label{fig:cens}
\end{figure}

For analyzing the small sample performance of our proposed methods, we conducted an extensive simulation study by using the computing environment R, version 4.2.1.\cite{R}
{\subsection{Simulation Setup}\label{ssec:SimuSetup}}
The simulation setup is based on Ditzhaus et al.\cite{RMST}
We simulated a factorial design with $k=4$ groups and utilized the three different contrast matrices introduced in Example~\ref{example}: the Dunnett-type, Tukey-type and Grand-mean-type contrast matrix.
Here, the local hypotheses were constructed by the rows of the contrast matrix, i.e., the blocks $\mathbf{H}_1,...,\mathbf{H}_R$ correspond to the rows of $\mathbf{H}$.\\
The survival times were always drawn from the same distribution for the first three groups. However, the survival distribution of the fourth group may differ.
As in Ditzhaus et al.\cite{RMST}, the data were generated from the following survival distributions:
\begin{itemize}
    \item Exponential distributions and early departures (\textit{exp early}): $T_{11},T_{21}, T_{31} \sim Exp(0.2)$ and $T_{41}$ with piece-wise constant hazard function $t\mapsto \lambda_{\delta,1} \cdot \mathbbm{1}\{t\leq 2\} + 0.2 \cdot \mathbbm{1}\{t > 2\}$,
    \item exponential distributions and late departures (\textit{exp late}): $T_{11},T_{21}, T_{31} \sim Exp(0.2)$ and $T_{41}$ with piece-wise constant hazard function $t\mapsto 0.2 \cdot \mathbbm{1}\{t\leq 2\} + \lambda_{\delta,2} \cdot \mathbbm{1}\{t > 2\}$,
    \item exponential distributions and proportional hazard alternative (\textit{exp prop}): $T_{11},T_{21}, T_{31} \sim Exp(0.2)$ and $T_{41} \sim Exp(\lambda_{\delta,3})$,
    \item lognormal distributions with scale alternatives (\textit{logn}): $T_{11},T_{21}, T_{31} \sim logN(2,0.25)$ and $T_{41}\sim logN(\lambda_{\delta,4},0.25)$,
    \item exponential distributions and piece-wise exponential distributions (\textit{pwExp}): $T_{11},T_{21}, T_{31} \sim Exp(0.2)$ and $T_{41}$ with piece-wise constant hazard function $t\mapsto 0.5\cdot \mathbbm{1}\{t\leq \lambda_{\delta,5} \} + 0.05 \cdot \mathbbm{1}\{t > \lambda_{\delta,5} \}$,
    \item Weibull distributions and late departures (\textit{Weib late}): $T_{11},T_{21}, T_{31} \sim Weib(3,8)$ and $T_{41}\sim Weib(3\cdot\lambda_{\delta,6},8/\lambda_{\delta,6})$,
    \item Weibull distributions and proportional hazard alternative (\textit{Weib prop}): $T_{11},T_{21}, T_{31} \sim Weib(3,8)$ and $T_{41}\sim Weib(3,\lambda_{\delta,7})$,
    \item Weibull distributions with crossing curves and scale alternatives (\textit{Weib scale}): $T_{11},T_{21}, T_{31} \sim Weib(3,8)$ and $T_{41}\sim Weib(1.5,\lambda_{\delta,8})$,
    \item Weibull distributions with crossing curves and shape alternatives (\textit{Weib shape}): $T_{11},T_{21}, T_{31} \sim Weib(3,8)$ and $T_{41}\sim Weib(\lambda_{\delta,9},14)$.    
\end{itemize}
Here, the parameters $\lambda_{\delta,1},...,\lambda_{\delta,9}$ were determined such that the RMST difference equals $\delta = \mu_1 - \mu_4$. This difference was set to $\delta = 0$ for simulating under the null and to $\delta = 1.5$ for simulating under the alternative hypothesis.
\\
Under the null hypothesis, note that the scenarios \textit{exp early}, \textit{exp late} and \textit{exp prop} as well as \textit{Weib late} and \textit{Weib prop} are respectively equal. Thus, we only included the results for these scenarios once in the figures and tables, respectively, by calculating the mean over the results whenever they differ.\\
For the censoring times, we chose the following three scenarios:
\begin{itemize}
    \item Equally Weibull distributed censoring times (\textit{equal}): $C_{11}, C_{21}, C_{31}, C_{41} \sim Weib(3,10)$,
    \item unequally Weibull distributed censoring times with high censoring rates (\textit{unequal, high}): $C_{11} \sim Weib(0.5,15), C_{21}\sim Weib(0.5,10), C_{31}\sim Weib(1,8)$ and $C_{41} \sim Weib(1,10)$,
    \item unequally Weibull distributed censoring times with low censoring rates (\textit{unequal, low}): $C_{11} \sim Weib(1,20), C_{21}\sim Weib(3,10), C_{31}\sim Weib(1,15)$ and $C_{41} \sim Weib(3,20)$.
\end{itemize}
The survival functions of these censoring times are illustrated in Figure~\ref{fig:cens}.
The resulting censoring rates of the different groups are presented in Table~\ref{tab:cens} in the supplement. The censoring rates ranged from 20\% up to 60\% in groups 1-3 and from 1\% up to 57\% in group 4. \\ 
We considered \textit{balanced} and \textit{unbalanced} designs with sample sizes $\mathbf{n}=(n_1,n_2,n_3,n_4)=K\cdot (15,15,15,15)$ and $\mathbf{n}=K\cdot (10,20,10,20)$, where $K\in\{1,2,4\}$ for \textit{small, medium} and \textit{large} samples.\\
Furthermore, $N_{sim} = 5000$ simulation runs with $B = 1999$ resampling iterations were generated. The level of significance was set to $\alpha = 5\%$ and the upper integration bound to $\tau = 10.$\\
{
The following methods were compared:
\begin{itemize}
\item \textit{asymptotic\_global}: The global Wald-type test as in Section~\ref{ssec:Wald},
\item \textit{permutation}: The global studentized permutation test as in Section~\ref{ssec:Perm},
\item \textit{asymptotic}: Multiple Wald-type tests, where the multivariate limit distribution in Theorem~\ref{MultiAsy} is approximated by using the estimator for $\boldsymbol\Sigma$ as defined in Section~\ref{ssec:Wald} and, then, applying the multiple testing procedure of Section~\ref{sec:Multiple},
\item \textit{wild, Rademacher}; \textit{wild, Gaussian}: Multiple wild bootstrap\cite{lin,parzen,wild_ursprung} tests as in Section~\ref{sec:Wild} in the supplement with Rademacher and Gaussian multipliers, respectively, by applying the muliple testing procedure of Section~\ref{sec:Multiple},
\item \textit{groupwise}: The multiple groupwise bootstrap test as in Section~\ref{sec:Multiple},
\item \textit{asymptotic\_bonf}: Global Wald-type tests as in Section~\ref{ssec:Wald} adjusted with the Bonferroni-correction,
\item \textit{permutation\_bonf}: Global studentized permutation tests as in Section~\ref{ssec:Perm} adjusted with the Bonferroni-correction.
\end{itemize}
Clearly, the first two methods (\textit{asymptotic\_global}, \textit{permutation}) can only be compared to multiple testing procedures for the global testing problem. However, by using a Bonferroni-correction (\textit{asymptotic\_bonf}, \textit{permutation\_bonf}), we can also obtain test decisions for the local hypotheses.}
\FloatBarrier
\subsection{{Simulation Results under the Null Hypothesis}}\label{ssec:SimuTypeI}
\begin{figure}[tb]
    \centering
    \includegraphics[scale=0.7]{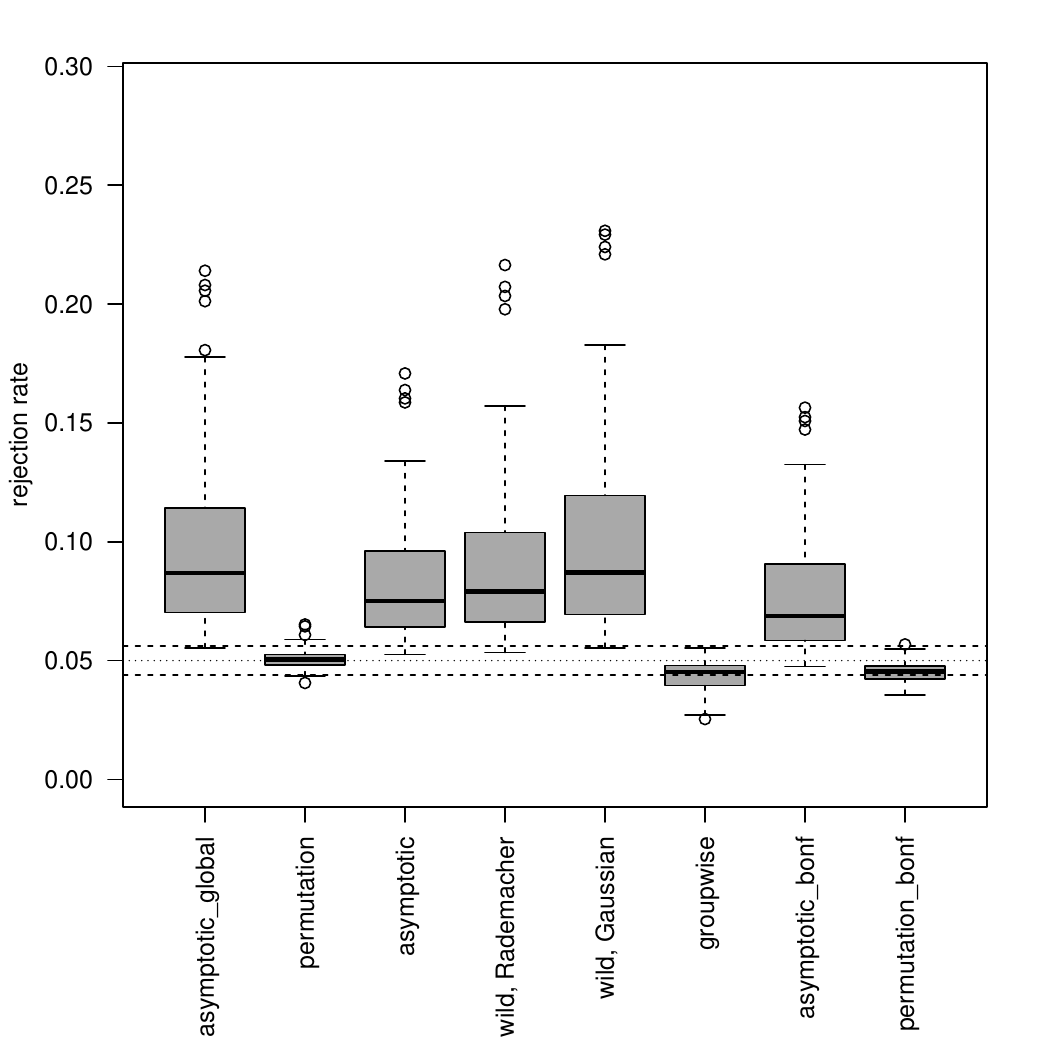}
    \caption{Rejection rates under $\mathcal{H}_0$  over all settings for the Dunnett-type contrast matrix. The dashed lines represent the borders of the binomial confidence interval $[4.4\%, 5.62\%]$.}
    \label{fig:Dunnett_rejection_rate}
\end{figure}

\begin{figure}[t]
    \centering
    \includegraphics[scale=0.7]{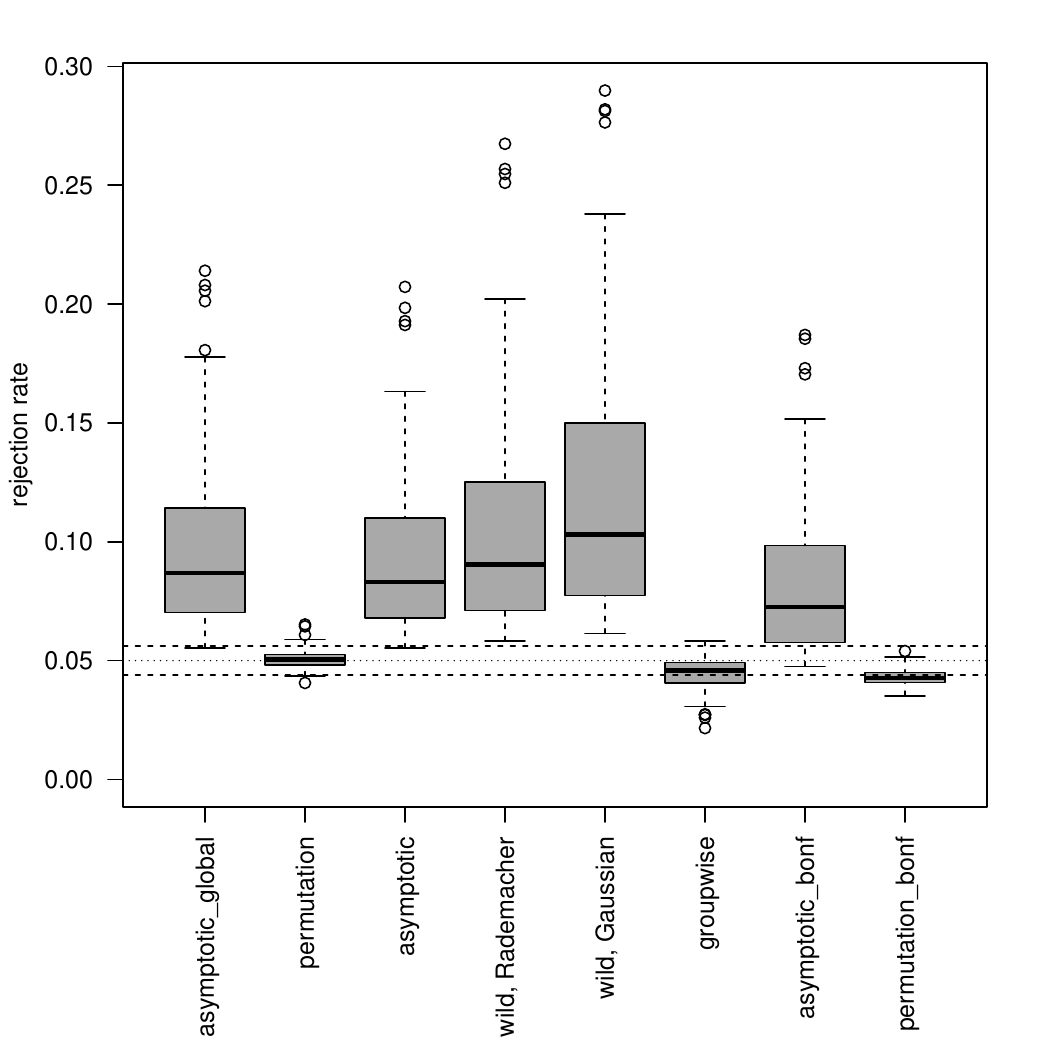}
    \caption{Rejection rates under $\mathcal{H}_0$  over all settings for the Tukey-type contrast matrix. The dashed lines represent the borders of the binomial confidence interval $[4.4\%, 5.62\%]$.}
    \label{fig:Tukey_rejection_rate}
\end{figure}

\begin{figure}[t]
    \centering
    \includegraphics[scale=0.7]{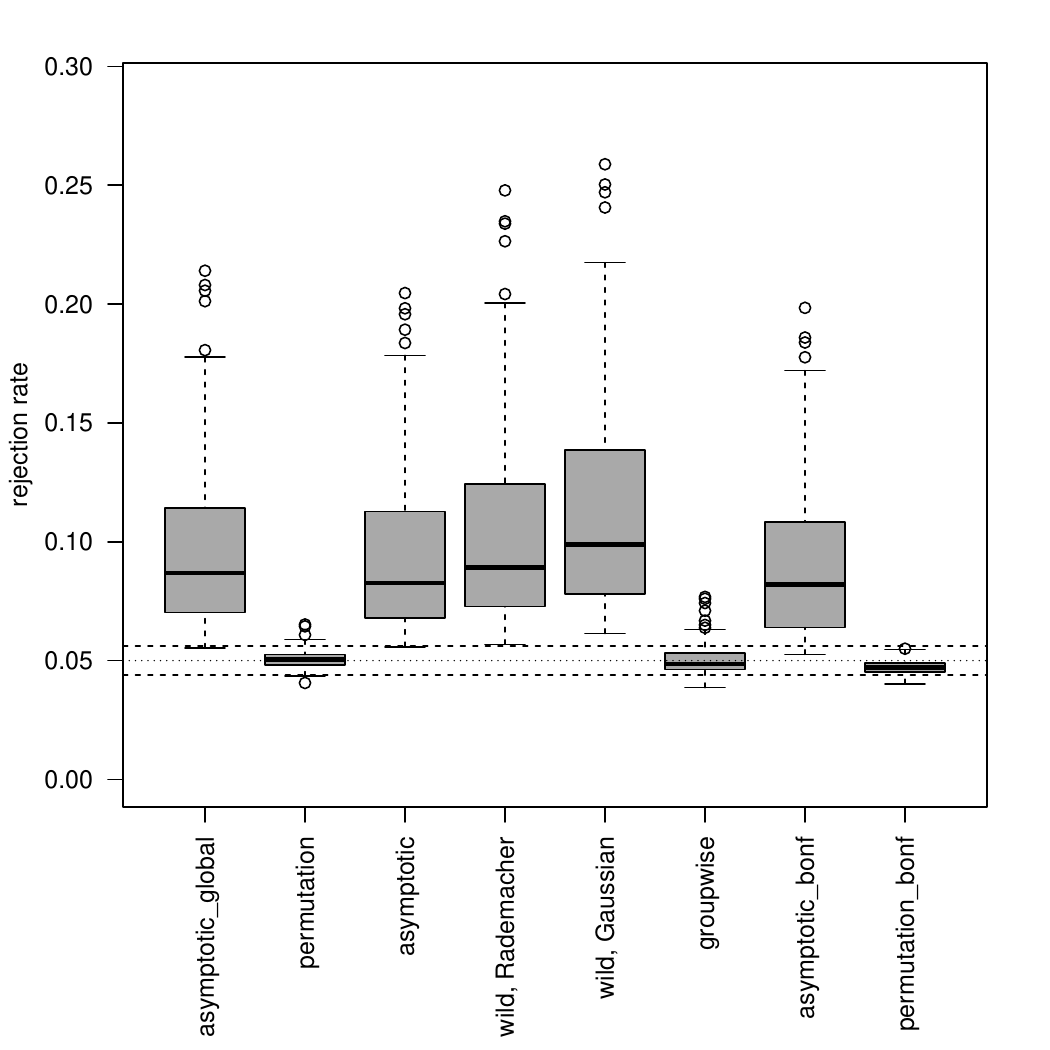}
    \caption{Rejection rates under $\mathcal{H}_0$  over all settings for the Grand-mean-type contrast matrix. The dashed lines represent the borders of the binomial confidence interval $[4.4\%, 5.62\%]$.}
    \label{fig:GrandMean_rejection_rate}
\end{figure}

Figures~\ref{fig:Dunnett_rejection_rate} to \ref{fig:GrandMean_rejection_rate} under $\mathcal{H}_0$ illustrate the global rejection rates, which coincide with the family wise error rates for the multiple tests, over all settings for the different contrast matrices. Here, the dotted line represents the $\alpha$-level of $5\%$ and the dashed lines represent the borders of the binomial confidence interval $[4.4\%, 5.62\%]$. 
\\ In all figures, one can see that only the permutation approach and the groupwise bootstrap seem to perform well over all simulation settings. Here, the permutation approach yields slightly better values than the groupwise bootstrap. 
Tables~\ref{tab:RR1} to \ref{tab:RR2} in the supplement show the global rejection rates of the different settings. Under the null hypothesis, all values in the binomial confidence interval are printed in bold type. The permutation method is exact under exchangeability and, thus, most of the values of the permutation method with equal survival distributions across the groups under the null (\textit{exp early, exp late, exp prop, logn, Weib late, Weib prop}) and \textit{equal} censoring distributions fall within that interval. Furthermore, when exchangeability is violated, the permutation method still seems to perform quite accurately in terms of type I error control for all sample sizes. The groupwise bootstrap approach also results in very accurate family-wise error rates, especially for medium and large sample sizes. 
 Moreover, we note that the {three} asymptotic approaches {(\textit{asymptotic\_global, asymptotic, asymptotic\_bonf})} and the wild bootstrap {approaches} are too liberal, as they exhibit too high rejection rates in nearly all settings. {In Figures~\ref{fig:Dunnett_asymptotic} to \ref{fig:GrandMean_asymptotic} in the supplement, it is observable that these methods exceed the desired level of significance particularly for settings with small sample sizes. By further analyzing the tables in the supplement, we observe that high censoring rates facilitate the liberality of the tests. Note that the highest rejection rates occur for small sample size settings, where at least 49\% of the data is censored.   }\\
It should be noted that the power of our multiple tests can be improved by using a stepwise procedure as described in Section~\ref{sec:Multiple}. The power of the Boferroni corrected methods can also be improved by a stepwise procedure, e.g., the Holm-correction\cite{Holm}. However, stepwise procedures cannot be used for the construction of confidence regions and, hence, we did not focus on these in the simulation study.

We proved that all approaches are asymptotically valid under the null hypothesis. Figures~\ref{fig:Dunnett_asymptotic} to \ref{fig:GrandMean_asymptotic} in the supplement confirm this empirically: all methods {seem to tend to} the desired level of significance of $5\%$ 
for {increasing} sample sizes. However, the convergence rates of the asymptotic and the wild bootstrap approaches appear to be very slow. {
This observation prompts an inquiry into analyzing how larger sample sizes might influence the type I error control for the naive methods, that are the three asymptotic approaches.
Therefore, further simulations under the null hypothesis were conducted in Section~\ref{ssec:LargeSimu} in the supplement. Specifically, we increased the scaling factor for sample sizes, that is $K\in\{6,8,10\}$, resulting in sample sizes ranging from 60 to 200 per group. 
}


\subsection{{Simulation Results under the Alternative Hypothesis}}\label{ssec:SimuTypeII}
In the power assessment, we observed small differences between the different methods. The global asymptotic approach leads to the highest power in most settings, followed by the wild bootstrap with Gaussian and with Rademacher multipliers. However, in view of the bad type I error control of these methods, we cannot recommend their use. 
\\
Let us now review the multiple testing problem.
Because of the bad type I error control of the wild bootstrap approaches and for the sake of clarity, we did not consider this method 
in the following. Moreover, the global approaches (\textit{asymptotic global} and \textit{permutation}) do not yield local decisions. Thus, we only compared the asymptotic, the groupwise bootstrap and the Bonferroni-corrected approaches for the multiple testing problem. Furthermore, only the settings under the alternative hypothesis are considered. Tables~\ref{tab:MultPower1} to \ref{tab:MultPower2} in the supplement  provide the rejection rates of the false local hypotheses across all settings for the different sample sizes; they are further illustrated in Figures~\ref{fig:Dunnett_H1_asymptotic} to \ref{fig:GrandMean_H1_asymptotic} in the supplement. Therein, it is apparent that the asymptotic approaches have a higher power for each false hypothesis than the groupwise bootstrap and the studentized permutation approach with the Bonferroni-correction. However, this difference is rather small, especially for large sample sizes.
Additionally, 
{by comparing the empirical power of the groupwise bootstrap test and of the studentized permutation test with Bonferroni-correction, the groupwise bootstrap test tends to be slightly more powerful for medium and large sample sizes. 
For small sample sizes, this trend reverses for the Dunnett-type and Tukey-type contrast matrix. However, it is important to note that the differences between the two methods regarding the empirical power are quite small and mainly not even visible in Figures~\ref{fig:Dunnett_H1_asymptotic} to \ref{fig:GrandMean_H1_asymptotic}.}\\
{Nevertheless, i}t is well-known that the Bonferroni-correction might lead to a loss of power.\cite{konietschke2013}
{In order to illustrate this, }
we conducted an additional simulation study under non-exchangeability; see Section~\ref{ssec:NonExSimu} in the supplement for details. Here, we saw that the groupwise bootstrap approach is able to outperform the permutation approach with Bonferroni-corrections in specific scenarios under non-exchangeability. This effect becomes particularly observable for the Tukey-type contrast matrix, where six hypotheses are tested simultaneously.
{\\
We conducted further investigations to assess the impact of censoring and sample sizes on the power. As expected, the power increases for larger sample sizes for each method. Additionally, settings with lower censoring rates tend to be more powerful. 
When comparing the power between the three false hypotheses $\mathcal H_{0,3}, \mathcal H_{0,5}$ and $\mathcal H_{0,6}$ of the Tukey-type contrast matrix, it becomes apparent that the fifth hypothesis  $\mathcal H_{0,5}$ can be rejected more often, see, e.g., Figure~\ref{fig:Tukey_H1_asymptotic}. The reason behind this can be attributed to the unequal sample sizes in the unbalanced design: Groups 1 and 3 contain only $K\cdot 10$ observations, respectively, while groups 2 and 4 contain $K\cdot 20$ observations each, for $K\in\{1,2,4\}$. Consequently, when comparing the RMSTs of groups 2 and 4, we have a larger dataset compared to other pairwise comparisons leading to more power. This exemplifies how an unbalanced design can boost the power of specific local hypotheses. However, depending on the contrast matrix, this is often done at the cost of a reduced power for testing other local hypotheses. 
\\
It should be noted that the empirical power is very low in some scenarios. This is particularly the case for the groupwise bootstrap and the studentized permutation approach with Bonferroni-correction and small sample sizes. Moreover, an increasing number of hypotheses decreases the power for the local hypotheses in general. Consequently, multiple tests based on the Tukey-type contrast matrix have even less power than multiple tests based on the Dunnett-type contrast matrix.
Furthermore, small differences to the null hypothesis are difficult to detect. This can be observed for the Grand-mean-type contrast matrix, see Figure~\ref{fig:GrandMean_H1_asymptotic} in the supplement, where the three null hypotheses $\mathcal H_{0,1}: \mu_1 = \overline{\mu} , \mathcal H_{0,2}: \mu_2 = \overline{\mu}$, and $\mathcal H_{0,3}: \mu_3 = \overline{\mu}$ have very low rejection rates under the alternative hypothesis due to a small difference of $\mu_i - \overline{\mu} = \delta/4 = 3/8$ for $i\in\{1,2,3\}$. 
}
\\
In conclusion, we recommend to use the studentized permutation method for the global testing problem.
For the multiple testing problem, the groupwise bootstrap test and the studentized permutation method with Bonferroni-correction perform similarly and quite well in terms of the type I error control and the empirical power across all simulation scenarios.
However, we recommend to use the groupwise bootstrap test for testing a large number of hypotheses since the Bonferroni-correction is known to have a lower power in this case.\cite{konietschke2013} 


\FloatBarrier

\section{{Application to Real Data about the Occurrence of Hay Fever}}\label{sec:Data}
In order to illustrate our novel methods on real data, we consider a data set with data about the occurrence of hay fever of boys and girls with and without contact to farming environments.\cite{genuneit2011gabriel,genuneit2014sex}
These data {derive from an observational study and} may be structured in a factorial 2-by-2 design: 
factor A represents whether the child was growing up on a farm; factor B represents the sex. 
The event of interest is the age at which hay fever occurred.
Ties are present in the data as each measured age was rounded (down) to full years.\\
The children were included in the survey via primary schools in 2006. Hence, their age has been mainly between six and ten years at the beginning of the study. The medical diagnoses of hay fever together with the age at initial diagnosis \emph{before} study entry were recorded retrospectively. The age at which the diagnosis was made is easy to remember so that no significant recall bias or inaccuracies were assumed here. Follow-up surveys took place in 2010 with retrospective recording of initial diagnoses since the last survey and from then on annually until 2016.
For simultaneous testing on a main effect of the two factors as well as on an interaction effect, we define $\mathbf{H} := [\mathbf{H}_A^{\prime}, \mathbf{H}_B^{\prime}, \mathbf{H}_{AB}^{\prime}]^{\prime}$ by using the notation of Section~\ref{ssec:Setup}. Furthermore, we set $\alpha = 5\%$ as the level of significance and chose $\tau = 15$ years.

\begin{figure}[b!]
    \centering
    \includegraphics[width=\textwidth]{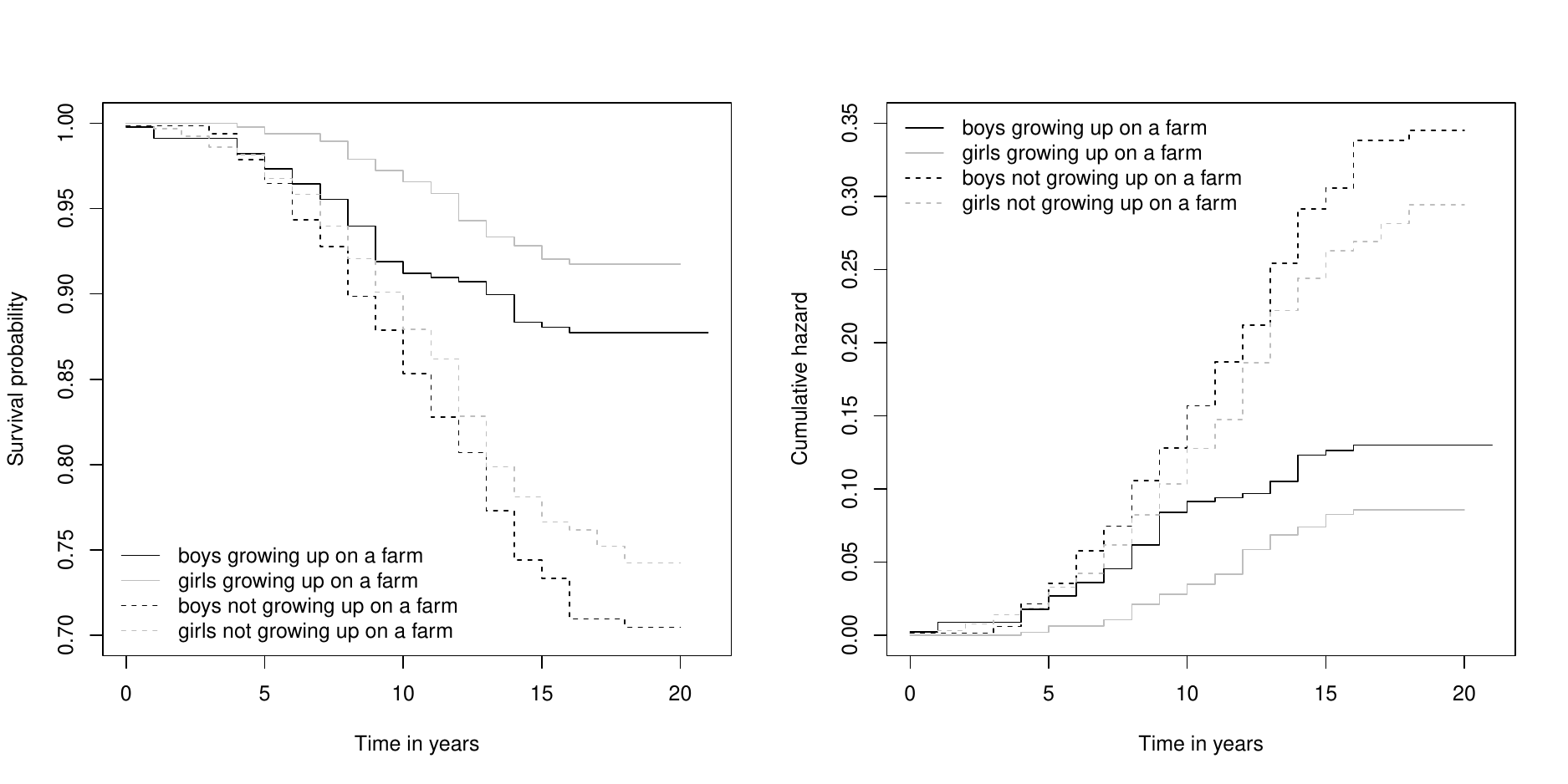}
    \caption{Kaplan-Meier and Nelson-Aalen curves of the different groups}
    \label{fig:KM}
\end{figure}

The data set consists of 2234 participants. In detail, 654 boys and 649 girls not growing up on a farm and
450 boys and 481 girls growing up on farms
were observed.
{Note that we did not adjust for any confounding variables in order to simplify this application of our method to real data. This comes with the limitation that the results may not fully reflect the causal effects of sex or growing up on a farm on the incidence of hay fever.}
The censoring rates in the different groups ranged from 74\% up to 93\%. The Kaplan-Meier and Nelson-Aalen curves of all groups are illustrated in Figure~\ref{fig:KM}. Here, it can be seen that the estimated cumulative hazard functions are crossing each other and, thus, the proportional hazards assumption is not justified.
If we would perform a Cox proportional hazards model nevertheless, the resulting (unadjusted) p-values of the existence of an impact on the occurrence of hay fever are $p_A < 10^{-8}$ for a main effect of factor A, $p_B = 0.112$ for a main effect of factor B and $p_{AB} = 0.235$ for an interaction effect. By using a Bonferroni- or Holm-correction of the p-values, we could only establish that factor A (growing up on a farm) has a main effect on the occurrence of hay fever at global level 5\%.

However, since the proportional hazards assumption seems violated, we aimed to compare the RMSTs in the different groups.
The estimated RMSTs respectively are 14.22 and 14.66 for boys and girls growing up on farms and 13.59 and 13.79 for boys and girls not growing up on a farm.
This indicates that boys tend to be more prone to hay fever than girls {until the age of 15}. Furthermore, growing up on a farm seems to reduce the risk of getting hay fever {until the age of 15}.
Performing the global asymptotic Wald-type test and its global studentized permutation version with $B=19999$ resampling iterations leads to p-values of $p<0.003$ and, thus, the existence of at least one main or the interaction effect on the occurrence of hay fever is highly significant. However, these tests cannot provide the information whether the sex and/or growing up on a farm and/or an interaction of these factors lead to a significant difference of hay fever occurrence. Therefore, we applied multiple testing procedures.
The resulting adjusted p-values of our proposed methods with $B=19999$ resampling iterations are shown in Table~\ref{tab:pValDataEx}. The p-values of the global asymptotic and permutation approach were adjusted by a Bonferroni-correction for enabling local test decisions. Here, we found that all methods rejected the local hypotheses of no main effect of the two factors simultaneously at the $\alpha=5\%$ level. However, the interaction effect of the two factors was not significant.

The data from this example do not fit perfectly to the simulation design in Section~\ref{sec:Simu} since, here, a 2-by-2 design with different hypothesis matrices and larger sample sizes and censoring rates is considered. Thus, additional simulation results inspired by this data example can be found in Section~\ref{ssec:DataSimu}.

\begin{table}[h!]
    \centering
    \begin{tabular}{l|cccccc}
         &  \textbf{asymptotic} & \textbf{wild} & \textbf{wild} & \textbf{groupwise} & \textbf{asymptotic} & \textbf{permutation}\\
         & & \textbf{Rademacher} & \textbf{Gaussian} & & \textbf{bonf} & \textbf{bonf}\\
         \hline
        \textbf{Farm }       & $<0.001$ &  $<0.001$    & $<0.001$ & $<0.001$ & $<0.001$ & $<0.001$\\
        \textbf{Sex}         & $0.005$ &  $0.006$    & $0.006$ & $0.007$ & $0.006$ & $0.006$\\
        \textbf{Interaction} & $0.605$ &  $0.599$    & $0.603$ & $0.597$ & $0.811$ & $0.800$\\
    \end{tabular}
    \caption{Adjusted p-values for the data example}
    \label{tab:pValDataEx}
\end{table}

\section{Discussion}\label{sec:Discussion}
In many applications, the proportional hazards assumption is not easy to detect by the naked eye or simply obviously not satisfied. In this case, the restricted mean survival time (RMST) can be used for summarizing the survival curve and, therefore, for comparing the survival curves of different groups in factorial designs. We considered a very general linear hypothesis testing problem for \mbox{RMSTs} in general factorial designs. To this end, we constructed a Wald-type test statistic and studied its asymptotic behaviour. Furthermore, we proposed resampling procedures for approximating the limiting distribution. This includes the studentized permutation approach and the groupwise bootstrap. In addition, we considered the multiple linear hypothesis testing problem for the RMST in general factorial designs, where several local hypotheses are tested simultaneously. Here, it turned out that the groupwise bootstrap can be used for approximating the joint limiting distribution of the test statistics.
However, the studentized permutation approach is not able to approximate the joint limiting distribution directly and, thus, a controlling procedure that works under any dependence structure of the individual test statistics has to be applied retrospectively, e.g., the Bonferroni correction. 
In an extensive simulation study, we analyzed the performance of the proposed methods. The results indicate that the groupwise bootstrap approach and the studentized permutation approach with Bonferroni correction perform best in terms of type I error control for the multiple testing problem and the global studentized permutation approach for the global testing problem. 
Finally, the proposed methods were applied to a real data set about hay fever.

It should be noted that the studentized permutation approach is finitely exact for the global testing problem under exchangeability. However, for the multiple testing problem, we cannot approximate the joint limiting distribution by the studentized permutation approach; see Section~\ref{sec:Multiple} for details. Hence, local test decisions can only be obtained by applying a correction procedure as the Bonferroni- or Holm-correction afterwards. These procedures are known to yield a lower power, particularly for a large number of hypotheses and positively correlated test statistics.\cite{konietschke2013}
The groupwise and wild bootstrap approach, on the other hand, can approximate the joint limiting distribution and, thus, the asymptotically exact dependence structure can be taken into account. A further advantage of the bootstrap approaches is that they also work for general hypotheses matrices and do not restrict to the case of contrast matrices.

A more flexible estimand than the usual RMST is the weighted version of the RMST.
That is, $\mu_{i,w_i} := \int_0^{\tau} w_i(t)S_i(t) \,\mathrm{d}t$ with estimator $\widehat{\mu}_{i,w_i} := \int_0^{\tau} w_i(t)\widehat{S}_i(t) \,\mathrm{d}t$ for some weight function $w_i\in\mathcal{L}_1([0,\tau])$ and $i\in\{1,...,k\}$ similar as in Zhao et al.\cite{wild_ursprung}. 
For the global testing problem, we get a similar statement as in Theorem~\ref{Asy} based on the weighted RMSTs. This can be shown analogously to the proof of Theorem~\ref{Asy} by using that the functional $D[0,\tau] \ni M \mapsto \int_{[0,\tau]} w_i(t)M(t) \,\mathrm{d}t$ is continuous for all $i\in\{1,...,k\}$.
It should be noted that the multivariate limiting distribution of the Wald-type test statistics for the multiple testing problem as in (\ref{eq:MultiAsymptotic}) based on the weighted RMSTs would also depend on the weight functions. 
Additionally, Zhao et al.\cite{wild_ursprung} already investigated the case of unknown weight functions for the two-sample case.
For future research, their result could be extended to complex factorial designs and to more general linear hypotheses. 

{Furthermore, it is important to note that our real data example derived from an observational study but that we did not account for potential confounding variables. These can significantly impact the survival times. The appropriate selection of confounding variables requires careful causal considerations; effective control for confounding also requires large enough sample size and recorded data on all confounding variables. Hence, it would be interesting to extend our methods in future research such that an adjustment for confounding variables is possible.}

\section*{Acknowledgements}
Merle Munko and Marc Ditzhaus gratefully acknowledge support from the \textit{Deutsche Forschungsgemeinschaft} (grant no. DI 2906/1-2). 
Part of the work has been done at Dennis Dobler's new affiliations: Department of Statistics at TU Dortmund University and Research Center Trustworthy Data Science and Security of the University Alliance Ruhr.


\section*{Conflict of interest}
The authors declare no potential conflict of interests.

\section*{Data availability statement}
The data use is restricted by informed consent of the study participants due to ethical requirements. Data access can be requested from author Jon Genuneit for collaborative research efforts.

\newpage




\bibliographystyle{abbrv} 
\bibliography{main}


\newpage
\appendix

\title { Supplementary materials to\\ RMST-Based Multiple Contrast Tests in General Factorial Designs }
\let\thefootnote\relax\footnote{*Corresponding author. Email address: \url{merle.munko@ovgu.de}}
\begin{center}
		Merle Munko\inst{1,*},
		Marc Ditzhaus\inst{1},
        Dennis Dobler\inst{2}
        and
        Jon Genuneit\inst{3}
\end{center}
	
	\institute{
		\inst{1} Otto-von-Guericke University Magdeburg; Magdeburg (Germany)\newline
        \inst{2} Vrije Universiteit Amsterdam; Amsterdam (Netherlands)\newline
        \inst{3} Leipzig University; Leipzig (Germany)
	}

\hrule

In this supplement, details on the wild bootstrap test are formulated in Section~\ref{sec:Wild}. The theoretical proofs of all stated theorems can be found in Section~\ref{sec:Proofs}. Furthermore, simulation results of an additional simulation study inspired by the data example are shown in Section~\ref{ssec:DataSimu}.
Moreover, we present the detailed simulation results in Section~\ref{sec:TabFig}.

\section{Wild Bootstrap Test}\label{sec:Wild}
In this section, we use the wild bootstrap approach similar as described in Zhao et al.\cite{wild_ursprung} for approximating the distribution of the Wald-type test statistic. Zhao et al.\cite{wild_ursprung} adopted the idea of Lin et al.\cite{lin} and Parzen et al.\cite{parzen} for developing this approach.

Parzen et al.\cite{parzen} proposed to replace $\sqrt{n}(\widehat{S}_i(t) - S_i(t))$ by
$$ \sqrt{n}\sum\limits_{j=1}^{n_i} G_{ij} \widehat{S}_i(t) \int\limits_{[0,t]} \dfrac{1}{Y_i(x)} \;\mathrm{d}N_{ij}(x) $$
for $t\geq 0$, where $G_{ij}, j\in\{1,...,n_i\}, i\in\{1,...,k\},$ are independent standard Gaussian random variables and $N_{ij}(x) := \delta_{ij}\mathbbm{1}\{X_{ij}\leq x \}$ for all $x\geq 0$. We modify this procedure analogously to Greenwood's formula\cite{Greenwood} and replace $\sqrt{n}(\widehat{S}_i(t) - S_i(t))$ by
\begin{align*}
    \sqrt{n}\sum\limits_{j=1}^{n_i} G_{ij} \widehat{S}_i(t) \int\limits_{[0,t]} \dfrac{1}{\sqrt{(Y_i(x)-\Delta N_i(x))Y_i(x)}} \;\mathrm{d}N_{ij}(x)
\end{align*}
for all $t>0$; also see related work by Dobler\cite{dobler2017disc}. 
For continuous survival functions $S_i$, this is asymptotically equivalent to the proposal of Parzen et al.\cite{parzen}.
However, the modification becomes important for the extension to discontinuous distribution functions $S_i$.
Moreover, we aim to weaken the assumption that $G_{ij}, j\in\{1,...,n_i\}, i\in\{1,...,k\},$ are standard Gaussian distributed. In fact, the multipliers only have to fulfill the following conditions:
\begin{enumerate}
    \item[(i)] $G_{ij}, j\in\{1,...,n_i\}, i\in\{1,...,k\},$  are independent and independent of the data $(\mathbf{X},\boldsymbol{\delta})$,
    \item[(ii)] $\E\left[G_{ij} \right]=0$,
    \item[(iii)] $\E\left[G_{ij}^2 \right] = 1$,
    \item[(iv)] $\E\left[G_{ij}^4 \right]\leq C$ for some constant $C < \infty$
\end{enumerate}
 for all $j\in\{1,...,n_i\}, i\in\{1,...,k\}$.
Hence, we replace $\sqrt{n}(\widehat{\mu}_i - \mu_i)$ by
\begin{align*}
    \sqrt{n} \widehat{\mu}_i^G :=
    \sqrt{n} \sum\limits_{j=1}^{n_i} G_{ij} \int\limits_0^{\tau} \widehat{S}_i(t) \int\limits_{[0,t]} \dfrac{1}{\sqrt{(Y_i(x)-\Delta N_i(x))Y_i(x)}}  \;\mathrm{d}N_{ij}(x) \;\mathrm{d}t
\end{align*} for all $i\in\{1,...,k\}.$
Furthermore, let
\begin{align*}
    W_n^G(\mathbf{H}) := n (\mathbf{H}\widehat{\boldsymbol\mu}^G)^{\prime} (\mathbf{H} \widehat{\boldsymbol{\Sigma}}^G \mathbf{H}^{\prime})^+ \mathbf{H}\widehat{\boldsymbol\mu}^G
\end{align*}
be the wild bootstrap counterpart of the Wald-type test statistic, where $\widehat{\boldsymbol\mu}^G := (\widehat{\mu}^G_1,...,\widehat{\mu}^G_k)^{\prime}$ and $\widehat{\boldsymbol\Sigma}^G := \text{diag}(\widehat{\sigma}_1^{G2},...,\widehat{\sigma}_k^{G2}) $ with
\begin{align*}
    \widehat{\sigma}_i^{G2} :=  n \sum\limits_{j=1}^{n_i} G_{ij}^2 \int\limits_{[0,{\tau}]} \left(\int\limits_x^{\tau} \widehat{S}_i(t) \;\mathrm{d}t \right)^2 \dfrac{1}{{(Y_i(x)-\Delta N_i(x))Y_i(x)}}  \;\mathrm{d}N_{ij}(x) .
\end{align*}

The following theorem ensures the wild bootstrap consistency.

\begin{theorem}\label{Wild}
    We have \begin{align*}
    W_n^G(\mathbf{H}) \xrightarrow{d} \chi^2_{\text{rank}(\mathbf{H})}
\end{align*}
almost surely as $n\to\infty$ given the data $(\mathbf{X},\boldsymbol{\delta})$. {Mathematically, this means
\begin{align*}
        \sup\limits_{z \in\R} \left|\P\left(W_n^{G}(\mathbf{H}) \leq z  \mid (\mathbf{X}, \boldsymbol\delta)\right) - \P\left(Z \leq z \right)\right|\xrightarrow{a.s.} 0
    \end{align*} as $n\to\infty$, where $Z\sim \chi^2_{\text{rank}(\mathbf{H})}.$
}
\end{theorem}

We define a wild bootstrap test by
\begin{align*}
    \varphi^G := \mathbbm{1}\{ W_n(\mathbf{H}, \mathbf{c}) > q_{1-\alpha}^G\},
\end{align*} where $q_{1-\alpha}^G$ denotes the $(1-\alpha)$-quantile of the conditional distribution of $W_n^G(\mathbf{H})$ given $(\mathbf{X},\boldsymbol\delta)$. The asymptotic validity of this test is provided by Lemma~1 of Janssen and Pauls\cite{janssenPauls2003}.

For the multiple testing problem, we get a similar result as in Theorem~\ref{Multigw} such that the procedure as in Section~\ref{sec:Multiple} can also be adopted with wild bootstrap counterparts instead of groupwise bootstrap counterparts.

\begin{theorem}\label{MultiWild}
    We have
    \begin{align*}
       (W_n^G(\mathbf{H}_{\ell}))_{\ell\in\{1,...,L\}} \xrightarrow{d}  \left( (\mathbf{H}_{\ell} \mathbf{Z})^{\prime} ( \mathbf{H}_{\ell}\boldsymbol{\Sigma} \mathbf{H}_{\ell}^{\prime}  )^+ \mathbf{H}_{\ell}\mathbf{Z}    \right)_{\ell\in\{1,...,L\}}
    \end{align*} 
    almost surely as $n\to\infty$ given the data $(\mathbf{X}, \boldsymbol\delta)$.
\end{theorem}

\section{Proofs}\label{sec:Proofs}
In this section, all stated theorems are proved.

\subsection{Proof of Theorem~\ref{Asy}}

\begin{myproof}[Proof of Theorem~\ref{Asy}]
    By Lemma~S.1 in the supplement of Ditzhaus et al.\cite{RMST}, it holds
\begin{align}\label{eq:clt}
   \sqrt{n}(\widehat{\boldsymbol\mu} - \boldsymbol\mu) \xrightarrow{d} \mathcal{N}_k(\mathbf{0}_k,\boldsymbol\Sigma)
\end{align}
as $n\to\infty$. 
Moreover, we have
\begin{align}\label{eq:varConv}
    \widehat{\sigma}_i^2 \xrightarrow{a.s.} \sigma_i^2
\end{align}
as $n\to\infty$ for all $i\in\{1,...,k\}$ by Section~S.5 in the supplement of Ditzhaus et al.\cite{RMST} {under $\P(X_{i1} \geq \tau) > 0$}.
Due to 
 $\P(T_{i1} {<} \tau) > 0$, it holds $\sigma_i^2 >0$ for all $i\in\{1,...,k\}.$
Hence, it follows $\text{rank}(\mathbf{H} {\boldsymbol{\Sigma}} \mathbf{H}^{\prime}) = \text{rank}(\mathbf{H} {\boldsymbol{\Sigma}}^{1/2}) = \text{rank}(\mathbf{H})$ and, analogously, $\text{rank}(\mathbf{H} \widehat{\boldsymbol{\Sigma}} \mathbf{H}^{\prime}) = \text{rank}(\mathbf{H})$ almost surely for sufficiently large $n$. Consequently, the Moore-Penrose inverse $(\mathbf{H} \widehat{\boldsymbol{\Sigma}} \mathbf{H}^{\prime})^+$ converges almost surely to $(\mathbf{H} {\boldsymbol{\Sigma}} \mathbf{H}^{\prime})^+$.
By Slutsky's lemma and Theorem 9.2.2
in Rao and Mitra\cite{rao1971generalized}, it follows
\begin{align*}
    W_n(\mathbf{H},\mathbf{c}) \xrightarrow{d} \chi^2_{\text{rank}(\mathbf{H})}
\end{align*}
as $n\to\infty$ under the null hypothesis in (\ref{eq:NullHyp}).
\end{myproof}

\subsection{Proof of Theorem~\ref{Perm}}
First of all, we introduce some notation. Let $\nu(t):= \sum_{i=1}^k \kappa_i\nu_i(t) $, $y(t) := \sum_{i=1}^k \kappa_i y_i(t)$, $A(t) := \int_{[0,t]} 1/y \;\mathrm{d}\nu$ and $S(t) := {\prodi\limits_{0 < s \leq t} (1-\mathrm{d}A(s))} $ for all $t\geq 0${, where $\prodi$ denotes the product integral as in Gill and Johansen\cite{GillJohansen}.}
Moreover, let $\widehat{S}$ denote the Kaplan-Meier estimator of the pooled survival function $S$, see Ditzhaus et al.\cite{RMST} for details,
and $\widehat{\mu} := \int_0^{\tau} \widehat{S}(t) \;\mathrm{d}t$ denote the estimator regarding the RMST of the pooled sample.

\begin{myproof}[Proof of Theorem~\ref{Perm}]
    As 
in the proof of Lemma~S.2 in the supplement of Ditzhaus et al.\cite{RMST}, it holds
\begin{align*}
    \sqrt{n} ( \widehat{\boldsymbol\mu}^{\pi} - \widehat{\mu}\mathbf{1}_k )
    \xrightarrow{d^{{*}}} \mathcal{N}_k(\mathbf{0}_k, \boldsymbol{\Sigma}^{\pi})
\end{align*}
  as $n\to\infty$, where
\begin{align*}
    (\boldsymbol{\Sigma}^{\pi})_{ii^{\prime}} := \left( \frac{1}{\kappa_i}\mathbbm{1}\{i=i^{\prime}\} - 1 \right) \sigma^{\pi 2}
\end{align*}
for all $i,i^{\prime}\in\{1,...,k\}$ and 
\begin{align*}
    \sigma^{\pi 2} := \int\limits_0^{\tau} \left(\int\limits_x^{\tau} {S}(t) \;\mathrm{d}t \right)^2
     \frac{1}{(1-\Delta{A}(x))y(x)}  \;\mathrm{d}{A}(x).
\end{align*}
Moreover, Ditzhaus et al.\cite{RMST} showed 
\begin{align*}
    \widehat{\sigma}_i^{\pi 2} \xrightarrow{\P} {\kappa_i}^{-1}\sigma^{\pi 2}
\end{align*}
as $n\to\infty$ for all $i\in\{1,...,k\}$  {under $\P(X_{i1} \geq \tau) > 0$} in the proof of Lemma~S.3 of the supplement.
Hence, it follows
\begin{align*}
    \widehat{\boldsymbol\Sigma}^{\pi} \xrightarrow{\P} \text{diag}\left( {\kappa_1}^{-1}\sigma^{\pi 2}, ..., {\kappa_k}^{-1}\sigma^{\pi 2} \right) = \boldsymbol{\Sigma}^{\pi} + \sigma^{\pi 2}\mathbf{J}_k
\end{align*}
as $n\to\infty$.
Since $\mathbf{H}\mathbf{J}_k = \mathbf{0}_r$, it holds 
$\mathbf{H}\widehat{\boldsymbol\Sigma}^{\pi} \xrightarrow{\P} \mathbf{H}\boldsymbol{\Sigma}^{\pi}$
as $n\to\infty$.
Moreover, 
$\P(T_{i1} {<} \tau) > 0$ implies $\sigma^{\pi 2} > 0$ and, thus, $\P(\widehat{\sigma}_i^{\pi 2} > 0 )\to 1$ as $n\to\infty$. Consequently, we have $(\mathbf{H}\widehat{\boldsymbol\Sigma}^{\pi}\mathbf{H}^{\prime})^+ \xrightarrow{P} (\mathbf{H}{\boldsymbol\Sigma}^{\pi}\mathbf{H}^{\prime})^+$ as $n\to\infty.$
Hence, it follows by Slutsky's lemma and Theorem 9.2.2 in Rao and Mitra\cite{rao1971generalized}
\begin{align*}
    W_n^{\pi}(\mathbf{H}) &= n (\mathbf{H}\widehat{\boldsymbol\mu}^{\pi})^{\prime} (\mathbf{H} \widehat{\boldsymbol{\Sigma}}^{\pi} \mathbf{H}^{\prime})^+ \mathbf{H}\widehat{\boldsymbol\mu}^{\pi} \\&= n [\mathbf{H}(\widehat{\boldsymbol\mu}^{\pi}-\widehat{\mu}\mathbf{1}_k)]^{\prime} (\mathbf{H} \widehat{\boldsymbol{\Sigma}}^{\pi} \mathbf{H}^{\prime})^+ \mathbf{H}(\widehat{\boldsymbol\mu}^{\pi}-\widehat{\mu}\mathbf{1}_k) \xrightarrow{d^{{*}}} \chi^2_{\text{rank}(\mathbf{H})}
\end{align*}
 as $n\to\infty$ with similar arguments as in the proof of Theorem~\ref{Asy}.
\end{myproof}

\subsection{Proof of Theorem~\ref{gwBS}}
In the following, let $\widehat{S}_i^*,  \widehat{A}_i^*, \widehat{\sigma}_i^{*2}, Y_i^*$ and $N_i^*$ denote the groupwise bootstrap counterparts of $\widehat{S}_i,  \widehat{A}_i, \widehat{\sigma}_i^{2}, Y_i$ and $N_i$, respectively, for all $i\in\{1,...,k\}$.

\begin{lemma} \label{gwCLT}
    We have
    \begin{align}\label{eq:gwCLT}
        \sqrt{n} (\widehat{\boldsymbol\mu}^* - \widehat{\boldsymbol\mu}) \xrightarrow{d^{{*}}} \mathbf{Z} = (Z_1,...,Z_k)^{\prime} \sim \mathcal{N}_k(\mathbf{0}_k, \boldsymbol\Sigma)
    \end{align}
    as $n\to\infty$.
    Mathematically, (\ref{eq:gwCLT}) means
    \begin{align*}
        \sup\limits_{z_1,...,z_k \in\R} \left|\P\left(\sqrt{n} (\widehat{\mu}^*_1 - \widehat{\mu}_1) \leq z_1, ..., \sqrt{n} (\widehat{\mu}^*_k - \widehat{\mu}_k) \leq z_k  \mid (\mathbf{X}, \boldsymbol\delta)\right) - \P\left(Z_1 \leq z_1, ..., Z_k \leq z_k \right)\right|\xrightarrow{P} 0
    \end{align*} as $n\to\infty$.
\end{lemma}
\begin{myproof}[Proof of Lemma~\ref{gwCLT}]
    By Theorem~4 in the supplement of Dobler and Pauly\cite{concordance2017}, it holds
\begin{align*}
    \sqrt{n_i} (\widehat{S}_i^* -  \widehat{S}_i) \xrightarrow{d^{{*}}} U_i \sim GP(0,\Gamma_i)
\end{align*}
 on $D[0,\tau]$ 
 as $n\to\infty$, where 
\begin{align*}
    \Gamma_i: (t,s) \mapsto S_i(t)S_i(s)\int\limits_{[0,\min\{t,s\}]} \dfrac{1}{(1-\Delta A_i(x)) y_{i}(x)} \;\mathrm{d}A_i(x)
\end{align*}
    for all $i\in\{1,...,k\}$ and $GP(0,\Gamma_i)$ denotes a centered Gaussian process with covariance function $\Gamma_i$. Since the samples are independent, it follows
\begin{align*}
    \sqrt{n} (\widehat{S}_i^* -  \widehat{S}_i)_{i\in\{1,...,k\}} \xrightarrow{d^{{*}}} \mathbb{G}^* = \left(\frac{1}{\sqrt{\kappa_i}} U_i\right)_{i\in\{1,...,k\}}\sim GP_k\left(\mathbf{0}_k, \text{diag}\left(\frac{1}{\kappa_1}\Gamma_1, ..., \frac{1}{\kappa_k}\Gamma_k\right)\right)
\end{align*}
    on $D[0,\tau]^k$ 
     as $n\to\infty$ by (\ref{eq:kappa}), where $GP_k\left(\mathbf{0}_k, \mathbf{D}\right)$ denotes a $k$-dimensional centered Gaussian process with covariance function $\mathbf{D}:[0,\tau]^2 \to \R^{k\times k}$. Hence, the continuous mapping theorem provides
\begin{align*}
    \sqrt{n} (\widehat{\boldsymbol\mu}^* - \widehat{\boldsymbol\mu}) \xrightarrow{d^{{*}}} \int\limits_0^{\tau} \mathbb{G}^*(t) \;\mathrm{d}t = \mathbf{Z}
\end{align*}
    as $n\to\infty$.
 The limiting variable $\mathbf{Z}$ is normally distributed as linear transformation of a Gaussian process and its moments can be calculated by using Fubini's theorem. Thus, we get
 $\E\left[\mathbf{Z}\right] = \mathbf{0}_k$ and 
 $\Cov\left(\mathbf{Z} \right) = \boldsymbol\Sigma$.
\end{myproof}

\begin{lemma}\label{gwCov}
    We have 
    \begin{align}\label{eq:gwCov}
        \widehat{\boldsymbol{\Sigma}}^* \xrightarrow{P} \boldsymbol{\Sigma}
    \end{align}
    as $n\to\infty$.
\end{lemma}
\begin{myproof}[Proof of Lemma~\ref{gwCov}]
Let $i\in\{1,...,k\}$ be arbitrary.
    Similarly as in the supplement of Dobler and Pauly\cite{concordance2017}, we consider the $\P^{(X_{i1},\delta_{i1})}$-Donsker classes
    \begin{align*}
        \mathcal{F}_1 := \left\{(x,\delta)\mapsto \mathbbm{1}\{x \leq t, \delta = 1\} \mid t\in[0,\tau]\right\} \text{ and } \mathcal{F}_2 := \left\{(x,\delta)\mapsto  \mathbbm{1}\{x \geq t\} \mid t\in[0,\tau]\right\}
    \end{align*}
     with finite envelope function $F \equiv 1.$ By Theorem~3.6.1 in van der Vaart and Wellner\cite{vaartWellner1996} and Slutsky's lemma, we obtain
     \begin{align*}
         \sup\limits_{t\in[0,\tau]} \left| \frac{1}{n_i} Y_i^*(t) - \frac{1}{n_i} Y_i(t) \right| \xrightarrow{P} 0 \quad\text{ and }\quad \sup\limits_{t\in[0,\tau]} \left| \frac{1}{n_i} N_i^*(t) -  \frac{1}{n_i} N_i(t) \right| \xrightarrow{P} 0
     \end{align*} as $n\to\infty$. Section~S.6 in the supplement of Ditzhaus et al.\cite{RMST} provides
     \begin{align*}
         \sup\limits_{t\in[0,\tau]} \left| \frac{1}{n_i} Y_i(t) - y_i(t) \right| \xrightarrow{P} 0 \quad\text{ and }\quad \sup\limits_{t\in[0,\tau]} \left| \frac{1}{n_i} N_i(t) -  \nu_i(t) \right| \xrightarrow{P} 0
     \end{align*} as $n\to\infty$.
It follows
\begin{align*}
        \sup\limits_{t\in[0,\tau]} \left| \widehat{S}_i^*(t) - S_i(t) \right| \xrightarrow{P} 0
\quad\text{ and }\quad    
        \sup\limits_{t\in[0,\tau]} \left| \widehat{A}_i^*(t) - A_i(t) \right| \xrightarrow{P} 0
    \end{align*}
    as $n\to\infty$  {under $\P(X_{i1} \geq \tau) > 0$}.
    Hence, we have
    $\widehat{\sigma}_i^{*2} \xrightarrow{P} {\sigma}_i^{2}$
    as $n\to\infty$. Since $i\in\{1,...,k\}$ was arbitrary, (\ref{eq:gwCov}) follows.
\end{myproof}

\begin{myproof}[Proof of Theorem~\ref{gwBS}]
    This statement follows with similar arguments as in the proofs of Theorem~\ref{Asy} and \ref{Perm} by Lemma~\ref{gwCLT} and \ref{gwCov}.
    Hereby, we apply Slutsky's lemma and Theorem 9.2.2 in Rao and Mitra\cite{rao1971generalized} again.
\end{myproof}

\subsection{Proof of Theorem~\ref{gwTest}}
First of all, we state a useful lemma, which provides the uniform convergence of quantile functions.

\begin{lemma}\label{ContLemma}
Let $\mathcal{F}_0:\R\to [0,1]$ be a distribution function that is continuous and strictly increasing on $[a,b] \subset \R$ and $\left(\mathcal{F}_n\right)_{n\in\mathbb N}$  denote a sequence of distribution functions with 
\begin{align}\label{eq:ptwGn}
      \mathcal{F}_n(w) \to \mathcal{F}_0(w) \quad \text{ for all } w\in [a,b]
\end{align}
 as $n\to\infty$.
Furthermore, let $\mathcal{F}_0(a) < p \leq q < \mathcal{F}_0(b).$ Then, we have
$$ \sup\limits_{r \in [p,q]} | \mathcal{F}_n^{-1}(r) - \mathcal{F}_0^{-1}(r) | \to 0 $$ as $n\to\infty$.
\end{lemma}
\begin{myproof}[Proof of Lemma~\ref{ContLemma}]
First of all, one can show 
\begin{align}\label{eq:unifGn}
     \sup\limits_{w \in [a,b]} | \mathcal{F}_n(w) - \mathcal{F}_0(w) | \to 0 
\end{align}
by (\ref{eq:ptwGn}) and the continuity of $\mathcal{F}_0$ on $[a,b]$. 
Now, we proceed similarly as in van der Vaart and Wellner\cite{vaartWellner1996}.
Let $(\delta_n)_{n\in\mathbb N}$ be a positive sequence with $\delta_n \to 0$ as $n\to\infty$.
By (\ref{eq:unifGn}), there exists an $N\in\mathbb N$ such that
$$ \mathcal{F}_0(b) - \mathcal{F}_n(b) \leq \mathcal{F}_0(b) - q \quad \text{and} \quad \mathcal{F}_n(a + \delta_n) - \mathcal{F}_0(a + \delta_n) < (p - \mathcal{F}_0(a))/2 $$ 
holds for all $n \geq N.$
Due to the continuity of $\mathcal{F}_0$, we can choose $N$ sufficiently large such that 
$\mathcal{F}_0(a + \delta_n) \leq \mathcal{F}_0(a) + (p-\mathcal{F}_0(a))/2$ holds for all $n \geq N$.
Hence, it follows that $\mathcal{F}_n(b) \geq q$ and $\mathcal{F}_n(a + \delta_n) < p$ for all $n \geq N$. Since $$\mathcal{F}_n^{-1}(r) \leq x  \quad \Leftrightarrow \quad  r \leq \mathcal{F}_n(x)$$ holds for all $r\in [p,q], x\in\R$ due to the definition of the inverse map, we have
$\mathcal{F}_n^{-1}(r) \leq b$ and $\mathcal{F}_n^{-1}(r) > \mathcal{F}_n^{-1}(r) - \delta_n > a $ for all $r\in [p,q]$ and $n\geq N$.
\\
  Moreover,
  it holds 
 $$ {\mathcal{F}}_n({\mathcal{F}}_n^{-1}(r) - \delta_n) \leq r \leq {\mathcal{F}}_n({\mathcal{F}}_n^{-1}(r)) $$ for all $r\in [p,q], n\in\mathbb N$ by the definition of the inverse map. Hence, it follows
 \begin{align}\label{eq:Absch}
     \mathcal{F}_{0}({\mathcal{F}}_n^{-1}(r)) - {\mathcal{F}}_n({\mathcal{F}}_n^{-1}(r)) \leq \mathcal{F}_{0}({\mathcal{F}}_n^{-1}(r)) - r \leq \mathcal{F}_{0}({\mathcal{F}}_n^{-1}(r)) - {\mathcal{F}}_n({\mathcal{F}}_n^{-1}(r) - \delta_n)
 \end{align}
 for all $r\in [p,q], n\in\mathbb N$.
 The left side of (\ref{eq:Absch}) is converging to $0$ uniformly in $r$ as $n\to\infty$ by (\ref{eq:unifGn}). Since $\mathcal{F}_0$ is continuous on the compact set $[a,b]$, it is also uniformly continuous. The right side of (\ref{eq:Absch}) can be rewritten as
 \begin{align*}
     \mathcal{F}_{0}({\mathcal{F}}_n^{-1}(r)) - {\mathcal{F}}_n({\mathcal{F}}_n^{-1}(r) - \delta_n) 
     = & \mathcal{F}_{0}({\mathcal{F}}_n^{-1}(r)) - \mathcal{F}_{0}({\mathcal{F}}_n^{-1}(r) - \delta_n) \\& + \mathcal{F}_{0}({\mathcal{F}}_n^{-1}(r) - \delta_n) - {\mathcal{F}}_n({\mathcal{F}}_n^{-1}(r) - \delta_n) ,
 \end{align*} where the first part vanishes asymptotically uniformly in $r$ due to the uniform continuity of $\mathcal{F}_{0}$ and the second part due to (\ref{eq:unifGn}).
 Thus, (\ref{eq:Absch}) implies  
 $$ \sup\limits_{r\in [p,q]} \left| \mathcal{F}_{0}({\mathcal{F}}_n^{-1}(r)) - r \right| \to 0 $$ as $n\to\infty$. By the strict monotony of $\mathcal{F}_{0}$ on 
 $[a,b]$, $\mathcal{F}_{0}^{-1}$ is continuous on $[(\mathcal{F}_0(a) + p)/2, \mathcal{F}_0(b)]$ and, thus, uniformly continuous on $[(\mathcal{F}_0(a) + p)/2, \mathcal{F}_0(b)]$. Let $\varepsilon > 0$ be arbitrary and $\delta \in (0,(p-\mathcal{F}_0(a))/2]$ such that $$ |\mathcal{F}_{0}^{-1}(x) - \mathcal{F}_{0}^{-1}(y)| < \varepsilon $$ holds for all $x,y\in [(\mathcal{F}_0(a) + p)/2, \mathcal{F}_0(b)]$ with $|x-y| < \delta$. There exists an $M \in\mathbb N$ such that  
 $$\sup\limits_{r\in [p,q]}  |\mathcal{F}_{0}({\mathcal{F}}_n^{-1}(r)) - r| < \delta $$ holds for all $n\geq M$.
 This further implies that $\mathcal{F}_{0}({\mathcal{F}}_n^{-1}(r)) > r - \delta \geq (\mathcal{F}_0(a) + p)/2 $ for all $r\in [p,q], n \geq M$. Since ${\mathcal{F}}_n^{-1}(r) \leq b$ for all $r\in [p,q], n\geq N$, we also have $\mathcal{F}_{0}({\mathcal{F}}_n^{-1}(r)) \leq \mathcal{F}_0(b)$ for all $r\in [p,q], n\geq N$.
 Hence, it follows  that
 $$ \sup\limits_{r\in [p,q]}|{\mathcal{F}}_n^{-1}(r) - {\mathcal{F}}_{0}^{-1}(r)| = \sup\limits_{r\in [p,q]}|\mathcal{F}_0^{-1}(\mathcal{F}_{0}({\mathcal{F}}_n^{-1}(r))) - \mathcal{F}_0^{-1}(r)| < \varepsilon  $$ for all $n\geq \max\{N,M\}$, which completes the proof.
\end{myproof}

The following lemma is a multivariate extension of Pólya's theorem.

\begin{lemma}\label{Polya}
    Let $W_\ell, \ell\in\{1,...,L\}$ denote $L$ random variables with continuous distribution functions $\mathcal{F}_\ell:\R\to[0,1], \ell\in\{1,...,L\}$ and joint distribution function $\mathcal{F}:\R^L \to [0,1]$. Furthermore, let $(\mathcal{F}_n:\R^L \to [0,1])_{n\in\mathbb N}$ denote a sequence of distribution functions with
    $\mathcal{F}_n(\mathbf{w}) \to \mathcal{F} (\mathbf{w})$ as $n\to\infty$ for all $\mathbf{w}\in\R^L$. Then, we have the uniform convergence, that is
$$ \sup\limits_{\mathbf{w}\in\R^L} |\mathcal{F}_n(\mathbf{w}) - \mathcal{F} (\mathbf{w})| \to 0 $$
   as $n\to\infty$.
\end{lemma}
\begin{myproof}[Proof of Lemma~\ref{Polya}]
    We aim to apply Proposition~2.1 in Bickel and Millar\cite{multivariatePolya}.
    Therefore, let 
    $$f_{\mathbf{w}}:\R^L \to \R, f_{\mathbf{w}}(x_1,...,x_L) = \mathbbm{1}\{ x_1 \leq w_1, ..., x_L \leq w_L \}$$ for all $\mathbf{w} = (w_1,...,w_L)^{\prime} \in (\R\cup \{-\infty,\infty\})^L$ and set
    $$\mathbf{F} := \left\{ f_{\mathbf{w}}  \mid \mathbf{w}  \in (\R\cup \{-\infty,\infty\})^L \right\}.$$
    In addition, let $\varepsilon>0$ be arbitrary, $m\in\mathbb{N}$ with $\varepsilon/L \geq 1/m$ and define $$-\infty =: w_{\ell, 0}  < w_{\ell, 1} < ...  < w_{\ell,m} := \infty$$ such that
    $\mathcal{F}_{\ell} (w_{\ell,i}) - \mathcal{F}_{\ell} (w_{\ell,i-1}) \leq \varepsilon/L$ for all $i\in\{1,...,m\}, \ell\in\{1,...,L\}$.
    Set $\mathbf{w}_{i_1,...,i_L} := (w_{1,i_1},...,w_{L,i_L})^{\prime}$ for all $i_1,...,i_L\in\{0,...,m\}$.
    Then, it holds that
    \begin{align*}
        \int f_{\mathbf{w}_{i_1,...,i_L}} - f_{\mathbf{w}_{i_1-1,...,i_L-1}}\;\mathrm{d}\P^{(W_1,...,W_L)} &= 
        \P\left(  \forall \ell\in\{1,...,L\}:  W_\ell\leq w_{\ell,i_\ell},
        \exists \ell\in\{1,...,L\}:  W_\ell > w_{\ell,i_\ell-1}\right)\\
        &\leq \sum\limits_{\ell =1}^L \P\left( W_\ell \in (w_{\ell,i_\ell-1}, w_{\ell,i_\ell}] \right)\\
        &= \sum\limits_{\ell =1}^L \left( \mathcal{F}_{\ell} (w_{\ell,i_\ell}) - \mathcal{F}_{\ell} (w_{\ell,i_\ell-1})\right)\\
        &\leq \varepsilon
    \end{align*}
    for all $i\in\{1,...,m\}$. Thus, the bracketing number is bounded by $(m+1)^L < \infty$. Proposition~2.1 in Bickel and Millar\cite{multivariatePolya} provides that $\mathbf{F}$ is a Pólya class. 
    By the definition of $\mathbf{F}$, the statement of the lemma follows.
\end{myproof}

For each $\ell\in\{1,...,L\}$, let $\mathcal{F}_{\ell}$ denote the distribution function of $W_{\ell} := (\mathbf{H}_{\ell} \mathbf{Z})^{\prime} ( \mathbf{H}_{\ell}\boldsymbol{\Sigma} \mathbf{H}_{\ell}^{\prime}  )^+ \mathbf{H}_{\ell}\mathbf{Z}  \sim\chi^2_{\text{rank}(\mathbf{H}_{\ell})}  $, ${\mathcal{F}}_{n,\ell}^*$ denote the distribution function of $W_n^{*}(\mathbf{H}_{\ell})$ given $(\mathbf{X}, \boldsymbol{\delta})$ and $\widehat{\mathcal{F}}_{n,\ell}^*$ denote the empirical distribution function of $W_n^{*,1}(\mathbf{H}_{\ell}),...,W_n^{*,B}(\mathbf{H}_{\ell})$ 
    in the following.
Furthermore, let $\mathcal{F}, {\mathcal{F}}_{n}^*$ denote the joint distribution functions of $(W_{\ell})_{\ell \in\{1,...,L\}} $ and $\left(W_n^{*}(\mathbf{H}_{\ell})\right)_{\ell \in\{1,...,L\}}$ given $(\mathbf{X}, \boldsymbol{\delta})$, respectively, and $\widehat{\mathcal{F}}_{n}^*$ denote the empirical joint distribution function of $\left(W_n^{*,b}(\mathbf{H}_{\ell})\right)_{\ell \in\{1,...,L\}}, b\in\{1,...,B\}.$
The $(1-\beta)$-quantile of the $\chi^2_{\text{rank}(\mathbf{H}_{\ell})}$-distribution is denoted by $ q_{\ell,1-\beta} := \mathcal{F}_{\ell}^{-1}(1 - \beta)$  in the following for all $\ell\in\{1,...,L\}, \beta\in (0,1).$
In addition, we define $$\mathrm{FWER}: \R \to [0,1], \quad\mathrm{FWER}(\beta) := 1- \mathcal{F}\left( (\mathcal{F}_{\ell}^{-1}(1 - \beta))_{\ell\in\{1,...,L\}} \right). $$

\begin{lemma}\label{UnifQuantLemma}
    For each $\ell\in\{1,...,L\}$, we have
    \begin{align*}
        \sup\limits_{\beta \in [\alpha/(2L), (1+\alpha)/2]} \left| q_{\ell,1-\beta}^* - q_{\ell,1-\beta} \right| \xrightarrow{P} 0
    \end{align*} as $n\to\infty$ under $ B = B(n)\to \infty$. 
\end{lemma}
\begin{myproof}[Proof of Lemma~\ref{UnifQuantLemma}]
    Let $\ell\in\{1,...,L\}$ be fixed.
    It is well known that 
    \begin{align*}
        \left| \mathcal{F}_{n,\ell}^*(w) - \mathcal{F}_{\ell}(w) \right| \xrightarrow{P} 0 
    \end{align*}
    as $n\to\infty$ for all $w\in\R$ due to Theorem~\ref{Multigw}. 
    Furthermore, one can show
\begin{align*}
    \left| \widehat{\mathcal{F}}_{n,\ell}^*(w) - {\mathcal{F}}_{n,\ell}^*(w) \right| \xrightarrow{P} 0 
\end{align*}
     as $B\to\infty$ for all $w\in\R$ by applying Chebyshev's inequality.
     Hence, it follows
     \begin{align*}
         \left| \widehat{\mathcal{F}}_{n,\ell}^*(w) - {\mathcal{F}}_{\ell}(w) \right| \xrightarrow{P} 0 
     \end{align*} as $n\to\infty$ for all $w\in\R$. 
     Applying Lemma~\ref{ContLemma} with $a = 0, b > \mathcal{F}_{\ell}^{-1}(1-\alpha/(2L)), p = (1-\alpha)/2, q = 1-\alpha/(2L)$ and $\mathcal{F}_0 = \mathcal{F}_{\ell}$ yields 
    \begin{align}\label{eq:uniformQuant}
      \sup\limits_{\beta\in [\alpha/(2L), (1+\alpha)/2]} \left| q_{\ell,1-\beta}^* -  q_{\ell,1-\beta} \right| = \sup\limits_{r\in [(1-\alpha)/2, 1-\alpha/(2L)]} \left| \widehat{\mathcal{F}}_{n,\ell}^{*,-1}(r) - \mathcal{F}_{\ell}^{-1}(r) \right|\xrightarrow{P} 0
    \end{align} as $n\to\infty$.  
\end{myproof}

\begin{lemma}\label{betaLemma}
    Under 
    $B \to \infty$
    as $n\to\infty$, we have
    \begin{align*}
        \beta_n(\alpha) \xrightarrow{P} \mathrm{FWER}^{-1}(\alpha)
    \end{align*} as $n\to\infty$.
\end{lemma}
\begin{myproof}[Proof of Lemma~\ref{betaLemma}]
    We first note that 
    $$ \left(\mathrm{FWER}^*_n\right)^{-1}(\alpha) \geq {\beta}_n(\alpha) \geq \left(\mathrm{FWER}^*_n\right)^{-1}(\alpha) - \frac{1}{B} $$
    holds by the definition of $\beta_n(\alpha)$. Additionally, one can show that $\mathrm{FWER}$ and $\mathrm{FWER}^*_n$ are distribution functions. Moreover,
    $\mathrm{FWER}$ is strictly increasing over $[0,1]$ and continuous. \\ 
    By Chebyshev's inequality, we have 
    $|\widehat{\mathcal{F}}_n^*(\mathbf w) - {\mathcal{F}}_n^*(\mathbf w)| \xrightarrow{P} 0$ as $B\to\infty$ for all $\mathbf w\in\R^L$.
  Theorem~\ref{Multigw} 
   implies
    $$ |{\mathcal{F}}_n^*(\mathbf w) - {\mathcal{F}}(\mathbf w)| \xrightarrow{P} 0$$ 
    as $n\to\infty$ for all $\mathbf w\in\R^L$.
    Putting the two previous equations together leads to
    $$ \sup\limits_{\mathbf w\in\R^L}|\widehat{\mathcal{F}}_n^*(\mathbf w) - {\mathcal{F}}(\mathbf w)| \xrightarrow{P} 0$$ as $n\to\infty$ by Lemma~\ref{Polya}.
    Hence, we have that 
    \begin{align*}
        \left|\mathrm{FWER}_n^*(\beta) - \mathrm{FWER}(\beta)\right| 
        &=\left| 1-\widehat{\mathcal{F}}_n^*\left(\left(  q_{\ell,1-\beta}^* \right)_{\ell\in\{1,...,L\}} \right)
        -1 + \mathcal{F}\left(\left(   q_{\ell,1-\beta} \right)_{\ell\in\{1,...,L\}} \right)\right|
        \\&\leq
        \sup\limits_{\mathbf w\in\R^L}|\widehat{\mathcal{F}}_n^*(\mathbf w) - {\mathcal{F}}(\mathbf w)|
       + \left|\mathcal{F}\left( \left(  q_{\ell,1-\beta}^*  \right)_{\ell\in\{1,...,L\}}\right) - \mathcal{F}\left( \left(  q_{\ell,1-\beta}\right)_{\ell\in\{1,...,L\}} \right) \right|
    \end{align*}
    converges to $0$ in probability as $n\to\infty$ uniformly for all $\beta\in [\alpha/(2L), (1+\alpha)/2] =: [a,b]$ by Lemma~\ref{UnifQuantLemma}.
    By applying Lemma~\ref{ContLemma} with $p = q = \alpha$ and $\mathcal{F}_0 = \mathrm{FWER}$, one can show
    $$ \left|(\mathrm{FWER}^*_n)^{-1}(\alpha) - \mathrm{FWER}^{-1}(\alpha)  \right| \xrightarrow{P} 0 $$
    as $n\to\infty$.
    Thus, it follows $\beta_n(\alpha) \xrightarrow{P} \mathrm{FWER}^{-1}(\alpha)$ as $n\to\infty$ since $B\to\infty$ as $n\to\infty$.
\end{myproof}

\begin{myproof}[Proof of Theorem~\ref{gwTest}]
     Lemma~\ref{UnifQuantLemma} and \ref{betaLemma} imply $q^*_{\ell,1-\beta_n(\alpha) }\xrightarrow{P} q_{\ell,1-\mathrm{FWER}^{-1}(\alpha)}$ as $n\to\infty$ for all $\ell\in\{1,...,L\}$ due to the continuity of $\mathcal{F}_{\ell}^{-1}$ on $(0,1).$
     Let $\T\subset \{1,...,L\}$ denote the set of true hypotheses in (\ref{eq:MultipleHypos}). 
     By Slutsky's lemma and Theorem~\ref{MultiAsy}, we have
     $$ \left( W_n(\mathbf{H}_{\ell},\mathbf{c}_{\ell}), q_{\ell,1-{\beta}_n(\alpha)}^{*} \right)_{\ell\in\T} \xrightarrow{d} \left(W_{\ell}, q_{\ell,1-\mathrm{FWER}^{-1}(\alpha)} \right)_{\ell\in\T} $$ as $n\to\infty$.
    Thus, one can show that
    \begin{align}\label{eq:ineqality}
        \P\left( \exists {\ell \in\T}: \: W_n(\mathbf{H}_{\ell},\mathbf{c}_{\ell}) >q_{\ell,1-{\beta}_n(\alpha)}^{*}\right)
        &=
        1 - \P\left( \forall {\ell \in\T}: \: W_n(\mathbf{H}_{\ell},\mathbf{c}_{\ell}) \leq q_{\ell,1-{\beta}_n(\alpha)}^{*}\right) \notag
       \\& \to 1 - P\left(\forall {\ell \in\T}: W_{\ell} \leq q_{\ell,1-\mathrm{FWER}^{-1}(\alpha)}\right) \notag
       \\& \leq 1 - P\left(\forall {\ell \in\{1,...,L\}}: W_{\ell} \leq q_{\ell,1-\mathrm{FWER}^{-1}(\alpha)}\right)
       \\&= \mathrm{FWER}(\mathrm{FWER}^{-1}(\alpha)) = \alpha \notag
    \end{align}
    as $n\to\infty$ under the null hypotheses $\mathcal{H}_{0,\ell}, \ell\in\T,$ in (\ref{eq:MultipleHypos}). 
    Note that the inequality in (\ref{eq:ineqality}) is an equality if $\T = \{1,...,L\}$.
\end{myproof}

\subsection{Proof of Proposition~\ref{pValues}}
   \begin{myproof}[Proof of Proposition~\ref{pValues}]
   Let $\ell\in\{1,...,L\}$ be fixed.
       Firstly, we aim to show $p_{\ell} \leq \alpha \quad \Rightarrow \quad w_n(\mathbf{H}_{\ell}, \mathbf{c}_{\ell}) > q_{l,1-{\beta_n}(\alpha)}^{*}$. Therefore, assume that $p_{\ell} \leq \alpha$ and $w_n(\mathbf{H}_{\ell}, \mathbf{c}_{\ell}) \leq q_{l,1-{\beta_n}(\alpha)}^{*}$ holds. Since $p_{\ell} \leq \alpha$ implies that $\beta_{n,\ell}$ satisfies
       $$ \mathrm{FWER}^*_n(\beta_{n,\ell}) \leq \alpha , $$
        it follows $\beta_{n,\ell} \leq \beta_n(\alpha)$ by the definition of $\beta_n(\alpha)$. On the other hand, $w_n(\mathbf{H}_{\ell}, \mathbf{c}_{\ell}) \leq q_{l,1-{\beta_n}(\alpha)}^{*}$ implies that  
        \begin{align*}
            \beta_{n,\ell} = \frac{1}{B} \sum\limits_{b=1}^B \mathbbm{1}\left\{ W_n^{*, b}(\mathbf{H}_{\ell})  \geq w_n(\mathbf{H}_{\ell}, \mathbf{c}_{\ell} ) \right\} \geq \frac{1}{B} \sum\limits_{b=1}^B \mathbbm{1}\left\{ W_n^{*, b}(\mathbf{H}_{\ell})  \geq q_{l,1-{\beta_n}(\alpha)}^{*} \right\} \geq \beta_n(\alpha) + \frac{1}{B}
        \end{align*}
        by the definition of quantiles. This yields the contradiction $\beta_n(\alpha) \geq \beta_n(\alpha) + 1/B$.\\
        Secondly,  we aim to proof $w_n(\mathbf{H}_{\ell}, \mathbf{c}_{\ell}) > q_{l,1-{\beta_n}(\alpha)}^{*} \quad \Rightarrow \quad p_{\ell} \leq \alpha$.
        By the definition of quantiles, $w_n(\mathbf{H}_{\ell}, \mathbf{c}_{\ell}) > q_{l,1-{\beta_n}(\alpha)}^{*}$ implies 
        \begin{align*}
            \beta_{n,\ell} = \frac{1}{B} \sum\limits_{b=1}^B \mathbbm{1}\left\{ W_n^{*, b}(\mathbf{H}_{\ell})  \geq w_n(\mathbf{H}_{\ell}, \mathbf{c}_{\ell} ) \right\} \leq \frac{1}{B} \sum\limits_{b=1}^B \mathbbm{1}\left\{ W_n^{*, b}(\mathbf{H}_{\ell})  > q_{l,1-{\beta_n}(\alpha)}^{*} \right\} \leq \beta_n(\alpha) .
        \end{align*}
        Thus, the definition of $\beta_n(\alpha)$ yields that $\beta_{n,\ell}$ fulfills
        $$ p_{\ell} = \mathrm{FWER}^*_n(\beta_{n,\ell}) = \frac{1}{B} \sum\limits_{b=1}^B \mathbbm{1}\left\{ \exists k\in\{ 1,...,L \} : W_n^{*, b}(\mathbf{H}_{k}) > q_{k,1-\beta_{n,\ell}}^{*} \right\} \leq \alpha . $$

        For (ii), we note that $p = \min\{p_1,...,p_L\} \leq \alpha$ if and only if there exists an $\ell\in\{1,...,L\}$ such that $p_{\ell} \leq \alpha$. Due to (i), this holds whenever there exists an $\ell\in\{1,...,L\}$ such that $w_n(\mathbf{H}_{\ell}, \mathbf{c}_{\ell}) > q_{l,1-{\beta_n}(\alpha)}^{*}$ or, equivalently, $\max\limits_{\ell \in\{ 1, \ldots,L\}} {w_n(\mathbf{H}_{\ell},\mathbf{c}_{\ell})}/{q_{\ell,1-{\beta_n}(\alpha)}^{*}} > 1$.
   \end{myproof}
\subsection{Proof of Theorem~\ref{Wild}}
\begin{lemma}\label{WildLemma1}
We have
    \begin{align*}
    \sqrt{n} \widehat{\boldsymbol\mu}^G \xrightarrow{d} \mathbf{Z} = (Z_1,...,Z_k)^{\prime} \sim \mathcal{N}_k(\mathbf{0}_k, \boldsymbol\Sigma)
    \end{align*}
almost surely as $n\to\infty$ given the data $(\mathbf{X}, \boldsymbol\delta)$.
\end{lemma}
\begin{myproof}[Proof of Lemma~\ref{WildLemma1}]
Let $i\in\{1,...,k\}$ be arbitrary. 
We aim to apply the Lindeberg-Feller theorem with $$Z_{j,n_i} := \sqrt{n_i} G_{ij} \int\limits_0^{\tau} \int\limits_x^{\tau} \widehat{S}_i(t)\;\mathrm{d}t \dfrac{1}{\sqrt{(Y_i(x)-\Delta N_i(x))Y_i(x)}}  \;\mathrm{d}N_{ij}(x)  
$$ for all $j\in\{1,...,n_i\}$. Then, $Z_{1,n_i},...,Z_{n_i,n_i}$ are independent conditionally on $(\mathbf{X}, \boldsymbol\delta)$. Moreover, we have
\begin{align*}
    \E^*\left[ Z_{j,n_i} \right] = \sqrt{n_i} \E^*\left[ G_{ij}  \right] \int\limits_0^{\tau} \int\limits_x^{\tau} \widehat{S}_i(t)\;\mathrm{d}t \dfrac{1}{\sqrt{(Y_i(x)-\Delta N_i(x))Y_i(x)}}  \;\mathrm{d}N_{ij}(x) = 0
\end{align*} almost surely, where here and throughout 
$\E^*$ denotes the conditional expectation given $(\mathbf{X}, \boldsymbol\delta)$. It should be noted that all following statements about conditional expectations hold just almost surely but we will not always add this throughout this paper, for the sake of clarity.
Furthermore, it holds
\begin{align*}
   s_n^2 := \sum\limits_{j=1}^{n_i}\Var^*\left( Z_{j,n_i} \right) &= n_i\sum\limits_{j=1}^{n_i} \Var^*\left( G_{ij}  \right) \left(\int\limits_0^{\tau}\int\limits_x^{\tau} \widehat{S}_i(t)\;\mathrm{d}t \dfrac{1}{\sqrt{(Y_i(x)-\Delta N_i(x))Y_i(x)}}  \;\mathrm{d}N_{ij}(x)\right)^2
    \\&= n_i \int\limits_0^{\tau} \left(\int\limits_x^{\tau} \widehat{S}_i(t)\;\mathrm{d}t\right)^2 \dfrac{1}{{(Y_i(x)-\Delta N_i(x))Y_i(x)}}  \;\mathrm{d}N_{i}(x)
    \\&= \frac{n_i}{n} \widehat{\sigma}_i^2 \xrightarrow{a.s.} \kappa_i\sigma_i^2
\end{align*} as $n\to\infty$  {under $\P(X_{i1} \geq \tau) > 0$} by Section~S.5 in the supplement of Ditzhaus et al.\cite{RMST}, where here and throughout $\Var^{{*}}$ denotes the conditional variance given $(\mathbf{X}, \boldsymbol\delta)$.
For showing Lindeberg's condition, let $\varepsilon>0$ be arbitrary.
Then, we have
\begin{align*}
&\frac{1}{s_n^2}\sum\limits_{j=1}^{n_i}
    \E^*\left[ Z_{j,n_i}^2 \mathbbm{1}\left\{ Z_{j,n_i}^2 > \varepsilon^2 s_n^2\right\}\right]
    \\&=
    \frac{n}{n_i \widehat{\sigma}_i^2}\sum\limits_{j=1}^{n_i}n_i
    \E^*\left[  G_{ij}^2  \mathbbm{1}\left\{ n_iG_{ij}^2 \int\limits_0^{\tau}  \dfrac{\left(\int_x^{\tau} \widehat{S}_i(t)\;\mathrm{d}t\right)^2}{{(Y_i(x)-\Delta N_i(x))Y_i(x)}}  \;\mathrm{d}N_{ij}(x) > \varepsilon^2 \frac{n_i}{n} \widehat{\sigma}_i^2\right\}\right]\cdot \\&\quad
    \int\limits_0^{\tau} \dfrac{\left(\int_x^{\tau} \widehat{S}_i(t)\;\mathrm{d}t\right)^2}{{(Y_i(x)-\Delta N_i(x))Y_i(x)}}  \;\mathrm{d}N_{ij}(x)
    \\&\leq  \frac{n}{ \widehat{\sigma}_i^2}\sum\limits_{j=1}^{n_i}  \E^*\left[  G_{ij}^2  \mathbbm{1}\left\{ G_{ij}^2 \sup\limits_{x\in [0,\tau]}\left\{  \dfrac{\left(\int_x^{\tau} \widehat{S}_i(t)\;\mathrm{d}t\right)^2}{{(Y_i(x)-\Delta N_i(x))Y_i(x)}} \right\} > \varepsilon^2 \frac{1}{n} \widehat{\sigma}_i^2\right\}\right]\cdot \\&\quad
    \int\limits_0^{\tau} \dfrac{\left(\int_x^{\tau} \widehat{S}_i(t)\;\mathrm{d}t\right)^2}{{(Y_i(x)-\Delta N_i(x))Y_i(x)}}  \;\mathrm{d}N_{ij}(x)
    \\&= \E^*\left[  G_{i1}^2  \mathbbm{1}\left\{ G_{i1}^2 \sup\limits_{x\in [0,\tau]}\left\{  \dfrac{\left(\int_x^{\tau} \widehat{S}_i(t)\;\mathrm{d}t\right)^2}{{(n_i^{-1}Y_i(x)-n_i^{-1}\Delta N_i(x))n_i^{-1}Y_i(x)}} \right\} > \varepsilon^2 \frac{n_i^2}{n} \widehat{\sigma}_i^2\right\}\right]
    \\&\xrightarrow{a.s.} 0
\end{align*} as $n\to\infty$ by the dominated convergence theorem with integrable majorant $G_{i1}^2$.
Here, we use that
\begin{align}\label{eq:konvergenzen}
    &
    \sup\limits_{x\in [0,\tau]} \left| n_i^{-1}Y_i(x) - y_i(x) \right| \xrightarrow{a.s.} 0, \quad \sup\limits_{x\in [0,\tau]} \left| n_i^{-1}N_i(x) - \nu_i(x) \right| \xrightarrow{a.s.} 0 \quad\text{ and } \quad \widehat{\sigma}_i^2 \xrightarrow{a.s.} {\sigma}_i^2
\end{align} as $n\to\infty$
holds  {under $\P(X_{i1} \geq \tau) > 0$} by Section~S.5 and S.6 in the supplement of Ditzhaus et al.\cite{RMST}
such that
\begin{align*}
    &\P\left(\mathbbm{1}\left\{ G_{i1}^2 \sup\limits_{x\in [0,\tau]}\left\{  \dfrac{\left(\int_x^{\tau} \widehat{S}_i(t)\;\mathrm{d}t\right)^2}{{(n_i^{-1}Y_i(x)-n_i^{-1}\Delta N_i(x))n_i^{-1}Y_i(x)}} \right\} > \varepsilon^2 \frac{n_i^2}{n} \widehat{\sigma}_i^2\right\} > \varepsilon \mid (\mathbf{X},\boldsymbol{\delta}) \right)
    \\&\leq\P\left( G_{i1}^2 \sup\limits_{x\in [0,\tau)}\left\{  \dfrac{\tau^2}{{(n_i^{-1}Y_i(x)-n_i^{-1}\Delta N_i(x))n_i^{-1}Y_i(x)}} \right\} > \varepsilon^2 \frac{n_i^2}{n} \widehat{\sigma}_i^2 \mid (\mathbf{X},\boldsymbol{\delta}) \right)
    \xrightarrow{a.s.} 0
\end{align*}  as $n\to\infty$ for all $\varepsilon>0$ follows.
Thus, the Lindeberg-Feller theorem implies \begin{align*}
   \sqrt{n_i} \widehat{\mu}_i^G = \sum\limits_{j=1}^{n_i} Z_{j,n_i} \xrightarrow{d} \mathcal{N}(0,\kappa_i\sigma_i^2)
\end{align*}
 almost surely as $n\to\infty$ given the data $(\mathbf{X}, \boldsymbol\delta)$.
 Hence, the statement of the lemma follows by Slutsky's lemma.
\end{myproof}

\begin{lemma}\label{WildLemma2}
    We have
    \begin{align*}
        \P\left(\left|\widehat{\sigma}_i^{G2} - \sigma_i^{2}\right| > \varepsilon \mid (\mathbf{X},\boldsymbol{\delta})\right)\xrightarrow{a.s.} 0 
    \end{align*} as $n\to\infty$ for all $i\in\{1,...,k\}$.
\end{lemma}
\begin{myproof}[Proof of Lemma~\ref{WildLemma2}]
    Let $i\in\{1,...,k\}$ be arbitrary. 
    Then, it holds
    \begin{align*}
        \E^*\left[ \widehat{\sigma}_i^{G2} \right]
        &= \sum\limits_{j=1}^{n_i} n\E^*\left[ G_{ij}^2 \right] \int\limits_{[0,{\tau}]} \left(\int\limits_x^{\tau} \widehat{S}_i(t) \;\mathrm{d}t \right)^2 \dfrac{1}{{(Y_i(x)-\Delta N_i(x))Y_i(x)}}  \;\mathrm{d}N_{ij}(x)
        \\&=   \widehat{\sigma}_i^2 \xrightarrow{a.s.} \sigma_i^{2}
    \end{align*}
    as $n\to\infty$ and, analogously,
    \begin{align*}
        \E^*\left[ (\widehat{\sigma}_i^{G2})^2 \right]
        &\leq 
        (\widehat{\sigma}_i^{2})^2 + (C-1)
        \sum\limits_{j=1}^{n_i} n^2\left(\int\limits_{[0,{\tau}]} \left(\int\limits_x^{\tau} \widehat{S}_i(t) \;\mathrm{d}t \right)^2 \dfrac{1}{{(Y_i(x)-\Delta N_i(x))Y_i(x)}}  \;\mathrm{d}N_{ij}(x)\right)^2
        \\&=\widehat{\sigma}_i^{4} + (C-1)
         n^2\int\limits_{[0,{\tau})} \left(\int\limits_x^{\tau} \widehat{S}_i(t) \;\mathrm{d}t \right)^4 \dfrac{1}{{(Y_i(x)-\Delta N_i(x))^2Y_i^2(x)}}  \;\mathrm{d}N_{i}(x)
         \\&\leq\widehat{\sigma}_i^{4} + (C-1)
         \frac{n^2}{n_i^{3}}\int\limits_{[0,{\tau})}  \dfrac{\tau^4}{{(n_i^{-1}Y_i(x)-n_i^{-1}\Delta N_i(x))^2n_i^{-1}Y_i(x)}}  \;\mathrm{d}\widehat{A}_{i}(x)
         \\&\xrightarrow{a.s.} \sigma_i^{4}
    \end{align*}
    as $n\to\infty$ by (\ref{eq:konvergenzen}).
    Thus, it follows
\begin{align*}
    \P\left(\left|\widehat{\sigma}_i^{G2} - \sigma_i^{2}\right| > \varepsilon \mid (\mathbf{X},\boldsymbol{\delta})\right)\xrightarrow{a.s.} 0 
\end{align*}
as $n\to\infty$ for all $\varepsilon > 0$ by Chebyshev's inequality. 
Hence, the statement of the lemma follows.
\end{myproof}
%

\begin{myproof}[Proof of Theorem~\ref{Wild}]
Lemma~\ref{WildLemma1} and \ref{WildLemma2} provide that there exists a measurable set $\Omega^{\prime} \subset \Omega$ with $\P(\Omega^{\prime})=1$ such that
\begin{align*}
    \sup\limits_{z_1,...,z_k \in\R} \left|\P\left(\sqrt{n} \widehat{\mu}^G_1 \leq z_1, ..., \sqrt{n} \widehat{\mu}^G_k \leq z_k  \mid (\mathbf{X}, \boldsymbol\delta)\right)(\omega) - \P\left(Z_1 \leq z_1, ..., Z_k \leq z_k \right)\right|\to 0
\end{align*}
    and
    \begin{align*}
        \P\left(\left|\widehat{\sigma}_i^{G2} - \sigma_i^{2}\right| > \varepsilon \mid (\mathbf{X},\boldsymbol{\delta})\right)(\omega) \to 0 
    \end{align*} as $n\to\infty$ for all $i\in\{1,...,k\}$, $\omega\in\Omega^{\prime}$.
    Then, by running through the same steps as in the proof of Theorem~\ref{Asy}, we get
    \begin{align*}
        \sup\limits_{z \in\R} \left|\P\left(W_n^{G}(\mathbf{H}) \leq z  \mid (\mathbf{X}, \boldsymbol\delta)\right)(\omega) - \P\left(Z \leq z \right)\right|\to 0
    \end{align*} as $n\to\infty$ for all $\omega\in\Omega^{\prime}$, where $Z\sim \chi^2_{\text{rank}(\mathbf{H})}$.
\end{myproof}

\subsection{Proofs of Theorem~\ref{MultiAsy}, \ref{Multigw} and \ref{MultiWild}}
The theorems about the joint convergences follow now easily from the previous results. Therefore, we apply Slutsky's lemma.
For Theorem~\ref{MultiAsy}, we combine (\ref{eq:clt}) and (\ref{eq:varConv}), for Theorem~\ref{Multigw} Lemma~\ref{gwCLT} and \ref{gwCov} and for Theorem~\ref{MultiWild} Lemma~\ref{WildLemma1} and \ref{WildLemma2}.
Then, we use the continuous mapping theorem with map
$$ \R^k \times \R^{k\times k} \ni (\mathbf{m},\mathbf{S}) \mapsto \left((\mathbf{H}_{\ell} \mathbf{m})^{\prime} (\mathbf{H}_{\ell}\mathbf{S}\mathbf{H}_{\ell}^{\prime})^+ \mathbf{H}_{\ell}\mathbf{m} \right)_{\ell \in\{1,...,L\}} \in\R^L. $$ The map is continuous on $\R^k \times \{\boldsymbol\Sigma\}$ due to $\sigma_i^2 > 0$ for all $i\in\{1,...,k\}.$
The three theorems follow, respectively.

\section{Additional Simulations}\label{sec:AddSimu}
In this section, the results of addition simulation studies are provided. First,  
{the simulation setup from Section~\ref{ssec:SimuSetup} is repeated for the asymptotic approaches by using larger sample sizes. Next,}
 we show a setup where the groupwise bootstrap outperforms the permutation approach with Bonferroni-correction in terms of empirical power. {Finally,} a simulation study inspired by the data example in Section~\ref{sec:Data} is investigated.

\subsection{{Simulations for Analyzing the Asymptotic Behaviour}}\label{ssec:LargeSimu}
\FloatBarrier
{
We have seen in Section~\ref{ssec:SimuTypeI} that the three asymptotic approaches (\textit{asymptotic\_global, asymptotic, asymptotic\_bonf}) do not lead to a good type I error control. Thus, one may be interested in how large the sample sizes should be to obtain a good control of the type I error for these naive methods.
Therefore, in this section we consider the simulation setup from Section~\ref{ssec:SimuSetup} again with an increased factor for the scaling of the sample sizes, that is $K\in\{6,8,10\}$, resulting in sample sizes from 60 up to 200 in the groups. Furthermore, only the three asymptotic approaches (\textit{asymptotic\_global, asymptotic, asymptotic\_bonf}) are considered under the null hypothesis. The performance of these methods regarding the power was already quite good for small and medium sample sizes, see Section~\ref{ssec:SimuTypeII} for details. This is why we did not analyze the power for larger sample sizes.
Note that the censoring rates for the different scenarios are as shown in Table~\ref{tab:Censrates}.
}

\begin{figure}[t!]
    \centering
    \includegraphics[width=0.9\textwidth]{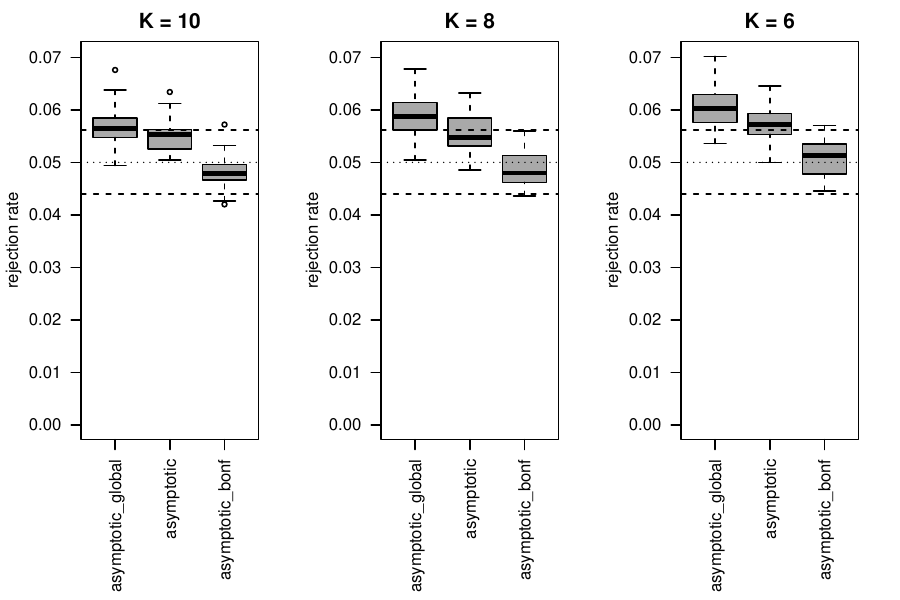}
    \caption{{Rejection rates over all settings under the null hypothesis for the Dunnett-type contrast matrix.}}
    \label{fig:DunnettLARGE}
\end{figure}

\begin{figure}[b!]
    \centering
    \includegraphics[width=0.9\textwidth]{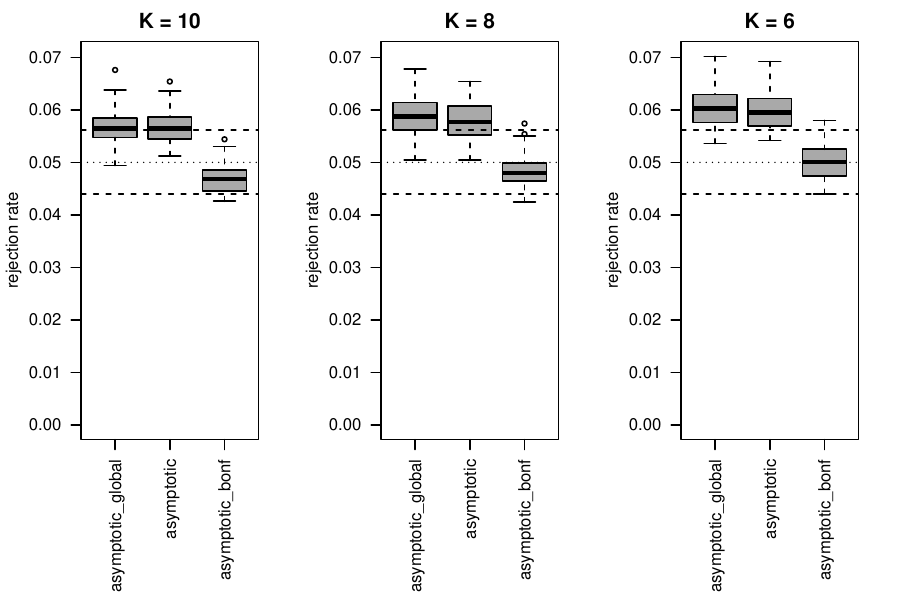}
    \caption{{Rejection rates over all settings under the null hypothesis for the Tukey-type contrast matrix.}}
    \label{fig:TukeyLARGE}
\end{figure}

\begin{figure}[t!]
    \centering
    \includegraphics[width=0.9\textwidth]{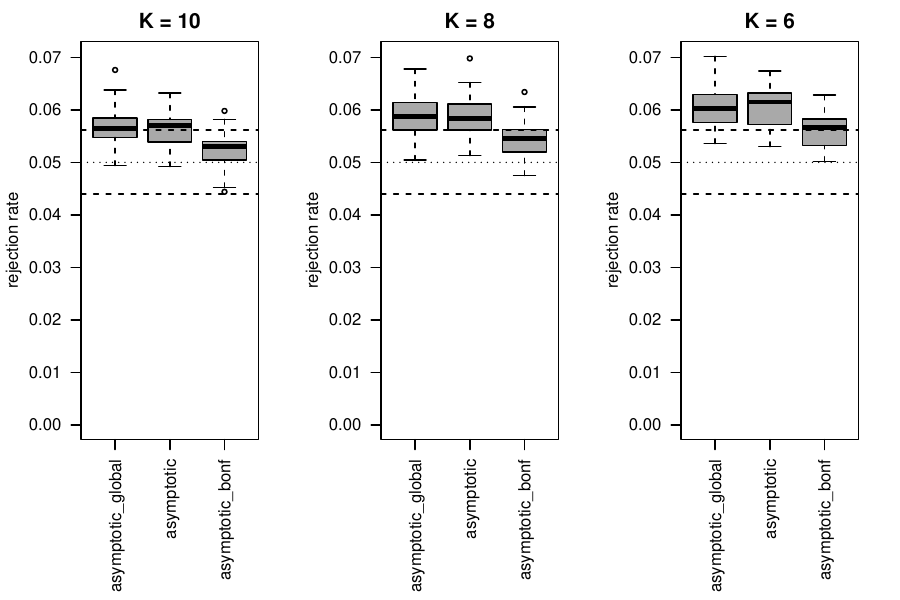}
    \caption{{Rejection rates over all settings under the null hypothesis for the Grand-mean-type contrast matrix.}}
    \label{fig:GrandmeanLARGE}
\end{figure}

\begin{figure}[H]
    \centering
    \includegraphics[height=0.4\textheight]{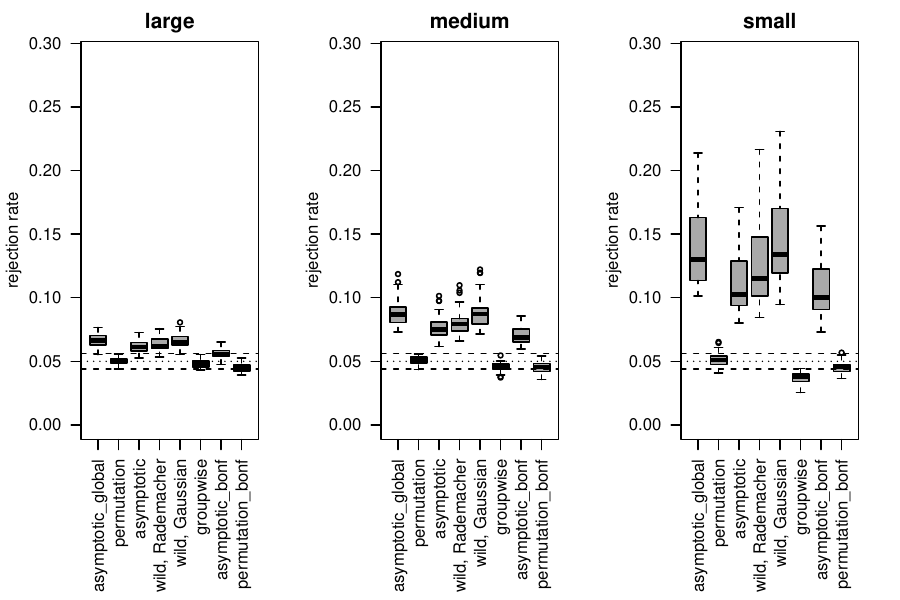}
    \caption{Rejection rates over all settings under the null hypothesis for the Dunnett-type contrast matrix.}
    \label{fig:Dunnett_asymptotic}
\end{figure}

\begin{figure}[H]
    \centering
    \includegraphics[height=0.4\textheight]{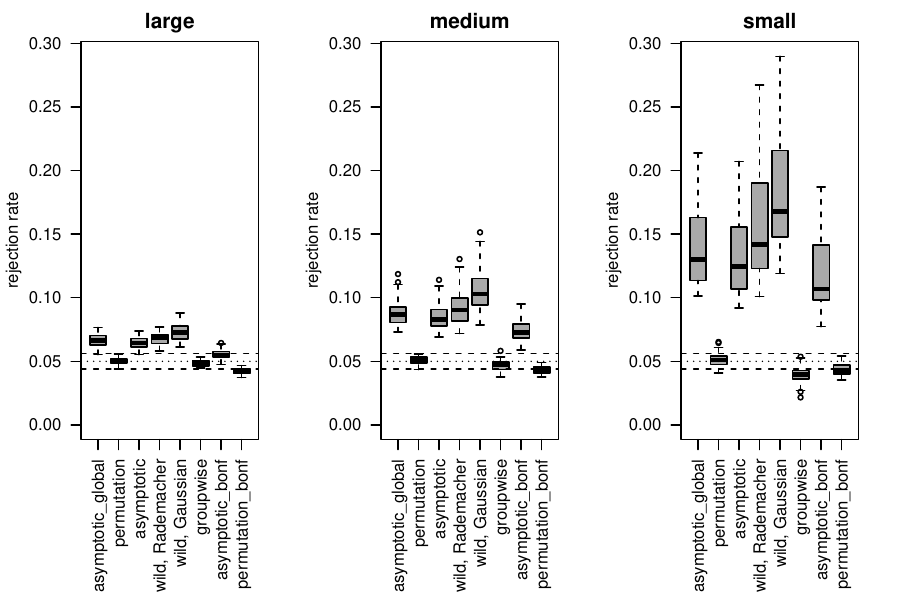}
    \caption{Rejection rates over all settings under the null hypothesis for the Tukey-type contrast matrix.}
    \label{fig:Tukey_asymptotic}
\end{figure}

\begin{figure}[H]
    \centering
    \includegraphics[height=0.4\textheight]{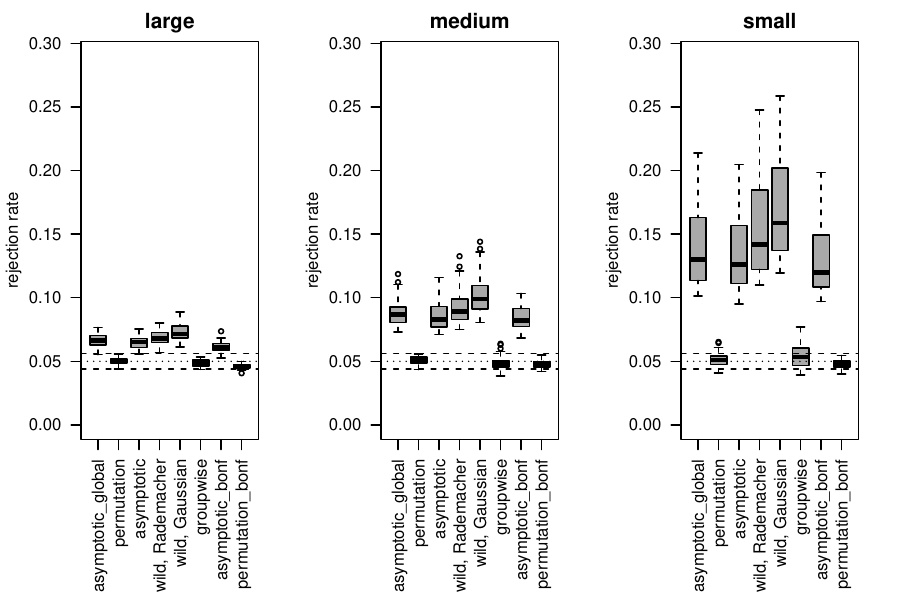}
    \caption{Rejection rates over all settings under the null hypothesis for the Grand-mean-type contrast matrix.}
    \label{fig:GrandMean_asymptotic}
\end{figure}

{
In Figures~\ref{fig:DunnettLARGE} to \ref{fig:GrandmeanLARGE}, the rejection rates across all settings are illustrated for the three different contrast matrices. It can be seen that the empirical type I error rates are quite close to the desired level of significance of 5\% for large sample sizes in all scenarios. The rejection rates seem to tend more and more to 5\% as the sample sizes increase. However, the difference between the rejection rates for different values of $K\in\{6,8,10\}$ is rather small, indicating that the convergence is relatively slow. \\
It can be observed that we need quite large sample sizes to obtain a good type I error control for the multiple asymptotic and the global asymptotic test without Bonferroni-correction. Even for $K = 10$, i.e. sample sizes between 100 and 200 in each group, these tests are still slightly liberal. The empirical type I error rates for the multiple asymptotic and the global asymptotic test without Bonferroni-correction reach up to 7.02\%. 
\\
By using a Bonferroni-correction, the asymptotic test does not need very large sample sizes to control the level of significance. Here, $K=6$, i.e. sample sizes between 60 and 120 in each group, or even $K=4$ seems to be enough as can be seen in Figures~\ref{fig:Dunnett_asymptotic} to \ref{fig:GrandMean_asymptotic}.
}


\FloatBarrier
\subsection{Additional Simulations under Non-Exchangeability}\label{ssec:NonExSimu}
\begin{figure}[bth]
    \centering
    \includegraphics[width=\textwidth]{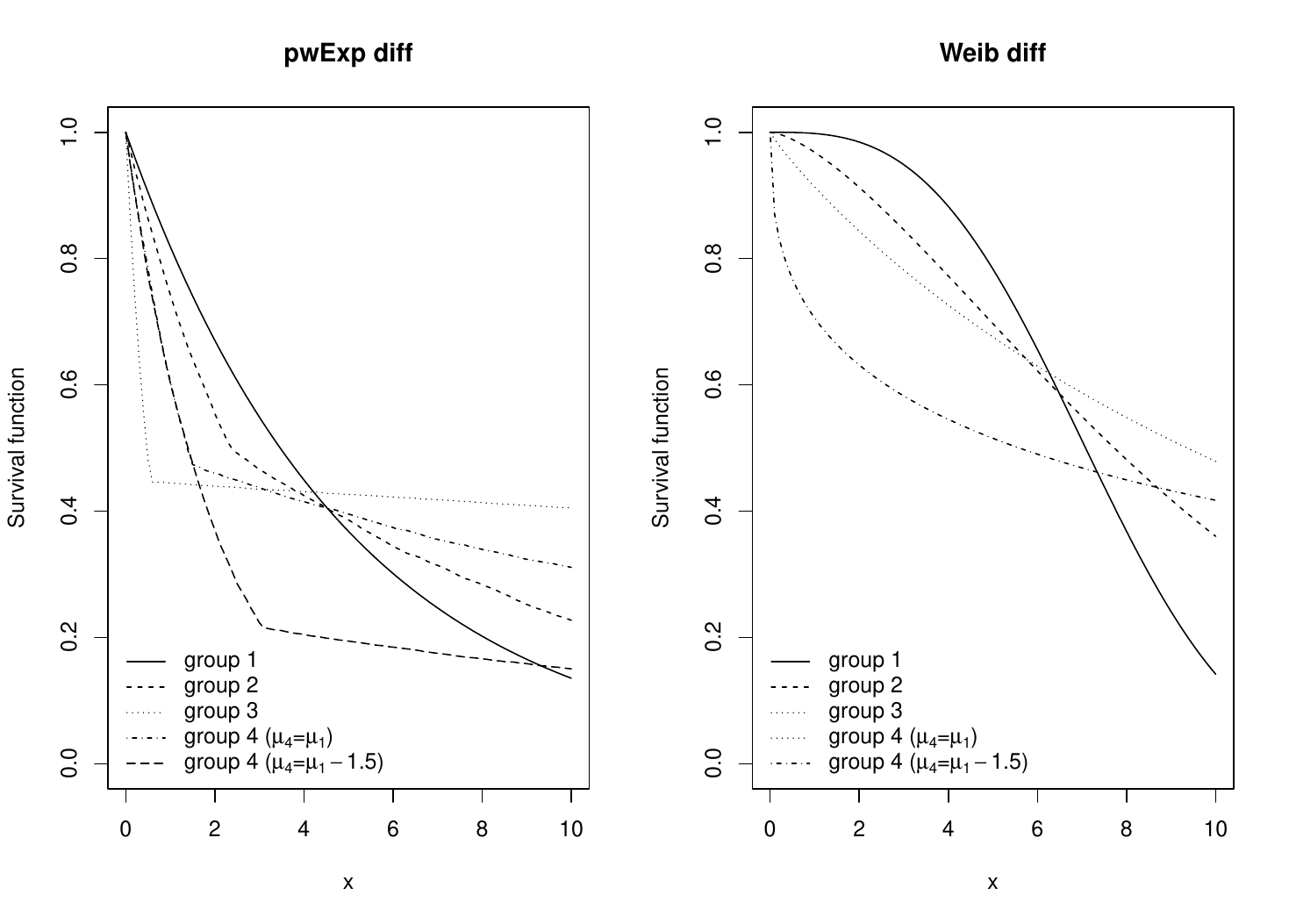}
    \caption{The survival functions of the two settings under the null hypothesis as well as under the alternative $\mu_4 = \mu_1 - 1.5$. Note that the survival functions of group 3 and 4 coincide under the null hypothesis for the setting \textit{Weib diff}.}
    \label{fig:AddSurv}
\end{figure}

In Section~\ref{sec:Simu}, the empirical power of the groupwise approach and the permutation approach with Bonferroni-correction seems comparable over all simulation setups. However, the Bo{n}ferroni-correction is known to have low power for a large number of hypotheses. 
Thus, we aim to motivate that the groupwise bootstrap approach may perform better than the permutation approach with Bonferroni-correction in specific setups in this section. Therefore, we consider again $k=4$ groups with sample sizes $\mathbf{n} = (40,80,40,80)$, hypotheses matrices as in Section~\ref{sec:Simu} and $\alpha = 5\%$. Furthermore, we generated $N_{sim} =5000$ simulation runs with $B=1999$ resampling iterations.
In contrast to the simulation study in Section~\ref{sec:Simu}, the survival times are drawn from different distributions for all groups as follows:
\begin{itemize}
    \item Different piece-wise exponential distributions (\textit{pwExp diff}): $T_{11}\sim Exp(0.2)$, \\$T_{21}$ with hazard function $t\mapsto 0.3\cdot\mathbbm{1}\{ t \leq \lambda_{10} \} + 0.1\cdot\mathbbm{1}\{ t > \lambda_{10} \}$, \\$T_{31}$ with hazard function $t\mapsto 1.5\cdot\mathbbm{1}\{ t \leq \lambda_{11} \} + 0.01\cdot\mathbbm{1}\{ t > \lambda_{11} \}$ and \\$T_{41}$ with hazard function $t\mapsto 0.5\cdot\mathbbm{1}\{ t \leq \lambda_{\delta, 5} \} + 0.05\cdot\mathbbm{1}\{ t > \lambda_{\delta, 5} \}$,
    \item Different Weibull distributions (\textit{Weib diff}): $T_{11} \sim Weib(3,8)$, $T_{21} \sim Weib(1.5,\lambda_{0,8})$, $T_{31} \sim Weib(\lambda_{0,9},14)$ and $T_{41} \sim Weib(\lambda_{\delta,9},14)$.
\end{itemize}
Here, the parameters $\lambda_{10}$ and $\lambda_{11}$ are determined such that the RMST equals $\mu_1$. Hence, note that only $\mu_4$ differs under the alternative hypothesis but the distributions of the survival times differ across the groups under the null and alternative hypothesis.
In Figure~\ref{fig:AddSurv}, the different survival functions are illustrated. For the censoring times, the same distributions as in Section~\ref{sec:Simu} are considered, i.e. \textit{equal; unequal, high} and \textit{unequal, low}. The resulting censoring rates can be found in Table~\ref{tab:AddCens} and reach from 11 up to 62\%.
\begin{table}[b!]
\centering
\begin{tabular}{lllrrrr}
  \hline
$\delta$ & distribution & censoring distribution & group 1 & group 2 & group 3 & group 4 \\ 
  \hline
0.0 & pwExp diff & equal & 0.21 & 0.27 & 0.40 & 0.33 \\ 
  0.0 & pwExp diff & unequal, high & 0.38 & 0.44 & 0.42 & 0.37 \\ 
  0.0 & pwExp diff & unequal, low & 0.20 & 0.27 & 0.39 & 0.22 \\ 
  0.0 & Weib diff & equal & 0.34 & 0.45 & 0.53 & 0.53 \\ 
  0.0 & Weib diff & unequal, high & 0.49 & 0.57 & 0.62 & 0.57 \\ 
  0.0 & Weib diff & unequal, low & 0.29 & 0.45 & 0.48 & 0.31 \\ 
  1.5 & pwExp diff & equal & 0.21 & 0.27 & 0.40 & 0.16 \\ 
  1.5 & pwExp diff & unequal, high & 0.38 & 0.44 & 0.42 & 0.25 \\ 
  1.5 & pwExp diff & unequal, low & 0.20 & 0.27 & 0.39 & 0.11 \\ 
  1.5 & Weib diff & equal & 0.34 & 0.45 & 0.53 & 0.44 \\ 
  1.5 & Weib diff & unequal, high & 0.49 & 0.57 & 0.62 & 0.48 \\ 
  1.5 & Weib diff & unequal, low & 0.29 & 0.45 & 0.48 & 0.34 \\ 
   \hline
\end{tabular}
\caption{Censoring rates for the additional simulation} 
\label{tab:AddCens}
\end{table}

\begin{figure}[bth]
    \centering
    \includegraphics[width=\textwidth]{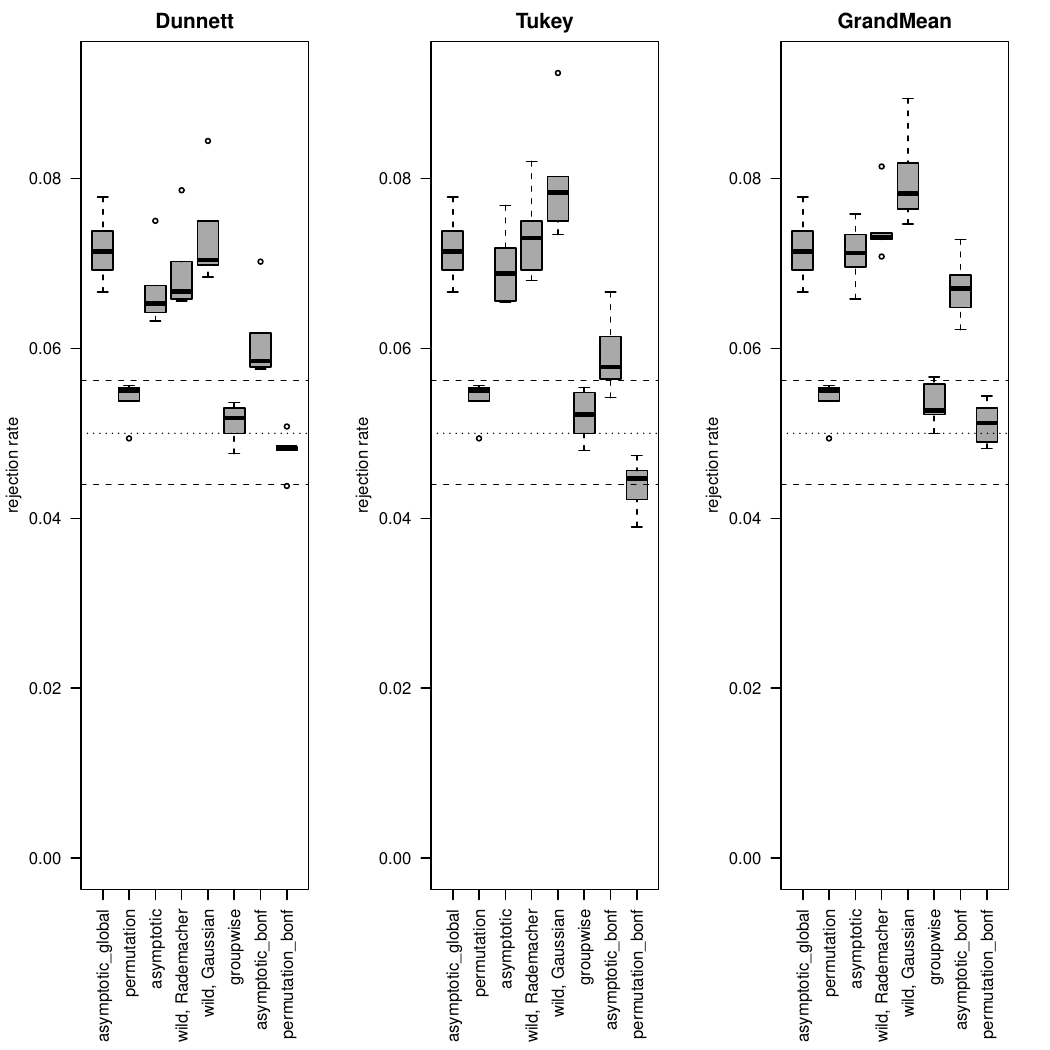}
    \caption{Rejection rates over all settings under the null hypothesis. The dashed lines represent the borders of the binomial confidence interval [4.4\%,5.62\%].}
    \label{fig:AddH0}
\end{figure}

\begin{figure}[ht]
    \centering
    \includegraphics[height=0.35\textheight]{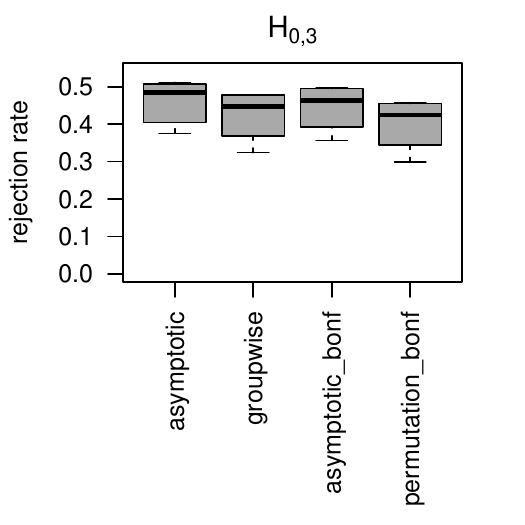}
    \caption{Rejection rates of the false local hypothesis over all settings under the alternative hypothesis for the Dunnett-type contrast matrix.}
    \label{fig:AddDunnett_H1}
\end{figure}

\begin{figure}[ht]
    \centering
    \includegraphics[height=0.5\textheight]{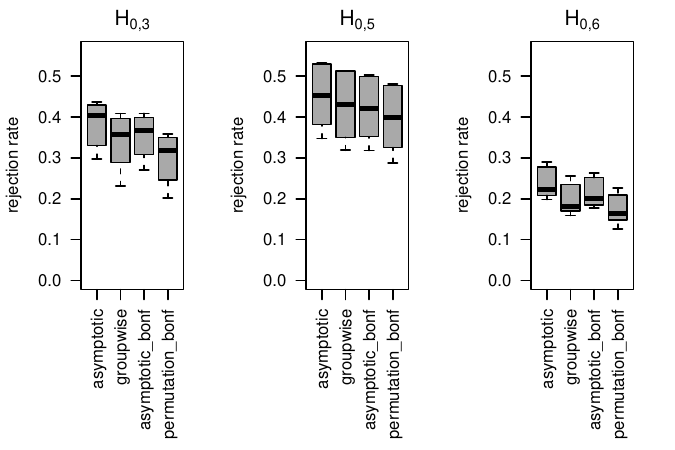}
    \caption{Rejection rates of all false local hypotheses over all settings under the alternative hypothesis for the Tukey-type contrast matrix.}
    \label{fig:AddTukey_H1}
\end{figure}

\begin{figure}[ht]
    \centering
    \includegraphics[width = \textwidth]{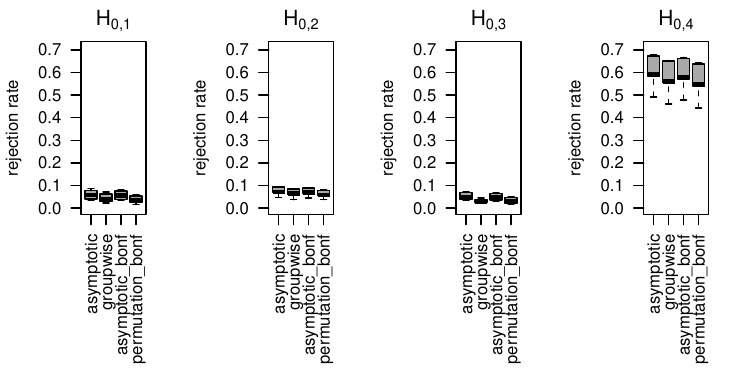}
    \caption{Rejection rates of the false local hypothesis over all settings under the alternative hypothesis for the Grand-mean-type contrast matrix.}
    \label{fig:AddGrandMean_H1}
\end{figure}

In Figure~\ref{fig:AddH0}, the rejection rates over all settings under the null hypothesis are presented. Here, it is observable that the groupwise bootstrap and the permutation approach with Bonferroni-correction perform well in terms of type I error control for the multiple testing problem. The permutation approach with Bonferroni-correction tends to be too conservative for the Tukey-type contrast matrix. Furthermore, the asymptotic approaches and the wild bootstrap are too liberal and, thus, they do not seem to control the family-wise type I error. However, the empirical power of the groupwise bootstrap is slightly higher than of the permutation approach with {B}onferroni{-}correction in most of the scenarios which can be seen in Table~\ref{tab:Dunn} to \ref{tab:GM}. Only for hypothesis $\mathcal{H}_{0,3}$ for the Grand-mean-type contrast matrix, the permutation approach with Bonferroni-correction has a higher power than the groupwise bootstrap in some scenarios. The empirical powers of the false hypotheses are also illustrated in Figure~\ref{fig:AddDunnett_H1} to \ref{fig:AddGrandMean_H1}, where it is observable that the groupwise bootstrap tends to have a higher empirical power than the permutation approach with Bonferroni-correction, particularly in Figure~\ref{fig:AddTukey_H1}. The asymptotic approaches even have higher empirical powers in several scenarios but, however, they can not control the family-wise error adequately which can be seen in Figure~\ref{fig:AddH0}.

\begin{table}[ht]
\centering
\begin{tabular}{lllcccc}
  \hline
hypothesis & distribution & censoring distribution & asymptotic & groupwise & asymptotic bonf & permutation bonf \\ 
  \hline
$\mathcal H_{0, 3 }$ & pwExp diff & equal & 0.495 & 0.461 & 0.474 & 0.437 \\ 
&  pwExp diff & unequal, high & 0.376 & 0.324 & 0.357 & 0.299 \\ 
&  pwExp diff & unequal, low & 0.474 & 0.435 & 0.451 & 0.412 \\ 
&  Weib diff & equal & 0.507 & 0.479 & 0.495 & 0.455 \\ 
&  Weib diff & unequal, high & 0.405 & 0.368 & 0.393 & 0.345 \\ 
&  Weib diff & unequal, low & 0.512 & 0.477 & 0.497 & 0.458 \\ 
   \hline
\end{tabular}
\caption{Rejection rates of the false hypothesis for the Dunnett-type contrast matrix with $\delta =$ 1.5} 
\label{tab:Dunn}
\end{table}
\begin{table}[ht]
\centering
\begin{tabular}{lllcccc}
  \hline
hypothesis & distribution & censoring distribution & asymptotic & groupwise & asymptotic bonf & permutation bonf \\ 
  \hline
$\mathcal H_{0, 3 }$ & pwExp diff & equal & 0.414 & 0.370 & 0.378 & 0.331 \\ 
&  pwExp diff & unequal, high & 0.298 & 0.231 & 0.270 & 0.201 \\ 
&  pwExp diff & unequal, low & 0.392 & 0.345 & 0.356 & 0.306 \\ 
&  Weib diff & equal & 0.430 & 0.397 & 0.399 & 0.350 \\ 
&  Weib diff & unequal, high & 0.331 & 0.288 & 0.308 & 0.245 \\ 
&  Weib diff & unequal, low & 0.436 & 0.408 & 0.409 & 0.358 \\ 
   \hline
$\mathcal H_{0, 5 }$ &pwExp diff & equal & 0.532 & 0.513 & 0.503 & 0.481 \\ 
&  pwExp diff & unequal, high & 0.382 & 0.350 & 0.352 & 0.325 \\ 
&  pwExp diff & unequal, low & 0.530 & 0.513 & 0.500 & 0.478 \\ 
&  Weib diff & equal & 0.444 & 0.423 & 0.412 & 0.391 \\ 
&  Weib diff & unequal, high & 0.348 & 0.319 & 0.318 & 0.288 \\ 
&  Weib diff & unequal, low & 0.460 & 0.440 & 0.428 & 0.407 \\ 
   \hline
$\mathcal H_{0, 6 }$ &pwExp diff & equal & 0.215 & 0.174 & 0.190 & 0.156 \\ 
&  pwExp diff & unequal, high & 0.198 & 0.159 & 0.177 & 0.127 \\ 
&  pwExp diff & unequal, low & 0.208 & 0.169 & 0.185 & 0.149 \\ 
&  Weib diff & equal & 0.290 & 0.256 & 0.262 & 0.226 \\ 
&  Weib diff & unequal, high & 0.231 & 0.188 & 0.211 & 0.171 \\ 
&  Weib diff & unequal, low & 0.277 & 0.235 & 0.252 & 0.209 \\ 
   \hline
\end{tabular}
\caption{Rejection rates of the false hypotheses for the Tukey-type contrast matrix with $\delta =$ 1.5} 
\end{table}
\begin{table}[ht]
\centering
\begin{tabular}{lllcccc}
  \hline
hypothesis & distribution & censoring distribution & asymptotic & groupwise & asymptotic bonf & permutation bonf \\ 
  \hline
$\mathcal H_{0, 1 }$ & pwExp diff & equal & 0.047 & 0.032 & 0.044 & 0.026 \\ 
&  pwExp diff & unequal, high & 0.037 & 0.024 & 0.036 & 0.017 \\ 
& pwExp diff & unequal, low & 0.041 & 0.031 & 0.039 & 0.025 \\ 
&  Weib diff & equal & 0.087 & 0.073 & 0.081 & 0.060 \\ 
&  Weib diff & unequal, high & 0.070 & 0.050 & 0.065 & 0.042 \\ 
&  Weib diff & unequal, low & 0.078 & 0.062 & 0.075 & 0.053 \\ 
   \hline
$\mathcal H_{0, 2 }$ & pwExp diff & equal & 0.071 & 0.065 & 0.067 & 0.059 \\ 
&  pwExp diff & unequal, high & 0.048 & 0.038 & 0.045 & 0.038 \\ 
&  pwExp diff & unequal, low & 0.071 & 0.063 & 0.067 & 0.058 \\ 
&  Weib diff & equal & 0.094 & 0.085 & 0.090 & 0.079 \\ 
&  Weib diff & unequal, high & 0.072 & 0.060 & 0.070 & 0.058 \\ 
&  Weib diff & unequal, low & 0.094 & 0.086 & 0.090 & 0.082 \\ 
   \hline
$\mathcal H_{0, 3 }$ &pwExp diff & equal & 0.035 & 0.024 & 0.033 & 0.021 \\ 
&  pwExp diff & unequal, high & 0.038 & 0.025 & 0.036 & 0.018 \\ 
&  pwExp diff & unequal, low & 0.038 & 0.025 & 0.035 & 0.022 \\ 
&  Weib diff & equal & 0.071 & 0.046 & 0.068 & 0.051 \\ 
&  Weib diff & unequal, high & 0.062 & 0.032 & 0.059 & 0.039 \\ 
&  Weib diff & unequal, low & 0.069 & 0.039 & 0.064 & 0.048 \\ 
   \hline
 $\mathcal H_{0, 4 }$ & pwExp diff & equal & 0.678 & 0.653 & 0.667 & 0.642 \\ 
&  pwExp diff & unequal, high & 0.590 & 0.556 & 0.580 & 0.540 \\ 
&  pwExp diff & unequal, low & 0.674 & 0.650 & 0.662 & 0.636 \\ 
&  Weib diff & equal & 0.585 & 0.558 & 0.570 & 0.543 \\ 
&  Weib diff & unequal, high & 0.492 & 0.460 & 0.479 & 0.442 \\ 
&  Weib diff & unequal, low & 0.590 & 0.560 & 0.579 & 0.553 \\ 
   \hline
\end{tabular}
\caption{Rejection rates of the false hypotheses for the Grand-mean-type contrast matrix with $\delta =$ 1.5} 
\label{tab:GM}
\end{table}


\FloatBarrier
\subsection{Simulation inspired by the Data Example}\label{ssec:DataSimu}
\begin{figure}[H]
    \centering
    \includegraphics[width=1.05\textwidth]{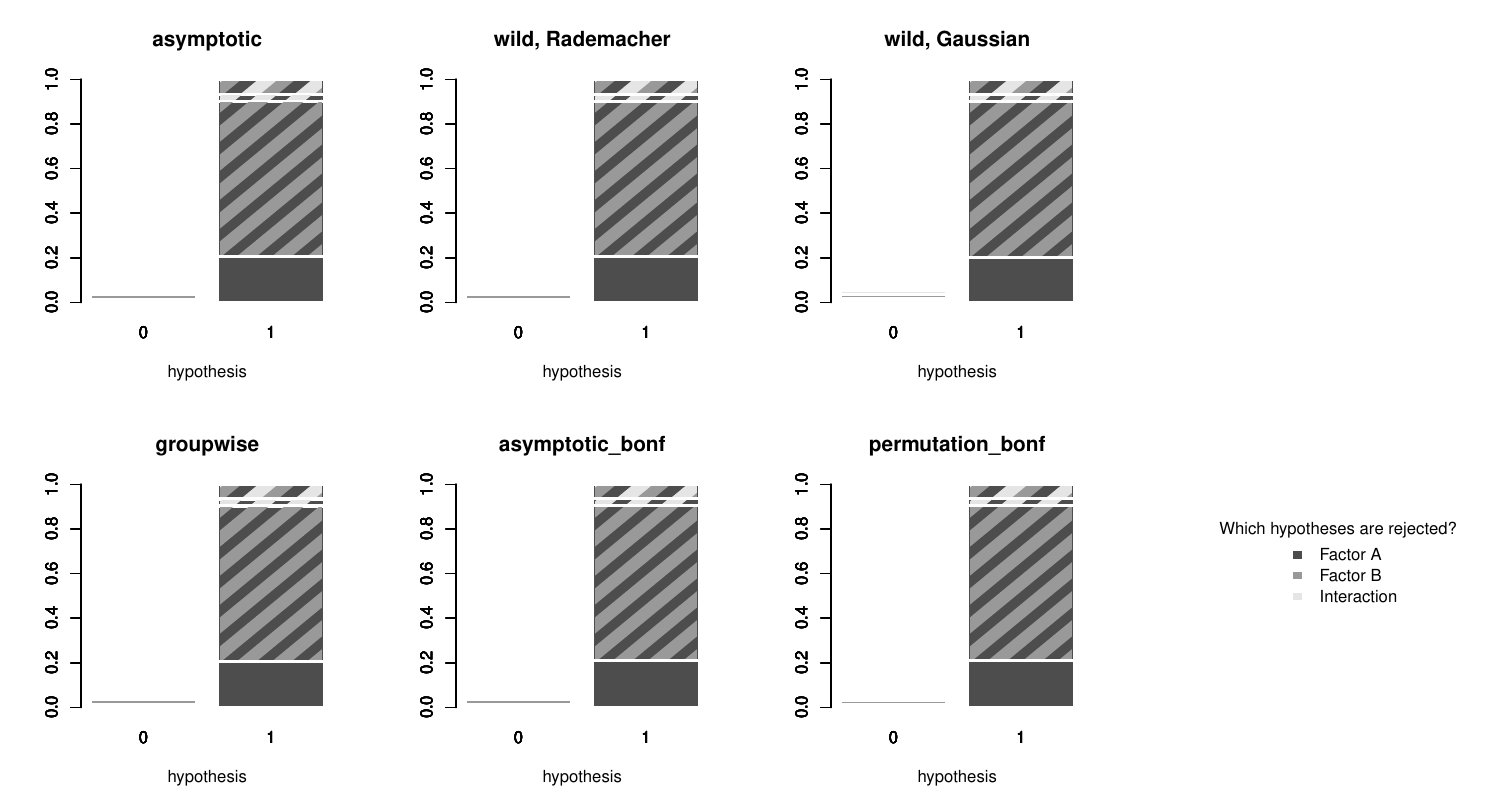}
    \caption{Rejection rates for the simulation inspired by the data example under the null (0) and under the alternative (1) hypothesis. The heights of the bars represent the rates of the rejections caused by the corresponding hypotheses. Two- and Three-colored bars indicate that the corresponding two or three hypotheses are rejected simultaneously. The overall height represents the rate of global rejections.}
    \label{fig:barplot}
\end{figure}

Since the Simulation study in Section~\ref{sec:Simu} does not fit perfectly to the data example about the occurrence of hay fever in Section~\ref{sec:Data}, we also considered a small simulation setup inspired by the data example. Therefore, we considered $k=4$ groups with sample sizes $\mathbf{n} = (450,481,654,649)$, the hypotheses matrices as in Section~\ref{sec:Data}, i.e. $\mathbf{H} := [\mathbf{H}_A^{\prime}, \mathbf{H}_B^{\prime}, \mathbf{H}_{AB}^{\prime}]^{\prime}$, and $\alpha = 5\%$. Moreover, $N_{sim} =5000$ simulation runs with $B=19999$ resampling iterations were generated.
The survival times of group $i$ were simulated from a distribution with the Kaplan-Meier estimator of the pooled sample under the null and of the $i$th sample under the alternative hypothesis as distribution function. Analogously, the Kaplan-Meier estimators for the censoring times of the different samples are used for the data generation of the censoring times. Proceeding as described leads to a censoring rate of 82\% in all groups under the null hypothesis and censoring rates from 72\% up to 89\% under the alternative hypothesis.
\\
In Table~\ref{tab:my_label}, the resulting rejection rates are shown. It is observable that all methods seem to control the global level of significance of 5\% quite accurately under the given scenario. However, the asymptotic and bootstrap approaches seem to be conservative with a family-wise error rate of less than 2\%.
Furthermore, all methods have a quite high empirical power under the alternative hypothesis. The power of the global approaches is around 90\% while all methods for the multiple testing problem have a power of 100\%. In Figure~\ref{fig:barplot}, it is shown how the rejection rates of the multiple testing procedures result from the local decisions. Here, it can be seen that the methods detect both of the main effects simultaneously in around 70\% of the simulation runs, only the main effect of factor A in 20\% and all main and interaction effects in 6\% under the alternative hypothesis. Furthermore, the methods seem to yield very similar local test decisions.
\begin{table}[bth]
    \centering
    \begin{tabular}{l|cccccccc}
   & \textbf{asymptotic }&\textbf{ permutation} & \textbf{asymptotic} & \textbf{wild} & \textbf{wild} & \textbf{groupwise} & \textbf{asymptotic} & \textbf{permutation} \\
   & \textbf{global} & & & \textbf{Rademacher} & \textbf{Gaussian} & & \textbf{bonf} & \textbf{bonf} \\ \hline 
        \textbf{Null hypothesis} & 0.0568 & 0.0566 & 0.0506 & 0.0514 & 0.0522 & 0.0516 & 0.0506 & 0.0498 \\
        \textbf{Alternative hypothesis} & 0.8984 & 0.8984 & 1.0000 & 1.0000 & 1.0000 & 1.0000 & 1.0000 & 1.0000 
    \end{tabular}
    \caption{Rejection rates for the simulation inspired by the data example.}
    \label{tab:my_label}
\end{table}

\section{Tables and Figures}\label{sec:TabFig}

In this section, more detailed results of the simulation study in Section~\ref{sec:Simu} can be found. This includes the censoring rates for the different settings, the rejection rates under the null and under the alternative hypothesis as well as the empirical powers for all false hypotheses.

\FloatBarrier
\subsection{Tables}

\begin{table}[H]\label{tab:Censrates}
\centering
\begin{tabular}{lllrrrr}
  \hline
$\delta$ & distribution & censoring distribution & group 1 & group 2 & group 3 & group 4 \\ 
  \hline
0.0 & exp early, late, prop & equal & 0.20 & 0.21 & 0.21 & 0.21 \\ 
  0.0 & exp early, late, prop & unequal, high & 0.38 & 0.44 & 0.38 & 0.33 \\ 
  0.0 & exp early, late, prop & unequal, low & 0.20 & 0.21 & 0.25 & 0.06 \\ 
  0.0 & logn & equal & 0.41 & 0.41 & 0.41 & 0.41 \\ 
  0.0 & logn & unequal, high & 0.51 & 0.58 & 0.60 & 0.53 \\ 
  0.0 & logn & unequal, low & 0.33 & 0.41 & 0.41 & 0.11 \\ 
  0.0 & pwExp & equal & 0.21 & 0.20 & 0.21 & 0.33 \\
\hline
\end{tabular}
\caption{Censoring rates for the different settings.} 
\label{tab:cens}
\end{table}

\newpage
\FloatBarrier
\subsubsection{Type-I error rates under the null hypothesis}
\begin{table}[ht]
\centering
\begin{tabular}{lp{0.6in}>{\centering\arraybackslash}p{0.45in}>{\centering\arraybackslash}p{0.45in}>{\centering\arraybackslash}p{0.45in}>{\centering\arraybackslash}p{0.45in}>{\centering\arraybackslash}p{0.45in}>{\centering\arraybackslash}p{0.45in}>{\centering\arraybackslash}p{0.45in}>{\centering\arraybackslash}p{0.45in}}
  \hline
distribution & censoring \newline distribution & asymptotic global & permutation & asymptotic & wild Rademacher & wild Gaussian & groupwise & asymptotic bonf & permutation bonf \\ 
  \hline
exp early,late,prop & equal & 0.064 & \textbf{0.052} & \textbf{0.056} & 0.060 & 0.059 & \textbf{0.050} & 0.057 & \textbf{0.048} \\ 
  exp early,late,prop & unequal, high & 0.069 & \textbf{0.051} & 0.059 & 0.059 & 0.061 & \textbf{0.047} & \textbf{0.056} & 0.043 \\ 
  exp early,late,prop & unequal, low & 0.061 & \textbf{0.051} & \textbf{0.054} & \textbf{0.054} & 0.059 & \textbf{0.045} & 0.057 & \textbf{0.047} \\ 
  logn & equal & 0.058 & \textbf{0.047} & 0.057 & 0.058 & 0.064 & \textbf{0.048} & \textbf{0.049} & 0.043 \\ 
  logn & unequal, high & 0.066 & \textbf{0.050} & 0.068 & 0.068 & 0.075 & \textbf{0.048} & 0.059 & \textbf{0.045} \\ 
  logn & unequal, low & 0.061 & \textbf{0.050} & 0.061 & 0.061 & 0.063 & \textbf{0.052} & \textbf{0.048} & 0.041 \\ 
  pwExp & equal & 0.065 & \textbf{0.053} & 0.063 & 0.064 & 0.066 & \textbf{0.055} & \textbf{0.051} & \textbf{0.046} \\
\hline
\end{tabular}
\caption{Rejection rates for the Dunnett-type contrast matrix with $\delta =$ 0.0 and sample size balanced large} \label{tab:RR1}
\end{table}
\begin{table}[ht]
\centering
\begin{tabular}{lp{0.6in}>{\centering\arraybackslash}p{0.45in}>{\centering\arraybackslash}p{0.45in}>{\centering\arraybackslash}p{0.45in}>{\centering\arraybackslash}p{0.45in}>{\centering\arraybackslash}p{0.45in}>{\centering\arraybackslash}p{0.45in}>{\centering\arraybackslash}p{0.45in}>{\centering\arraybackslash}p{0.45in}}
  \hline
distribution & censoring \newline distribution & asymptotic global & permutation & asymptotic & wild Rademacher & wild Gaussian & groupwise & asymptotic bonf & permutation bonf \\ 
  \hline
exp early,late,prop & equal & 0.073 & \textbf{0.048} & 0.067 & 0.070 & 0.074 & \textbf{0.048} & 0.061 & 0.043 \\ 
  exp early,late,prop & unequal, high & 0.094 & \textbf{0.053} & 0.082 & 0.083 & 0.092 & \textbf{0.048} & 0.069 & \textbf{0.048} \\ 
  exp early,late,prop & unequal, low & 0.078 & \textbf{0.052} & 0.074 & 0.076 & 0.080 & \textbf{0.050} & 0.062 & \textbf{0.048} \\ 
  logn & equal & 0.073 & \textbf{0.044} & 0.062 & 0.066 & 0.073 & 0.040 & 0.064 & \textbf{0.044} \\ 
  logn & unequal, high & 0.096 & \textbf{0.050} & 0.083 & 0.090 & 0.101 & 0.042 & 0.079 & \textbf{0.049} \\ 
  logn & unequal, low & 0.081 & \textbf{0.055} & 0.073 & 0.076 & 0.084 & \textbf{0.050} & 0.068 & \textbf{0.048} \\ 
  pwExp & equal & 0.076 & \textbf{0.050} & 0.071 & 0.071 & 0.076 & \textbf{0.048} & 0.060 & \textbf{0.047} \\
\hline
\end{tabular}
\caption{Rejection rates for the Dunnett-type contrast matrix with $\delta =$ 0.0 and sample size balanced medium} 
\end{table}
\begin{table}[ht]
\centering
\begin{tabular}{lp{0.6in}>{\centering\arraybackslash}p{0.45in}>{\centering\arraybackslash}p{0.45in}>{\centering\arraybackslash}p{0.45in}>{\centering\arraybackslash}p{0.45in}>{\centering\arraybackslash}p{0.45in}>{\centering\arraybackslash}p{0.45in}>{\centering\arraybackslash}p{0.45in}>{\centering\arraybackslash}p{0.45in}}
  \hline
distribution & censoring \newline distribution & asymptotic global & permutation & asymptotic & wild Rademacher & wild Gaussian & groupwise & asymptotic bonf & permutation bonf \\ 
  \hline
exp early,late,prop & equal & 0.101 & \textbf{0.051} & 0.080 & 0.085 & 0.095 & 0.040 & 0.073 & \textbf{0.045} \\ 
  exp early,late,prop & unequal, high & 0.147 & \textbf{0.049} & 0.102 & 0.112 & 0.131 & 0.025 & 0.102 & \textbf{0.046} \\ 
  exp early,late,prop & unequal, low & 0.112 & \textbf{0.053} & 0.087 & 0.091 & 0.101 & 0.038 & 0.082 & \textbf{0.047} \\ 
  logn & equal & 0.115 & \textbf{0.047} & 0.092 & 0.103 & 0.119 & 0.038 & 0.086 & \textbf{0.046} \\ 
  logn & unequal, high & 0.170 & \textbf{0.052} & 0.130 & 0.156 & 0.181 & 0.028 & 0.128 & 0.041 \\ 
  logn & unequal, low & 0.121 & \textbf{0.055} & 0.101 & 0.108 & 0.124 & 0.043 & 0.092 & \textbf{0.050} \\ 
  pwExp & equal & 0.102 & \textbf{0.050} & 0.084 & 0.088 & 0.100 & 0.040 & 0.079 & 0.043 \\
\hline
\end{tabular}
\caption{Rejection rates for the Dunnett-type contrast matrix with $\delta =$ 0.0 and sample size balanced small} 
\end{table}
\begin{table}[ht]
\centering
\begin{tabular}{lp{0.6in}>{\centering\arraybackslash}p{0.45in}>{\centering\arraybackslash}p{0.45in}>{\centering\arraybackslash}p{0.45in}>{\centering\arraybackslash}p{0.45in}>{\centering\arraybackslash}p{0.45in}>{\centering\arraybackslash}p{0.45in}>{\centering\arraybackslash}p{0.45in}>{\centering\arraybackslash}p{0.45in}}
  \hline
distribution & censoring \newline distribution & asymptotic global & permutation & asymptotic & wild Rademacher & wild Gaussian & groupwise & asymptotic bonf & permutation bonf \\ 
  \hline
exp early,late,prop & equal & 0.065 & \textbf{0.048} & 0.062 & 0.063 & 0.065 & \textbf{0.051} & 0.060 & \textbf{0.050} \\ 
  exp early,late,prop & unequal, high & 0.071 & \textbf{0.049} & 0.068 & 0.068 & 0.072 & \textbf{0.050} & 0.061 & \textbf{0.045} \\ 
  exp early,late,prop & unequal, low & 0.064 & \textbf{0.047} & 0.064 & 0.066 & 0.066 & \textbf{0.052} & 0.058 & \textbf{0.047} \\ 
  logn & equal & 0.065 & \textbf{0.048} & 0.063 & 0.064 & 0.066 & \textbf{0.049} & \textbf{0.052} & 0.040 \\ 
  logn & unequal, high & 0.077 & \textbf{0.049} & 0.072 & 0.076 & 0.081 & \textbf{0.046} & 0.057 & 0.039 \\ 
  logn & unequal, low & 0.065 & \textbf{0.049} & 0.062 & 0.063 & 0.069 & \textbf{0.048} & \textbf{0.052} & 0.042 \\ 
  pwExp & equal & 0.069 & \textbf{0.052} & 0.059 & 0.061 & 0.063 & \textbf{0.050} & \textbf{0.052} & 0.042 \\
\hline
\end{tabular}
\caption{Rejection rates for the Dunnett-type contrast matrix with $\delta =$ 0.0 and sample size unbalanced large} 
\end{table}
\begin{table}[ht]
\centering
\begin{tabular}{lp{0.6in}>{\centering\arraybackslash}p{0.45in}>{\centering\arraybackslash}p{0.45in}>{\centering\arraybackslash}p{0.45in}>{\centering\arraybackslash}p{0.45in}>{\centering\arraybackslash}p{0.45in}>{\centering\arraybackslash}p{0.45in}>{\centering\arraybackslash}p{0.45in}>{\centering\arraybackslash}p{0.45in}}
  \hline
distribution & censoring \newline distribution & asymptotic global & permutation & asymptotic & wild Rademacher & wild Gaussian & groupwise & asymptotic bonf & permutation bonf \\ 
  \hline
exp early,late,prop & equal & 0.084 & \textbf{0.047} & 0.070 & 0.072 & 0.077 & \textbf{0.046} & 0.060 & 0.036 \\ 
  exp early,late,prop & unequal, high & 0.109 & \textbf{0.056} & 0.089 & 0.095 & 0.104 & \textbf{0.048} & 0.074 & 0.039 \\ 
  exp early,late,prop & unequal, low & 0.090 & \textbf{0.053} & 0.074 & 0.074 & 0.082 & \textbf{0.049} & 0.061 & 0.037 \\ 
  logn & equal & 0.088 & \textbf{0.051} & 0.077 & 0.081 & 0.090 & \textbf{0.048} & 0.069 & 0.042 \\ 
  logn & unequal, high & 0.119 & \textbf{0.055} & 0.101 & 0.110 & 0.122 & 0.041 & 0.085 & 0.043 \\ 
  logn & unequal, low & 0.092 & \textbf{0.055} & 0.077 & 0.082 & 0.091 & \textbf{0.045} & 0.074 & \textbf{0.045} \\ 
  pwExp & equal & 0.076 & 0.043 & 0.067 & 0.070 & 0.075 & \textbf{0.047} & 0.062 & 0.040 \\
\hline
\end{tabular}
\caption{Rejection rates for the Dunnett-type contrast matrix with $\delta =$ 0.0 and sample size unbalanced medium} 
\end{table}
\begin{table}[ht]
\centering
\begin{tabular}{lp{0.6in}>{\centering\arraybackslash}p{0.45in}>{\centering\arraybackslash}p{0.45in}>{\centering\arraybackslash}p{0.45in}>{\centering\arraybackslash}p{0.45in}>{\centering\arraybackslash}p{0.45in}>{\centering\arraybackslash}p{0.45in}>{\centering\arraybackslash}p{0.45in}>{\centering\arraybackslash}p{0.45in}}
  \hline
distribution & censoring \newline distribution & asymptotic global & permutation & asymptotic & wild Rademacher & wild Gaussian & groupwise & asymptotic bonf & permutation bonf \\ 
  \hline
exp early,late,prop & equal & 0.123 & \textbf{0.048} & 0.096 & 0.100 & 0.115 & 0.041 & 0.083 & 0.040 \\ 
  exp early,late,prop & unequal, high & 0.181 & \textbf{0.052} & 0.128 & 0.143 & 0.164 & 0.034 & 0.122 & \textbf{0.044} \\ 
  exp early,late,prop & unequal, low & 0.138 & \textbf{0.053} & 0.100 & 0.109 & 0.124 & 0.041 & 0.096 & \textbf{0.046} \\ 
  logn & equal & 0.127 & \textbf{0.047} & 0.108 & 0.120 & 0.138 & 0.037 & 0.094 & 0.040 \\ 
  logn & unequal, high & 0.214 & \textbf{0.045} & 0.171 & 0.216 & 0.231 & 0.037 & 0.156 & 0.040 \\ 
  logn & unequal, low & 0.145 & 0.064 & 0.111 & 0.129 & 0.145 & 0.040 & 0.108 & \textbf{0.047} \\ 
  pwExp & equal & 0.113 & \textbf{0.046} & 0.086 & 0.093 & 0.106 & 0.037 & 0.091 & 0.043 \\
\hline
\end{tabular}
\caption{Rejection rates for the Dunnett-type contrast matrix with $\delta =$ 0.0 and sample size unbalanced small} 
\end{table}
\begin{table}[ht]
\centering
\begin{tabular}{lp{0.6in}>{\centering\arraybackslash}p{0.45in}>{\centering\arraybackslash}p{0.45in}>{\centering\arraybackslash}p{0.45in}>{\centering\arraybackslash}p{0.45in}>{\centering\arraybackslash}p{0.45in}>{\centering\arraybackslash}p{0.45in}>{\centering\arraybackslash}p{0.45in}>{\centering\arraybackslash}p{0.45in}}
  \hline
distribution & censoring \newline distribution & asymptotic global & permutation & asymptotic & wild Rademacher & wild Gaussian & groupwise & asymptotic bonf & permutation bonf \\ 
  \hline
exp early,late,prop & equal & 0.064 & \textbf{0.052} & 0.063 & 0.063 & 0.066 & \textbf{0.052} & \textbf{0.049} & 0.040 \\ 
  exp early,late,prop & unequal, high & 0.069 & \textbf{0.051} & 0.065 & 0.066 & 0.072 & \textbf{0.047} & 0.057 & \textbf{0.045} \\ 
  exp early,late,prop & unequal, low & 0.061 & \textbf{0.051} & 0.059 & 0.060 & 0.061 & \textbf{0.050} & \textbf{0.055} & \textbf{0.044} \\ 
  logn & equal & 0.058 & \textbf{0.047} & 0.060 & 0.063 & 0.065 & \textbf{0.049} & \textbf{0.050} & 0.041 \\ 
  logn & unequal, high & 0.066 & \textbf{0.050} & 0.069 & 0.072 & 0.080 & \textbf{0.045} & 0.056 & 0.039 \\ 
  logn & unequal, low & 0.061 & \textbf{0.050} & 0.062 & 0.064 & 0.069 & \textbf{0.049} & \textbf{0.048} & 0.037 \\ 
  pwExp & equal & 0.065 & \textbf{0.053} & 0.058 & 0.061 & 0.064 & \textbf{0.046} & \textbf{0.051} & 0.040 \\
\hline
\end{tabular}
\caption{Rejection rates for the Tukey-type contrast matrix with $\delta =$ 0.0 and sample size balanced large} 
\end{table}
\begin{table}[ht]
\centering
\begin{tabular}{lp{0.6in}>{\centering\arraybackslash}p{0.45in}>{\centering\arraybackslash}p{0.45in}>{\centering\arraybackslash}p{0.45in}>{\centering\arraybackslash}p{0.45in}>{\centering\arraybackslash}p{0.45in}>{\centering\arraybackslash}p{0.45in}>{\centering\arraybackslash}p{0.45in}>{\centering\arraybackslash}p{0.45in}}
  \hline
distribution & censoring \newline distribution & asymptotic global & permutation & asymptotic & wild Rademacher & wild Gaussian & groupwise & asymptotic bonf & permutation bonf \\ 
  \hline
exp early,late,prop & equal & 0.073 & \textbf{0.048} & 0.069 & 0.072 & 0.078 & \textbf{0.047} & 0.060 & 0.043 \\ 
  exp early,late,prop & unequal, high & 0.094 & \textbf{0.053} & 0.083 & 0.089 & 0.100 & \textbf{0.046} & 0.075 & \textbf{0.045} \\ 
  exp early,late,prop & unequal, low & 0.078 & \textbf{0.052} & 0.076 & 0.081 & 0.086 & \textbf{0.047} & 0.063 & 0.042 \\ 
  logn & equal & 0.073 & \textbf{0.044} & 0.070 & 0.075 & 0.085 & 0.041 & 0.062 & 0.041 \\ 
  logn & unequal, high & 0.096 & \textbf{0.050} & 0.092 & 0.100 & 0.117 & 0.038 & 0.081 & 0.041 \\ 
  logn & unequal, low & 0.081 & \textbf{0.055} & 0.077 & 0.082 & 0.094 & \textbf{0.048} & 0.062 & 0.039 \\ 
  pwExp & equal & 0.076 & \textbf{0.050} & 0.071 & 0.075 & 0.083 & \textbf{0.050} & 0.059 & 0.041 \\
\hline
\end{tabular}
\caption{Rejection rates for the Tukey-type contrast matrix with $\delta =$ 0.0 and sample size balanced medium} 
\end{table}
\begin{table}[ht]
\centering
\begin{tabular}{lp{0.6in}>{\centering\arraybackslash}p{0.45in}>{\centering\arraybackslash}p{0.45in}>{\centering\arraybackslash}p{0.45in}>{\centering\arraybackslash}p{0.45in}>{\centering\arraybackslash}p{0.45in}>{\centering\arraybackslash}p{0.45in}>{\centering\arraybackslash}p{0.45in}>{\centering\arraybackslash}p{0.45in}}
  \hline
distribution & censoring \newline distribution & asymptotic global & permutation & asymptotic & wild Rademacher & wild Gaussian & groupwise & asymptotic bonf & permutation bonf \\ 
  \hline
exp early,late,prop & equal & 0.101 & \textbf{0.051} & 0.092 & 0.101 & 0.119 & 0.040 & 0.077 & 0.042 \\ 
  exp early,late,prop & unequal, high & 0.147 & \textbf{0.049} & 0.127 & 0.146 & 0.175 & 0.022 & 0.104 & 0.043 \\ 
  exp early,late,prop & unequal, low & 0.112 & \textbf{0.053} & 0.104 & 0.115 & 0.131 & 0.042 & 0.085 & \textbf{0.045} \\ 
  logn & equal & 0.115 & \textbf{0.047} & 0.111 & 0.125 & 0.151 & 0.037 & 0.100 & 0.041 \\ 
  logn & unequal, high & 0.170 & \textbf{0.052} & 0.162 & 0.202 & 0.238 & 0.027 & 0.152 & 0.042 \\ 
  logn & unequal, low & 0.121 & \textbf{0.055} & 0.117 & 0.132 & 0.160 & 0.040 & 0.101 & \textbf{0.046} \\ 
  pwExp & equal & 0.102 & \textbf{0.050} & 0.093 & 0.104 & 0.122 & 0.035 & 0.081 & 0.041 \\
\hline
\end{tabular}
\caption{Rejection rates for the Tukey-type contrast matrix with $\delta =$ 0.0 and sample size balanced small} 
\end{table}
\begin{table}[ht]
\centering
\begin{tabular}{lp{0.6in}>{\centering\arraybackslash}p{0.45in}>{\centering\arraybackslash}p{0.45in}>{\centering\arraybackslash}p{0.45in}>{\centering\arraybackslash}p{0.45in}>{\centering\arraybackslash}p{0.45in}>{\centering\arraybackslash}p{0.45in}>{\centering\arraybackslash}p{0.45in}>{\centering\arraybackslash}p{0.45in}}
  \hline
distribution & censoring \newline distribution & asymptotic global & permutation & asymptotic & wild Rademacher & wild Gaussian & groupwise & asymptotic bonf & permutation bonf \\ 
  \hline
exp early,late,prop & equal & 0.065 & \textbf{0.048} & 0.061 & 0.064 & 0.068 & \textbf{0.047} & 0.060 & \textbf{0.046} \\ 
  exp early,late,prop & unequal, high & 0.071 & \textbf{0.049} & 0.069 & 0.074 & 0.079 & \textbf{0.052} & 0.062 & 0.042 \\ 
  exp early,late,prop & unequal, low & 0.064 & \textbf{0.047} & 0.066 & 0.070 & 0.071 & \textbf{0.049} & 0.062 & \textbf{0.047} \\ 
  logn & equal & 0.065 & \textbf{0.048} & 0.065 & 0.068 & 0.074 & \textbf{0.050} & \textbf{0.052} & 0.038 \\ 
  logn & unequal, high & 0.077 & \textbf{0.049} & 0.071 & 0.077 & 0.088 & \textbf{0.046} & 0.063 & 0.039 \\ 
  logn & unequal, low & 0.065 & \textbf{0.049} & 0.064 & 0.067 & 0.074 & \textbf{0.046} & \textbf{0.053} & 0.041 \\ 
  pwExp & equal & 0.069 & \textbf{0.052} & 0.067 & 0.069 & 0.071 & \textbf{0.053} & \textbf{0.051} & 0.039 \\
\hline
\end{tabular}
\caption{Rejection rates for the Tukey-type contrast matrix with $\delta =$ 0.0 and sample size unbalanced large} 
\end{table}
\begin{table}[ht]
\centering
\begin{tabular}{lp{0.6in}>{\centering\arraybackslash}p{0.45in}>{\centering\arraybackslash}p{0.45in}>{\centering\arraybackslash}p{0.45in}>{\centering\arraybackslash}p{0.45in}>{\centering\arraybackslash}p{0.45in}>{\centering\arraybackslash}p{0.45in}>{\centering\arraybackslash}p{0.45in}>{\centering\arraybackslash}p{0.45in}}
  \hline
distribution & censoring \newline distribution & asymptotic global & permutation & asymptotic & wild Rademacher & wild Gaussian & groupwise & asymptotic bonf & permutation bonf \\ 
  \hline
exp early,late,prop & equal & 0.084 & \textbf{0.047} & 0.080 & 0.084 & 0.094 & \textbf{0.046} & 0.065 & 0.038 \\ 
  exp early,late,prop & unequal, high & 0.109 & \textbf{0.056} & 0.100 & 0.109 & 0.124 & \textbf{0.053} & 0.088 & 0.042 \\ 
  exp early,late,prop & unequal, low & 0.090 & \textbf{0.053} & 0.084 & 0.087 & 0.096 & \textbf{0.052} & 0.069 & 0.041 \\ 
  logn & equal & 0.088 & \textbf{0.051} & 0.084 & 0.093 & 0.104 & \textbf{0.047} & 0.072 & 0.041 \\ 
  logn & unequal, high & 0.119 & \textbf{0.055} & 0.114 & 0.130 & 0.151 & 0.043 & 0.094 & 0.040 \\ 
  logn & unequal, low & 0.092 & \textbf{0.055} & 0.092 & 0.100 & 0.114 & \textbf{0.046} & 0.076 & \textbf{0.046} \\ 
  pwExp & equal & 0.076 & 0.043 & 0.074 & 0.078 & 0.083 & \textbf{0.046} & 0.067 & 0.038 \\
\hline
\end{tabular}
\caption{Rejection rates for the Tukey-type contrast matrix with $\delta =$ 0.0 and sample size unbalanced medium} 
\end{table}
\begin{table}[ht]
\centering
\begin{tabular}{lp{0.6in}>{\centering\arraybackslash}p{0.45in}>{\centering\arraybackslash}p{0.45in}>{\centering\arraybackslash}p{0.45in}>{\centering\arraybackslash}p{0.45in}>{\centering\arraybackslash}p{0.45in}>{\centering\arraybackslash}p{0.45in}>{\centering\arraybackslash}p{0.45in}>{\centering\arraybackslash}p{0.45in}}
  \hline
distribution & censoring \newline distribution & asymptotic global & permutation & asymptotic & wild Rademacher & wild Gaussian & groupwise & asymptotic bonf & permutation bonf \\ 
  \hline
exp early,late,prop & equal & 0.123 & \textbf{0.048} & 0.113 & 0.125 & 0.149 & \textbf{0.047} & 0.094 & 0.036 \\ 
  exp early,late,prop & unequal, high & 0.181 & \textbf{0.052} & 0.163 & 0.194 & 0.214 & 0.040 & 0.142 & 0.043 \\ 
  exp early,late,prop & unequal, low & 0.138 & \textbf{0.053} & 0.126 & 0.139 & 0.164 & \textbf{0.052} & 0.114 & \textbf{0.048} \\ 
  logn & equal & 0.127 & \textbf{0.047} & 0.127 & 0.148 & 0.172 & \textbf{0.044} & 0.103 & 0.038 \\ 
  logn & unequal, high & 0.214 & \textbf{0.045} & 0.207 & 0.267 & 0.290 & \textbf{0.047} & 0.187 & \textbf{0.045} \\ 
  logn & unequal, low & 0.145 & 0.064 & 0.140 & 0.168 & 0.190 & \textbf{0.054} & 0.124 & \textbf{0.051} \\ 
  pwExp & equal & 0.113 & \textbf{0.046} & 0.103 & 0.113 & 0.135 & 0.042 & 0.099 & 0.039 \\
\hline
\end{tabular}
\caption{Rejection rates for the Tukey-type contrast matrix with $\delta =$ 0.0 and sample size unbalanced small} 
\end{table}
\begin{table}[ht]
\centering
\begin{tabular}{lp{0.6in}>{\centering\arraybackslash}p{0.45in}>{\centering\arraybackslash}p{0.45in}>{\centering\arraybackslash}p{0.45in}>{\centering\arraybackslash}p{0.45in}>{\centering\arraybackslash}p{0.45in}>{\centering\arraybackslash}p{0.45in}>{\centering\arraybackslash}p{0.45in}>{\centering\arraybackslash}p{0.45in}}
  \hline
distribution & censoring \newline distribution & asymptotic global & permutation & asymptotic & wild Rademacher & wild Gaussian & groupwise & asymptotic bonf & permutation bonf \\ 
  \hline
exp early,late,prop & equal & 0.064 & \textbf{0.052} & 0.065 & 0.068 & 0.068 & \textbf{0.052} & 0.058 & \textbf{0.047} \\ 
  exp early,late,prop & unequal, high & 0.069 & \textbf{0.051} & 0.067 & 0.069 & 0.073 & \textbf{0.049} & 0.065 & \textbf{0.048} \\ 
  exp early,late,prop & unequal, low & 0.061 & \textbf{0.051} & 0.061 & 0.063 & 0.066 & \textbf{0.049} & 0.061 & \textbf{0.048} \\ 
  logn & equal & 0.058 & \textbf{0.047} & \textbf{0.056} & 0.060 & 0.063 & \textbf{0.045} & \textbf{0.056} & \textbf{0.044} \\ 
  logn & unequal, high & 0.066 & \textbf{0.050} & 0.068 & 0.073 & 0.078 & \textbf{0.047} & 0.061 & \textbf{0.045} \\ 
  logn & unequal, low & 0.061 & \textbf{0.050} & 0.057 & 0.058 & 0.062 & \textbf{0.046} & \textbf{0.053} & 0.041 \\ 
  pwExp & equal & 0.065 & \textbf{0.053} & 0.067 & 0.068 & 0.070 & \textbf{0.053} & \textbf{0.055} & \textbf{0.045} \\
\hline
\end{tabular}
\caption{Rejection rates for the Grand-mean-type contrast matrix with $\delta =$ 0.0 and sample size balanced large} 
\end{table}
\begin{table}[ht]
\centering
\begin{tabular}{lp{0.6in}>{\centering\arraybackslash}p{0.45in}>{\centering\arraybackslash}p{0.45in}>{\centering\arraybackslash}p{0.45in}>{\centering\arraybackslash}p{0.45in}>{\centering\arraybackslash}p{0.45in}>{\centering\arraybackslash}p{0.45in}>{\centering\arraybackslash}p{0.45in}>{\centering\arraybackslash}p{0.45in}}
  \hline
distribution & censoring \newline distribution & asymptotic global & permutation & asymptotic & wild Rademacher & wild Gaussian & groupwise & asymptotic bonf & permutation bonf \\ 
  \hline
exp early,late,prop & equal & 0.073 & \textbf{0.048} & 0.074 & 0.077 & 0.081 & \textbf{0.048} & 0.078 & \textbf{0.050} \\ 
  exp early,late,prop & unequal, high & 0.094 & \textbf{0.053} & 0.095 & 0.099 & 0.107 & 0.060 & 0.092 & \textbf{0.048} \\ 
  exp early,late,prop & unequal, low & 0.078 & \textbf{0.052} & 0.075 & 0.080 & 0.089 & \textbf{0.052} & 0.074 & \textbf{0.049} \\ 
  logn & equal & 0.073 & \textbf{0.044} & 0.075 & 0.080 & 0.086 & \textbf{0.046} & 0.071 & \textbf{0.045} \\ 
  logn & unequal, high & 0.096 & \textbf{0.050} & 0.093 & 0.101 & 0.112 & 0.039 & 0.092 & \textbf{0.049} \\ 
  logn & unequal, low & 0.081 & \textbf{0.055} & 0.076 & 0.082 & 0.093 & \textbf{0.045} & 0.072 & \textbf{0.047} \\ 
  pwExp & equal & 0.076 & \textbf{0.050} & 0.079 & 0.082 & 0.086 & \textbf{0.050} & 0.068 & \textbf{0.044} \\
\hline
\end{tabular}
\caption{Rejection rates for the Grand-mean-type contrast matrix with $\delta =$ 0.0 and sample size balanced medium} 
\end{table}
\begin{table}[ht]
\centering
\begin{tabular}{lp{0.6in}>{\centering\arraybackslash}p{0.45in}>{\centering\arraybackslash}p{0.45in}>{\centering\arraybackslash}p{0.45in}>{\centering\arraybackslash}p{0.45in}>{\centering\arraybackslash}p{0.45in}>{\centering\arraybackslash}p{0.45in}>{\centering\arraybackslash}p{0.45in}>{\centering\arraybackslash}p{0.45in}}
  \hline
distribution & censoring \newline distribution & asymptotic global & permutation & asymptotic & wild Rademacher & wild Gaussian & groupwise & asymptotic bonf & permutation bonf \\ 
  \hline
exp early,late,prop & equal & 0.101 & \textbf{0.051} & 0.102 & 0.110 & 0.119 & 0.059 & 0.097 & 0.043 \\ 
  exp early,late,prop & unequal, high & 0.147 & \textbf{0.049} & 0.150 & 0.160 & 0.178 & 0.074 & 0.143 & \textbf{0.050} \\ 
  exp early,late,prop & unequal, low & 0.112 & \textbf{0.053} & 0.115 & 0.122 & 0.132 & 0.062 & 0.106 & \textbf{0.045} \\ 
  logn & equal & 0.115 & \textbf{0.047} & 0.111 & 0.125 & 0.143 & \textbf{0.044} & 0.112 & \textbf{0.047} \\ 
  logn & unequal, high & 0.170 & \textbf{0.052} & 0.161 & 0.193 & 0.209 & \textbf{0.054} & 0.162 & \textbf{0.049} \\ 
  logn & unequal, low & 0.121 & \textbf{0.055} & 0.115 & 0.128 & 0.143 & \textbf{0.047} & 0.108 & \textbf{0.049} \\ 
  pwExp & equal & 0.102 & \textbf{0.050} & 0.103 & 0.111 & 0.125 & \textbf{0.054} & 0.101 & 0.043 \\
\hline
\end{tabular}
\caption{Rejection rates for the Grand-mean-type contrast matrix with $\delta =$ 0.0 and sample size balanced small} 
\end{table}
\begin{table}[ht]
\centering
\begin{tabular}{lp{0.6in}>{\centering\arraybackslash}p{0.45in}>{\centering\arraybackslash}p{0.45in}>{\centering\arraybackslash}p{0.45in}>{\centering\arraybackslash}p{0.45in}>{\centering\arraybackslash}p{0.45in}>{\centering\arraybackslash}p{0.45in}>{\centering\arraybackslash}p{0.45in}>{\centering\arraybackslash}p{0.45in}}
  \hline
distribution & censoring \newline distribution & asymptotic global & permutation & asymptotic & wild Rademacher & wild Gaussian & groupwise & asymptotic bonf & permutation bonf \\ 
  \hline
exp early,late,prop & equal & 0.065 & \textbf{0.048} & 0.060 & 0.065 & 0.066 & \textbf{0.046} & 0.066 & \textbf{0.050} \\ 
  exp early,late,prop & unequal, high & 0.071 & \textbf{0.049} & 0.075 & 0.076 & 0.084 & \textbf{0.052} & 0.074 & \textbf{0.049} \\ 
  exp early,late,prop & unequal, low & 0.064 & \textbf{0.047} & 0.063 & 0.066 & 0.071 & \textbf{0.045} & 0.065 & \textbf{0.046} \\ 
  logn & equal & 0.065 & \textbf{0.048} & 0.061 & 0.063 & 0.069 & \textbf{0.047} & 0.059 & \textbf{0.045} \\ 
  logn & unequal, high & 0.077 & \textbf{0.049} & 0.075 & 0.080 & 0.089 & \textbf{0.044} & 0.068 & \textbf{0.046} \\ 
  logn & unequal, low & 0.065 & \textbf{0.049} & 0.062 & 0.065 & 0.071 & \textbf{0.047} & 0.059 & \textbf{0.046} \\ 
  pwExp & equal & 0.069 & \textbf{0.052} & 0.065 & 0.068 & 0.071 & \textbf{0.052} & 0.062 & \textbf{0.045} \\
\hline
\end{tabular}
\caption{Rejection rates for the Grand-mean-type contrast matrix with $\delta =$ 0.0 and sample size unbalanced large} 
\end{table}
\begin{table}[ht]
\centering
\begin{tabular}{lp{0.6in}>{\centering\arraybackslash}p{0.45in}>{\centering\arraybackslash}p{0.45in}>{\centering\arraybackslash}p{0.45in}>{\centering\arraybackslash}p{0.45in}>{\centering\arraybackslash}p{0.45in}>{\centering\arraybackslash}p{0.45in}>{\centering\arraybackslash}p{0.45in}>{\centering\arraybackslash}p{0.45in}}
  \hline
distribution & censoring \newline distribution & asymptotic global & permutation & asymptotic & wild Rademacher & wild Gaussian & groupwise & asymptotic bonf & permutation bonf \\ 
  \hline
exp early,late,prop & equal & 0.084 & \textbf{0.047} & 0.079 & 0.084 & 0.091 & \textbf{0.047} & 0.073 & 0.043 \\ 
  exp early,late,prop & unequal, high & 0.109 & \textbf{0.056} & 0.112 & 0.120 & 0.127 & 0.064 & 0.101 & \textbf{0.046} \\ 
  exp early,late,prop & unequal, low & 0.090 & \textbf{0.053} & 0.085 & 0.089 & 0.095 & \textbf{0.056} & 0.074 & \textbf{0.045} \\ 
  logn & equal & 0.088 & \textbf{0.051} & 0.081 & 0.089 & 0.097 & 0.043 & 0.079 & \textbf{0.044} \\ 
  logn & unequal, high & 0.119 & \textbf{0.055} & 0.116 & 0.133 & 0.144 & \textbf{0.045} & 0.101 & \textbf{0.045} \\ 
  logn & unequal, low & 0.092 & \textbf{0.055} & 0.093 & 0.099 & 0.114 & 0.044 & 0.081 & \textbf{0.051} \\ 
  pwExp & equal & 0.076 & 0.043 & 0.076 & 0.078 & 0.087 & \textbf{0.049} & 0.077 & 0.043 \\
\hline
\end{tabular}
\caption{Rejection rates for the Grand-mean-type contrast matrix with $\delta =$ 0.0 and sample size unbalanced medium} 
\end{table}
\begin{table}[ht]
\centering
\begin{tabular}{lp{0.6in}>{\centering\arraybackslash}p{0.45in}>{\centering\arraybackslash}p{0.45in}>{\centering\arraybackslash}p{0.45in}>{\centering\arraybackslash}p{0.45in}>{\centering\arraybackslash}p{0.45in}>{\centering\arraybackslash}p{0.45in}>{\centering\arraybackslash}p{0.45in}>{\centering\arraybackslash}p{0.45in}}
  \hline
distribution & censoring \newline distribution & asymptotic global & permutation & asymptotic & wild Rademacher & wild Gaussian & groupwise & asymptotic bonf & permutation bonf \\ 
  \hline
exp early,late,prop & equal & 0.123 & \textbf{0.048} & 0.121 & 0.132 & 0.148 & 0.062 & 0.109 & 0.043 \\ 
  exp early,late,prop & unequal, high & 0.181 & \textbf{0.052} & 0.184 & 0.204 & 0.217 & 0.076 & 0.167 & \textbf{0.047} \\ 
  exp early,late,prop & unequal, low & 0.138 & \textbf{0.053} & 0.132 & 0.144 & 0.160 & 0.063 & 0.123 & \textbf{0.052} \\ 
  logn & equal & 0.127 & \textbf{0.047} & 0.124 & 0.147 & 0.164 & \textbf{0.047} & 0.111 & \textbf{0.044} \\ 
  logn & unequal, high & 0.214 & \textbf{0.045} & 0.205 & 0.248 & 0.259 & 0.065 & 0.198 & \textbf{0.044} \\ 
  logn & unequal, low & 0.145 & 0.064 & 0.135 & 0.162 & 0.177 & \textbf{0.053} & 0.130 & \textbf{0.054} \\ 
  pwExp & equal & 0.113 & \textbf{0.046} & 0.111 & 0.121 & 0.135 & 0.057 & 0.114 & \textbf{0.045} \\
\hline
\end{tabular}
\caption{Rejection rates for the Grand-mean-type contrast matrix with $\delta =$ 0.0 and sample size unbalanced small} 
\end{table}

\FloatBarrier
\subsubsection{Power under the alternative hypothesis}
{
\begin{table}[ht]{
\centering
\begin{tabular}{lp{0.6in}>{\centering\arraybackslash}p{0.45in}>{\centering\arraybackslash}p{0.45in}>{\centering\arraybackslash}p{0.45in}>{\centering\arraybackslash}p{0.45in}>{\centering\arraybackslash}p{0.45in}>{\centering\arraybackslash}p{0.45in}>{\centering\arraybackslash}p{0.45in}>{\centering\arraybackslash}p{0.45in}}
  \hline
distribution & censoring \newline distribution & asymptotic global & permutation & asymptotic & wild Rademacher & wild Gaussian & groupwise & asymptotic bonf & permutation bonf \\ 
  \hline
exp early & equal & 0.719 & 0.686 & 0.575 & 0.577 & 0.584 & 0.550 & 0.573 & 0.550 \\ 
  exp early & unequal, high & 0.630 & 0.570 & 0.472 & 0.478 & 0.483 & 0.431 & 0.464 & 0.420 \\ 
  exp early & unequal, low & 0.733 & 0.696 & 0.570 & 0.570 & 0.576 & 0.543 & 0.570 & 0.539 \\ 
  exp late & equal & 0.938 & 0.926 & 0.756 & 0.756 & 0.762 & 0.727 & 0.732 & 0.710 \\ 
  exp late & unequal, high & 0.857 & 0.814 & 0.615 & 0.620 & 0.623 & 0.557 & 0.588 & 0.543 \\ 
  exp late & unequal, low & 0.940 & 0.926 & 0.735 & 0.740 & 0.743 & 0.702 & 0.719 & 0.692 \\ 
  exp prop & equal & 0.839 & 0.812 & 0.669 & 0.669 & 0.670 & 0.645 & 0.647 & 0.622 \\
\hline
\end{tabular}
\caption{Rejection rates for the Dunnett-type contrast matrix with $\delta =$ 1.5 and sample size balanced large} 
}\end{table}
\begin{table}[ht]{
\centering
\begin{tabular}{lp{0.6in}>{\centering\arraybackslash}p{0.45in}>{\centering\arraybackslash}p{0.45in}>{\centering\arraybackslash}p{0.45in}>{\centering\arraybackslash}p{0.45in}>{\centering\arraybackslash}p{0.45in}>{\centering\arraybackslash}p{0.45in}>{\centering\arraybackslash}p{0.45in}>{\centering\arraybackslash}p{0.45in}}
  \hline
distribution & censoring \newline distribution & asymptotic global & permutation & asymptotic & wild Rademacher & wild Gaussian & groupwise & asymptotic bonf & permutation bonf \\ 
  \hline
exp early & equal & 0.457 & 0.380 & 0.341 & 0.344 & 0.354 & 0.286 & 0.318 & 0.271 \\ 
  exp early & unequal, high & 0.407 & 0.293 & 0.284 & 0.291 & 0.302 & 0.209 & 0.266 & 0.207 \\ 
  exp early & unequal, low & 0.466 & 0.378 & 0.321 & 0.327 & 0.338 & 0.267 & 0.306 & 0.257 \\ 
  exp late & equal & 0.680 & 0.598 & 0.439 & 0.441 & 0.453 & 0.361 & 0.405 & 0.352 \\ 
  exp late & unequal, high & 0.560 & 0.418 & 0.318 & 0.329 & 0.346 & 0.216 & 0.303 & 0.234 \\ 
  exp late & unequal, low & 0.687 & 0.595 & 0.421 & 0.427 & 0.444 & 0.341 & 0.387 & 0.333 \\ 
  exp prop & equal & 0.534 & 0.457 & 0.368 & 0.372 & 0.379 & 0.308 & 0.353 & 0.303 \\
\hline
\end{tabular}
\caption{Rejection rates for the Dunnett-type contrast matrix with $\delta =$ 1.5 and sample size balanced medium} 
}\end{table}
\begin{table}[ht]{
\centering
\begin{tabular}{lp{0.6in}>{\centering\arraybackslash}p{0.45in}>{\centering\arraybackslash}p{0.45in}>{\centering\arraybackslash}p{0.45in}>{\centering\arraybackslash}p{0.45in}>{\centering\arraybackslash}p{0.45in}>{\centering\arraybackslash}p{0.45in}>{\centering\arraybackslash}p{0.45in}>{\centering\arraybackslash}p{0.45in}}
  \hline
distribution & censoring \newline distribution & asymptotic global & permutation & asymptotic & wild Rademacher & wild Gaussian & groupwise & asymptotic bonf & permutation bonf \\ 
  \hline
exp early & equal & 0.320 & 0.204 & 0.233 & 0.241 & 0.257 & 0.143 & 0.218 & 0.153 \\ 
  exp early & unequal, high & 0.314 & 0.155 & 0.233 & 0.244 & 0.269 & 0.079 & 0.207 & 0.113 \\ 
  exp early & unequal, low & 0.316 & 0.189 & 0.214 & 0.225 & 0.242 & 0.123 & 0.215 & 0.145 \\ 
  exp late & equal & 0.423 & 0.273 & 0.245 & 0.258 & 0.281 & 0.138 & 0.231 & 0.156 \\ 
  exp late & unequal, high & 0.382 & 0.172 & 0.224 & 0.240 & 0.273 & 0.073 & 0.197 & 0.103 \\ 
  exp late & unequal, low & 0.422 & 0.256 & 0.247 & 0.258 & 0.283 & 0.123 & 0.224 & 0.149 \\ 
  exp prop & equal & 0.343 & 0.221 & 0.223 & 0.229 & 0.245 & 0.133 & 0.211 & 0.148 \\
\hline
\end{tabular}
\caption{Rejection rates for the Dunnett-type contrast matrix with $\delta =$ 1.5 and sample size balanced small} 
}\end{table}
\begin{table}[ht]{
\centering
\begin{tabular}{lp{0.6in}>{\centering\arraybackslash}p{0.45in}>{\centering\arraybackslash}p{0.45in}>{\centering\arraybackslash}p{0.45in}>{\centering\arraybackslash}p{0.45in}>{\centering\arraybackslash}p{0.45in}>{\centering\arraybackslash}p{0.45in}>{\centering\arraybackslash}p{0.45in}>{\centering\arraybackslash}p{0.45in}}
  \hline
distribution & censoring \newline distribution & asymptotic global & permutation & asymptotic & wild Rademacher & wild Gaussian & groupwise & asymptotic bonf & permutation bonf \\ 
  \hline
exp early & equal & 0.818 & 0.779 & 0.548 & 0.549 & 0.556 & 0.503 & 0.512 & 0.467 \\ 
  exp early & unequal, high & 0.708 & 0.631 & 0.432 & 0.439 & 0.443 & 0.361 & 0.396 & 0.322 \\ 
  exp early & unequal, low & 0.811 & 0.770 & 0.518 & 0.521 & 0.529 & 0.472 & 0.485 & 0.440 \\ 
  exp late & equal & 0.957 & 0.945 & 0.652 & 0.656 & 0.663 & 0.592 & 0.597 & 0.548 \\ 
  exp late & unequal, high & 0.882 & 0.823 & 0.507 & 0.508 & 0.525 & 0.402 & 0.454 & 0.376 \\ 
  exp late & unequal, low & 0.961 & 0.946 & 0.631 & 0.637 & 0.642 & 0.552 & 0.568 & 0.513 \\ 
  exp prop & equal & 0.885 & 0.861 & 0.584 & 0.587 & 0.592 & 0.533 & 0.546 & 0.498 \\
\hline
\end{tabular}
\caption{Rejection rates for the Dunnett-type contrast matrix with $\delta =$ 1.5 and sample size unbalanced large} 
}\end{table}
\begin{table}[ht]{
\centering
\begin{tabular}{lp{0.6in}>{\centering\arraybackslash}p{0.45in}>{\centering\arraybackslash}p{0.45in}>{\centering\arraybackslash}p{0.45in}>{\centering\arraybackslash}p{0.45in}>{\centering\arraybackslash}p{0.45in}>{\centering\arraybackslash}p{0.45in}>{\centering\arraybackslash}p{0.45in}>{\centering\arraybackslash}p{0.45in}}
  \hline
distribution & censoring \newline distribution & asymptotic global & permutation & asymptotic & wild Rademacher & wild Gaussian & groupwise & asymptotic bonf & permutation bonf \\ 
  \hline
exp early & equal & 0.515 & 0.410 & 0.290 & 0.298 & 0.309 & 0.222 & 0.261 & 0.194 \\ 
  exp early & unequal, high & 0.455 & 0.283 & 0.251 & 0.256 & 0.275 & 0.146 & 0.210 & 0.121 \\ 
  exp early & unequal, low & 0.517 & 0.408 & 0.281 & 0.290 & 0.302 & 0.205 & 0.253 & 0.180 \\ 
  exp late & equal & 0.727 & 0.618 & 0.355 & 0.361 & 0.378 & 0.247 & 0.299 & 0.225 \\ 
  exp late & unequal, high & 0.591 & 0.388 & 0.262 & 0.275 & 0.301 & 0.126 & 0.224 & 0.129 \\ 
  exp late & unequal, low & 0.718 & 0.608 & 0.329 & 0.341 & 0.362 & 0.221 & 0.284 & 0.206 \\ 
  exp prop & equal & 0.596 & 0.489 & 0.311 & 0.320 & 0.334 & 0.224 & 0.278 & 0.201 \\
\hline
\end{tabular}
\caption{Rejection rates for the Dunnett-type contrast matrix with $\delta =$ 1.5 and sample size unbalanced medium} 
}\end{table}
\begin{table}[ht]{
\centering
\begin{tabular}{lp{0.6in}>{\centering\arraybackslash}p{0.45in}>{\centering\arraybackslash}p{0.45in}>{\centering\arraybackslash}p{0.45in}>{\centering\arraybackslash}p{0.45in}>{\centering\arraybackslash}p{0.45in}>{\centering\arraybackslash}p{0.45in}>{\centering\arraybackslash}p{0.45in}>{\centering\arraybackslash}p{0.45in}}
  \hline
distribution & censoring \newline distribution & asymptotic global & permutation & asymptotic & wild Rademacher & wild Gaussian & groupwise & asymptotic bonf & permutation bonf \\ 
  \hline
exp early & equal & 0.346 & 0.174 & 0.199 & 0.212 & 0.235 & 0.086 & 0.188 & 0.099 \\ 
  exp early & unequal, high & 0.337 & 0.106 & 0.206 & 0.233 & 0.263 & 0.036 & 0.192 & 0.072 \\ 
  exp early & unequal, low & 0.340 & 0.157 & 0.195 & 0.207 & 0.237 & 0.070 & 0.189 & 0.093 \\ 
  exp late & equal & 0.448 & 0.230 & 0.205 & 0.220 & 0.251 & 0.075 & 0.196 & 0.101 \\ 
  exp late & unequal, high & 0.401 & 0.122 & 0.216 & 0.247 & 0.282 & 0.030 & 0.192 & 0.079 \\ 
  exp late & unequal, low & 0.464 & 0.230 & 0.210 & 0.227 & 0.261 & 0.058 & 0.196 & 0.097 \\ 
  exp prop & equal & 0.378 & 0.190 & 0.197 & 0.210 & 0.241 & 0.079 & 0.176 & 0.090 \\
\hline
\end{tabular}
\caption{Rejection rates for the Dunnett-type contrast matrix with $\delta =$ 1.5 and sample size unbalanced small} 
}\end{table}
\begin{table}[ht]{
\centering
\begin{tabular}{lp{0.6in}>{\centering\arraybackslash}p{0.45in}>{\centering\arraybackslash}p{0.45in}>{\centering\arraybackslash}p{0.45in}>{\centering\arraybackslash}p{0.45in}>{\centering\arraybackslash}p{0.45in}>{\centering\arraybackslash}p{0.45in}>{\centering\arraybackslash}p{0.45in}>{\centering\arraybackslash}p{0.45in}}
  \hline
distribution & censoring \newline distribution & asymptotic global & permutation & asymptotic & wild Rademacher & wild Gaussian & groupwise & asymptotic bonf & permutation bonf \\ 
  \hline
exp early & equal & 0.719 & 0.686 & 0.710 & 0.711 & 0.717 & 0.679 & 0.692 & 0.662 \\ 
  exp early & unequal, high & 0.630 & 0.570 & 0.609 & 0.613 & 0.624 & 0.558 & 0.582 & 0.532 \\ 
  exp early & unequal, low & 0.733 & 0.696 & 0.711 & 0.715 & 0.725 & 0.685 & 0.694 & 0.662 \\ 
  exp late & equal & 0.938 & 0.926 & 0.909 & 0.910 & 0.912 & 0.886 & 0.885 & 0.864 \\ 
  exp late & unequal, high & 0.857 & 0.814 & 0.795 & 0.801 & 0.815 & 0.734 & 0.758 & 0.711 \\ 
  exp late & unequal, low & 0.940 & 0.926 & 0.907 & 0.910 & 0.911 & 0.883 & 0.882 & 0.857 \\ 
  exp prop & equal & 0.839 & 0.812 & 0.808 & 0.807 & 0.820 & 0.784 & 0.788 & 0.764 \\
\hline
\end{tabular}
\caption{Rejection rates for the Tukey-type contrast matrix with $\delta =$ 1.5 and sample size balanced large} 
}\end{table}
\begin{table}[ht]{
\centering
\begin{tabular}{lp{0.6in}>{\centering\arraybackslash}p{0.45in}>{\centering\arraybackslash}p{0.45in}>{\centering\arraybackslash}p{0.45in}>{\centering\arraybackslash}p{0.45in}>{\centering\arraybackslash}p{0.45in}>{\centering\arraybackslash}p{0.45in}>{\centering\arraybackslash}p{0.45in}>{\centering\arraybackslash}p{0.45in}}
  \hline
distribution & censoring \newline distribution & asymptotic global & permutation & asymptotic & wild Rademacher & wild Gaussian & groupwise & asymptotic bonf & permutation bonf \\ 
  \hline
exp early & equal & 0.457 & 0.380 & 0.434 & 0.448 & 0.460 & 0.373 & 0.408 & 0.352 \\ 
  exp early & unequal, high & 0.407 & 0.293 & 0.378 & 0.389 & 0.412 & 0.262 & 0.345 & 0.260 \\ 
  exp early & unequal, low & 0.466 & 0.378 & 0.441 & 0.446 & 0.466 & 0.368 & 0.404 & 0.341 \\ 
  exp late & equal & 0.680 & 0.598 & 0.603 & 0.617 & 0.636 & 0.509 & 0.559 & 0.489 \\ 
  exp late & unequal, high & 0.560 & 0.418 & 0.464 & 0.482 & 0.510 & 0.293 & 0.428 & 0.322 \\ 
  exp late & unequal, low & 0.687 & 0.595 & 0.597 & 0.609 & 0.634 & 0.490 & 0.554 & 0.476 \\ 
  exp prop & equal & 0.534 & 0.457 & 0.495 & 0.503 & 0.519 & 0.423 & 0.474 & 0.402 \\
\hline
\end{tabular}
\caption{Rejection rates for the Tukey-type contrast matrix with $\delta =$ 1.5 and sample size balanced medium} 
}\end{table}
\begin{table}[ht]{
\centering
\begin{tabular}{lp{0.6in}>{\centering\arraybackslash}p{0.45in}>{\centering\arraybackslash}p{0.45in}>{\centering\arraybackslash}p{0.45in}>{\centering\arraybackslash}p{0.45in}>{\centering\arraybackslash}p{0.45in}>{\centering\arraybackslash}p{0.45in}>{\centering\arraybackslash}p{0.45in}>{\centering\arraybackslash}p{0.45in}}
  \hline
distribution & censoring \newline distribution & asymptotic global & permutation & asymptotic & wild Rademacher & wild Gaussian & groupwise & asymptotic bonf & permutation bonf \\ 
  \hline
exp early & equal & 0.320 & 0.204 & 0.291 & 0.309 & 0.342 & 0.166 & 0.274 & 0.185 \\ 
  exp early & unequal, high & 0.314 & 0.155 & 0.284 & 0.311 & 0.360 & 0.083 & 0.258 & 0.136 \\ 
  exp early & unequal, low & 0.316 & 0.189 & 0.288 & 0.306 & 0.332 & 0.152 & 0.273 & 0.173 \\ 
  exp late & equal & 0.423 & 0.273 & 0.350 & 0.370 & 0.412 & 0.181 & 0.315 & 0.203 \\ 
  exp late & unequal, high & 0.382 & 0.172 & 0.303 & 0.346 & 0.402 & 0.064 & 0.270 & 0.127 \\ 
  exp late & unequal, low & 0.422 & 0.256 & 0.345 & 0.371 & 0.422 & 0.150 & 0.315 & 0.194 \\ 
  exp prop & equal & 0.343 & 0.221 & 0.308 & 0.323 & 0.358 & 0.170 & 0.287 & 0.193 \\
\hline
\end{tabular}
\caption{Rejection rates for the Tukey-type contrast matrix with $\delta =$ 1.5 and sample size balanced small} 
}\end{table}
\begin{table}[ht]{
\centering
\begin{tabular}{lp{0.6in}>{\centering\arraybackslash}p{0.45in}>{\centering\arraybackslash}p{0.45in}>{\centering\arraybackslash}p{0.45in}>{\centering\arraybackslash}p{0.45in}>{\centering\arraybackslash}p{0.45in}>{\centering\arraybackslash}p{0.45in}>{\centering\arraybackslash}p{0.45in}>{\centering\arraybackslash}p{0.45in}}
  \hline
distribution & censoring \newline distribution & asymptotic global & permutation & asymptotic & wild Rademacher & wild Gaussian & groupwise & asymptotic bonf & permutation bonf \\ 
  \hline
exp early & equal & 0.818 & 0.779 & 0.784 & 0.786 & 0.794 & 0.753 & 0.756 & 0.714 \\ 
  exp early & unequal, high & 0.708 & 0.631 & 0.661 & 0.668 & 0.682 & 0.594 & 0.618 & 0.549 \\ 
  exp early & unequal, low & 0.811 & 0.770 & 0.780 & 0.781 & 0.790 & 0.746 & 0.751 & 0.710 \\ 
  exp late & equal & 0.957 & 0.945 & 0.918 & 0.919 & 0.925 & 0.890 & 0.898 & 0.880 \\ 
  exp late & unequal, high & 0.882 & 0.823 & 0.799 & 0.806 & 0.820 & 0.720 & 0.771 & 0.703 \\ 
  exp late & unequal, low & 0.961 & 0.946 & 0.922 & 0.924 & 0.929 & 0.894 & 0.901 & 0.876 \\ 
  exp prop & equal & 0.885 & 0.861 & 0.851 & 0.852 & 0.856 & 0.827 & 0.824 & 0.792 \\
\hline
\end{tabular}
\caption{Rejection rates for the Tukey-type contrast matrix with $\delta =$ 1.5 and sample size unbalanced large} 
}\end{table}
\begin{table}[ht]{
\centering
\begin{tabular}{lp{0.6in}>{\centering\arraybackslash}p{0.45in}>{\centering\arraybackslash}p{0.45in}>{\centering\arraybackslash}p{0.45in}>{\centering\arraybackslash}p{0.45in}>{\centering\arraybackslash}p{0.45in}>{\centering\arraybackslash}p{0.45in}>{\centering\arraybackslash}p{0.45in}>{\centering\arraybackslash}p{0.45in}}
  \hline
distribution & censoring \newline distribution & asymptotic global & permutation & asymptotic & wild Rademacher & wild Gaussian & groupwise & asymptotic bonf & permutation bonf \\ 
  \hline
exp early & equal & 0.515 & 0.410 & 0.471 & 0.481 & 0.499 & 0.397 & 0.442 & 0.355 \\ 
  exp early & unequal, high & 0.455 & 0.283 & 0.404 & 0.421 & 0.449 & 0.265 & 0.355 & 0.234 \\ 
  exp early & unequal, low & 0.517 & 0.408 & 0.475 & 0.486 & 0.503 & 0.391 & 0.438 & 0.352 \\ 
  exp late & equal & 0.727 & 0.618 & 0.628 & 0.642 & 0.663 & 0.515 & 0.578 & 0.482 \\ 
  exp late & unequal, high & 0.591 & 0.388 & 0.477 & 0.499 & 0.538 & 0.279 & 0.435 & 0.298 \\ 
  exp late & unequal, low & 0.718 & 0.608 & 0.627 & 0.636 & 0.661 & 0.503 & 0.574 & 0.476 \\ 
  exp prop & equal & 0.596 & 0.489 & 0.535 & 0.542 & 0.566 & 0.445 & 0.500 & 0.414 \\
\hline
\end{tabular}
\caption{Rejection rates for the Tukey-type contrast matrix with $\delta =$ 1.5 and sample size unbalanced medium} 
}\end{table}
\begin{table}[ht]{
\centering
\begin{tabular}{lp{0.6in}>{\centering\arraybackslash}p{0.45in}>{\centering\arraybackslash}p{0.45in}>{\centering\arraybackslash}p{0.45in}>{\centering\arraybackslash}p{0.45in}>{\centering\arraybackslash}p{0.45in}>{\centering\arraybackslash}p{0.45in}>{\centering\arraybackslash}p{0.45in}>{\centering\arraybackslash}p{0.45in}}
  \hline
distribution & censoring \newline distribution & asymptotic global & permutation & asymptotic & wild Rademacher & wild Gaussian & groupwise & asymptotic bonf & permutation bonf \\ 
  \hline
exp early & equal & 0.346 & 0.174 & 0.314 & 0.334 & 0.368 & 0.165 & 0.296 & 0.181 \\ 
  exp early & unequal, high & 0.337 & 0.106 & 0.295 & 0.343 & 0.386 & 0.063 & 0.271 & 0.125 \\ 
  exp early & unequal, low & 0.340 & 0.157 & 0.310 & 0.332 & 0.365 & 0.151 & 0.303 & 0.175 \\ 
  exp late & equal & 0.448 & 0.230 & 0.365 & 0.392 & 0.443 & 0.166 & 0.338 & 0.210 \\ 
  exp late & unequal, high & 0.401 & 0.122 & 0.327 & 0.401 & 0.450 & {0.053} & 0.288 & 0.135 \\ 
  exp late & unequal, low & 0.464 & 0.230 & 0.384 & 0.418 & 0.470 & 0.161 & 0.343 & 0.209 \\ 
  exp prop & equal & 0.378 & 0.190 & 0.328 & 0.350 & 0.400 & 0.157 & 0.298 & 0.179 \\
\hline
\end{tabular}
\caption{Rejection rates for the Tukey-type contrast matrix with $\delta =$ 1.5 and sample size unbalanced small} 
}\end{table}
\begin{table}[ht]{
\centering
\begin{tabular}{lp{0.6in}>{\centering\arraybackslash}p{0.45in}>{\centering\arraybackslash}p{0.45in}>{\centering\arraybackslash}p{0.45in}>{\centering\arraybackslash}p{0.45in}>{\centering\arraybackslash}p{0.45in}>{\centering\arraybackslash}p{0.45in}>{\centering\arraybackslash}p{0.45in}>{\centering\arraybackslash}p{0.45in}}
  \hline
distribution & censoring \newline distribution & asymptotic global & permutation & asymptotic & wild Rademacher & wild Gaussian & groupwise & asymptotic bonf & permutation bonf \\ 
  \hline
exp early & equal & 0.719 & 0.686 & 0.733 & 0.737 & 0.743 & 0.692 & 0.742 & 0.709 \\ 
  exp early & unequal, high & 0.630 & 0.570 & 0.651 & 0.654 & 0.663 & 0.581 & 0.644 & 0.586 \\ 
  exp early & unequal, low & 0.733 & 0.696 & 0.750 & 0.751 & 0.759 & 0.711 & 0.745 & 0.709 \\ 
  exp late & equal & 0.938 & 0.926 & 0.943 & 0.944 & 0.946 & 0.935 & 0.937 & 0.925 \\ 
  exp late & unequal, high & 0.857 & 0.814 & 0.871 & 0.872 & 0.876 & 0.849 & 0.863 & 0.819 \\ 
  exp late & unequal, low & 0.940 & 0.926 & 0.947 & 0.948 & 0.949 & 0.941 & 0.945 & 0.930 \\ 
  exp prop & equal & 0.839 & 0.812 & 0.853 & 0.852 & 0.856 & 0.826 & 0.842 & 0.810 \\
\hline
\end{tabular}
\caption{Rejection rates for the Grand-mean-type contrast matrix with $\delta =$ 1.5 and sample size balanced large} 
}\end{table}
\begin{table}[ht]{
\centering
\begin{tabular}{lp{0.6in}>{\centering\arraybackslash}p{0.45in}>{\centering\arraybackslash}p{0.45in}>{\centering\arraybackslash}p{0.45in}>{\centering\arraybackslash}p{0.45in}>{\centering\arraybackslash}p{0.45in}>{\centering\arraybackslash}p{0.45in}>{\centering\arraybackslash}p{0.45in}>{\centering\arraybackslash}p{0.45in}}
  \hline
distribution & censoring \newline distribution & asymptotic global & permutation & asymptotic & wild Rademacher & wild Gaussian & groupwise & asymptotic bonf & permutation bonf \\ 
  \hline
exp early & equal & 0.457 & 0.380 & 0.463 & 0.469 & 0.479 & 0.370 & 0.460 & 0.382 \\ 
  exp early & unequal, high & 0.407 & 0.293 & 0.414 & 0.422 & 0.432 & 0.308 & 0.397 & 0.288 \\ 
  exp early & unequal, low & 0.466 & 0.378 & 0.476 & 0.481 & 0.491 & 0.379 & 0.460 & 0.378 \\ 
  exp late & equal & 0.680 & 0.598 & 0.698 & 0.699 & 0.709 & 0.650 & 0.690 & 0.605 \\ 
  exp late & unequal, high & 0.560 & 0.418 & 0.580 & 0.584 & 0.596 & 0.505 & 0.569 & 0.422 \\ 
  exp late & unequal, low & 0.687 & 0.595 & 0.705 & 0.708 & 0.715 & 0.659 & 0.692 & 0.598 \\ 
  exp prop & equal & 0.534 & 0.457 & 0.548 & 0.551 & 0.562 & 0.479 & 0.553 & 0.464 \\
\hline
\end{tabular}
\caption{Rejection rates for the Grand-mean-type contrast matrix with $\delta =$ 1.5 and sample size balanced medium} 
}\end{table}
\begin{table}[ht]{
\centering
\begin{tabular}{lp{0.6in}>{\centering\arraybackslash}p{0.45in}>{\centering\arraybackslash}p{0.45in}>{\centering\arraybackslash}p{0.45in}>{\centering\arraybackslash}p{0.45in}>{\centering\arraybackslash}p{0.45in}>{\centering\arraybackslash}p{0.45in}>{\centering\arraybackslash}p{0.45in}>{\centering\arraybackslash}p{0.45in}}
  \hline
distribution & censoring \newline distribution & asymptotic global & permutation & asymptotic & wild Rademacher & wild Gaussian & groupwise & asymptotic bonf & permutation bonf \\ 
  \hline
exp early & equal & 0.320 & 0.204 & 0.319 & 0.330 & 0.347 & 0.212 & 0.318 & 0.200 \\ 
  exp early & unequal, high & 0.314 & 0.155 & 0.322 & 0.339 & 0.362 & 0.200 & 0.320 & 0.147 \\ 
  exp early & unequal, low & 0.316 & 0.189 & 0.322 & 0.332 & 0.351 & 0.203 & 0.320 & 0.184 \\ 
  exp late & equal & 0.423 & 0.273 & 0.437 & 0.447 & 0.462 & 0.339 & 0.428 & 0.267 \\ 
  exp late & unequal, high & 0.382 & 0.172 & 0.399 & 0.414 & 0.434 & 0.257 & 0.383 & 0.155 \\ 
  exp late & unequal, low & 0.422 & 0.256 & 0.441 & 0.453 & 0.471 & 0.336 & 0.432 & 0.254 \\ 
  exp prop & equal & 0.343 & 0.221 & 0.354 & 0.366 & 0.387 & 0.258 & 0.358 & 0.225 \\
\hline
\end{tabular}
\caption{Rejection rates for the Grand-mean-type contrast matrix with $\delta =$ 1.5 and sample size balanced small} 
}\end{table}
\begin{table}[ht]{
\centering
\begin{tabular}{lp{0.6in}>{\centering\arraybackslash}p{0.45in}>{\centering\arraybackslash}p{0.45in}>{\centering\arraybackslash}p{0.45in}>{\centering\arraybackslash}p{0.45in}>{\centering\arraybackslash}p{0.45in}>{\centering\arraybackslash}p{0.45in}>{\centering\arraybackslash}p{0.45in}>{\centering\arraybackslash}p{0.45in}}
  \hline
distribution & censoring \newline distribution & asymptotic global & permutation & asymptotic & wild Rademacher & wild Gaussian & groupwise & asymptotic bonf & permutation bonf \\ 
  \hline
exp early & equal & 0.818 & 0.779 & 0.819 & 0.819 & 0.825 & 0.794 & 0.800 & 0.777 \\ 
  exp early & unequal, high & 0.708 & 0.631 & 0.714 & 0.716 & 0.720 & 0.675 & 0.695 & 0.648 \\ 
  exp early & unequal, low & 0.811 & 0.770 & 0.809 & 0.809 & 0.812 & 0.779 & 0.797 & 0.772 \\ 
  exp late & equal & 0.957 & 0.945 & 0.961 & 0.960 & 0.962 & 0.953 & 0.956 & 0.949 \\ 
  exp late & unequal, high & 0.882 & 0.823 & 0.889 & 0.891 & 0.890 & 0.871 & 0.883 & 0.854 \\ 
  exp late & unequal, low & 0.961 & 0.946 & 0.960 & 0.961 & 0.963 & 0.952 & 0.952 & 0.941 \\ 
  exp prop & equal & 0.885 & 0.861 & 0.891 & 0.890 & 0.892 & 0.876 & 0.884 & 0.866 \\
\hline
\end{tabular}
\caption{Rejection rates for the Grand-mean-type contrast matrix with $\delta =$ 1.5 and sample size unbalanced large} 
}\end{table}
\begin{table}[ht]{
\centering
\begin{tabular}{lp{0.6in}>{\centering\arraybackslash}p{0.45in}>{\centering\arraybackslash}p{0.45in}>{\centering\arraybackslash}p{0.45in}>{\centering\arraybackslash}p{0.45in}>{\centering\arraybackslash}p{0.45in}>{\centering\arraybackslash}p{0.45in}>{\centering\arraybackslash}p{0.45in}>{\centering\arraybackslash}p{0.45in}}
  \hline
distribution & censoring \newline distribution & asymptotic global & permutation & asymptotic & wild Rademacher & wild Gaussian & groupwise & asymptotic bonf & permutation bonf \\ 
  \hline
exp early & equal & 0.515 & 0.410 & 0.512 & 0.516 & 0.527 & 0.446 & 0.500 & 0.443 \\ 
  exp early & unequal, high & 0.455 & 0.283 & 0.460 & 0.469 & 0.478 & 0.368 & 0.440 & 0.341 \\ 
  exp early & unequal, low & 0.517 & 0.408 & 0.511 & 0.516 & 0.525 & 0.450 & 0.499 & 0.435 \\ 
  exp late & equal & 0.727 & 0.618 & 0.727 & 0.729 & 0.737 & 0.684 & 0.707 & 0.652 \\ 
  exp late & unequal, high & 0.591 & 0.388 & 0.601 & 0.601 & 0.616 & 0.506 & 0.589 & 0.482 \\ 
  exp late & unequal, low & 0.718 & 0.608 & 0.718 & 0.718 & 0.727 & 0.661 & 0.702 & 0.643 \\ 
  exp prop & equal & 0.596 & 0.489 & 0.595 & 0.603 & 0.611 & 0.543 & 0.589 & 0.524 \\
\hline
\end{tabular}
\caption{Rejection rates for the Grand-mean-type contrast matrix with $\delta =$ 1.5 and sample size unbalanced medium} 
}\end{table}
\begin{table}[ht]{
\centering
\begin{tabular}{lp{0.6in}>{\centering\arraybackslash}p{0.45in}>{\centering\arraybackslash}p{0.45in}>{\centering\arraybackslash}p{0.45in}>{\centering\arraybackslash}p{0.45in}>{\centering\arraybackslash}p{0.45in}>{\centering\arraybackslash}p{0.45in}>{\centering\arraybackslash}p{0.45in}>{\centering\arraybackslash}p{0.45in}}
  \hline
distribution & censoring \newline distribution & asymptotic global & permutation & asymptotic & wild Rademacher & wild Gaussian & groupwise & asymptotic bonf & permutation bonf \\ 
  \hline
exp early & equal & 0.346 & 0.174 & 0.344 & 0.356 & 0.372 & 0.236 & 0.337 & 0.234 \\ 
  exp early & unequal, high & 0.337 & 0.106 & 0.340 & 0.359 & 0.381 & 0.181 & 0.334 & 0.173 \\ 
  exp early & unequal, low & 0.340 & 0.157 & 0.337 & 0.349 & 0.366 & 0.220 & 0.342 & 0.226 \\ 
  exp late & equal & 0.448 & 0.230 & 0.445 & 0.455 & 0.477 & 0.336 & 0.436 & 0.321 \\ 
  exp late & unequal, high & 0.401 & 0.122 & 0.417 & 0.436 & 0.460 & 0.209 & 0.396 & 0.204 \\ 
  exp late & unequal, low & 0.464 & 0.230 & 0.456 & 0.467 & 0.491 & 0.317 & 0.445 & 0.311 \\ 
  exp prop & equal & 0.378 & 0.190 & 0.378 & 0.390 & 0.408 & 0.268 & 0.365 & 0.259 \\
\hline
\end{tabular}
\caption{Rejection rates for the Grand-mean-type contrast matrix with $\delta =$ 1.5 and sample size unbalanced small}
\label{tab:RR2} 
}\end{table}
}

\FloatBarrier
\subsubsection{Power under the alternative hypothesis for the local hypotheses}\label{sssec:MultPower}

\begin{table}[ht]
\centering
\begin{tabular}{llcccc}
  \hline
distribution & censoring distribution & asymptotic & groupwise & asymptotic bonf & permutation bonf \\ 
  \hline
exp early & equal & 0.557 & 0.532 & 0.558 & 0.537 \\ 
  exp early & unequal, high & 0.447 & 0.411 & 0.445 & 0.406 \\ 
  exp early & unequal, low & 0.552 & 0.528 & 0.551 & 0.525 \\ 
  exp late & equal & 0.739 & 0.711 & 0.719 & 0.698 \\ 
  exp late & unequal, high & 0.592 & 0.539 & 0.572 & 0.532 \\ 
  exp late & unequal, low & 0.718 & 0.686 & 0.703 & 0.679 \\ 
  exp prop & equal & 0.654 & 0.631 & 0.630 & 0.607 \\ 
  exp prop & unequal, high & 0.511 & 0.471 & 0.493 & 0.459 \\ 
  exp prop & unequal, low & 0.631 & 0.604 & 0.612 & 0.588 \\ 
  logn & equal & 0.789 & 0.766 & 0.782 & 0.759 \\ 
  logn & unequal, high & 0.621 & 0.577 & 0.620 & 0.584 \\ 
  logn & unequal, low & 0.797 & 0.776 & 0.791 & 0.769 \\ 
  pwExp & equal & 0.532 & 0.508 & 0.518 & 0.496 \\ 
  pwExp & unequal, high & 0.435 & 0.394 & 0.413 & 0.377 \\ 
  pwExp & unequal, low & 0.518 & 0.496 & 0.512 & 0.489 \\ 
  Weib late & equal & 0.940 & 0.932 & 0.934 & 0.925 \\ 
  Weib late & unequal, high & 0.823 & 0.792 & 0.804 & 0.777 \\ 
  Weib late & unequal, low & 0.935 & 0.925 & 0.929 & 0.918 \\ 
  Weib prop & equal & 0.895 & 0.882 & 0.887 & 0.869 \\ 
  Weib prop & unequal, high & 0.747 & 0.709 & 0.737 & 0.701 \\ 
  Weib prop & unequal, low & 0.893 & 0.876 & 0.884 & 0.870 \\ 
  Weib scale & equal & 0.713 & 0.685 & 0.687 & 0.666 \\ 
  Weib scale & unequal, high & 0.558 & 0.518 & 0.542 & 0.501 \\ 
  Weib scale & unequal, low & 0.707 & 0.685 & 0.700 & 0.675 \\ 
  Weib shape & equal & 0.486 & 0.453 & 0.468 & 0.443 \\ 
  Weib shape & unequal, high & 0.400 & 0.363 & 0.385 & 0.355 \\ 
  Weib shape & unequal, low & 0.485 & 0.453 & 0.465 & 0.439 \\ 
   \hline
\end{tabular}
\caption{Rejection rates for hypothesis $\mathcal H_{0, 3 }$ of the Dunnett-type contrast matrix with $\delta =$ 1.5 and with sample size balanced large} 
\label{tab:MultPower1}
\end{table}
\begin{table}[ht]
\centering
%
\caption{Rejection rates for hypothesis $\mathcal H_{0, 3 }$ of the Dunnett-type contrast matrix with $\delta =$ 1.5 and with sample size balanced medium} 
\end{table}
\begin{table}[ht]
\centering
%
\caption{Rejection rates for hypothesis $\mathcal H_{0, 3 }$ of the Dunnett-type contrast matrix with $\delta =$ 1.5 and with sample size balanced small} 
\end{table}
\begin{table}[ht]
\centering
%
\caption{Rejection rates for hypothesis $\mathcal H_{0, 3 }$ of the Dunnett-type contrast matrix with $\delta =$ 1.5 and with sample size unbalanced large} 
\end{table}
\begin{table}[ht]
\centering
%
\caption{Rejection rates for hypothesis $\mathcal H_{0, 3 }$ of the Dunnett-type contrast matrix with $\delta =$ 1.5 and with sample size unbalanced medium} 
\end{table}
\begin{table}[ht]
\centering
%
\caption{Rejection rates for hypothesis $\mathcal H_{0, 3 }$ of the Dunnett-type contrast matrix with $\delta =$ 1.5 and with sample size unbalanced small} 
\end{table}
\begin{table}[ht]
\centering
%
\caption{Rejection rates for hypothesis $\mathcal H_{0, 3 }$ of the Tukey-type contrast matrix with $\delta =$ 1.5 and with sample size balanced large} 
\end{table}
\begin{table}[ht]
\centering
%
\caption{Rejection rates for hypothesis $\mathcal H_{0, 5 }$ of the Tukey-type contrast matrix with $\delta =$ 1.5 and with sample size balanced large} 
\end{table}
\begin{table}[ht]
\centering
%
\caption{Rejection rates for hypothesis $\mathcal H_{0, 6 }$ of the Tukey-type contrast matrix with $\delta =$ 1.5 and with sample size balanced large} 
\end{table}
\begin{table}[ht]
\centering
%
\caption{Rejection rates for hypothesis $\mathcal H_{0, 3 }$ of the Tukey-type contrast matrix with $\delta =$ 1.5 and with sample size balanced medium} 
\end{table}
\begin{table}[ht]
\centering
%
\caption{Rejection rates for hypothesis $\mathcal H_{0, 5 }$ of the Tukey-type contrast matrix with $\delta =$ 1.5 and with sample size balanced medium} 
\end{table}
\begin{table}[ht]
\centering
%
\caption{Rejection rates for hypothesis $\mathcal H_{0, 6 }$ of the Tukey-type contrast matrix with $\delta =$ 1.5 and with sample size balanced medium} 
\end{table}
\begin{table}[ht]
\centering
%
\caption{Rejection rates for hypothesis $\mathcal H_{0, 3 }$ of the Tukey-type contrast matrix with $\delta =$ 1.5 and with sample size balanced small} 
\end{table}
\begin{table}[ht]
\centering
%
\caption{Rejection rates for hypothesis $\mathcal H_{0, 5 }$ of the Tukey-type contrast matrix with $\delta =$ 1.5 and with sample size balanced small} 
\end{table}
\begin{table}[ht]
\centering
%
\caption{Rejection rates for hypothesis $\mathcal H_{0, 6 }$ of the Tukey-type contrast matrix with $\delta =$ 1.5 and with sample size balanced small} 
\end{table}
\begin{table}[ht]
\centering
%
\caption{Rejection rates for hypothesis $\mathcal H_{0, 3 }$ of the Tukey-type contrast matrix with $\delta =$ 1.5 and with sample size unbalanced large} 
\end{table}
\begin{table}[ht]
\centering
%
\caption{Rejection rates for hypothesis $\mathcal H_{0, 5 }$ of the Tukey-type contrast matrix with $\delta =$ 1.5 and with sample size unbalanced large} 
\end{table}
\begin{table}[ht]
\centering
%
\caption{Rejection rates for hypothesis $\mathcal H_{0, 6 }$ of the Tukey-type contrast matrix with $\delta =$ 1.5 and with sample size unbalanced large} 
\end{table}
\begin{table}[ht]
\centering
%
\caption{Rejection rates for hypothesis $\mathcal H_{0, 3 }$ of the Tukey-type contrast matrix with $\delta =$ 1.5 and with sample size unbalanced medium} 
\end{table}
\begin{table}[ht]
\centering
%
\caption{Rejection rates for hypothesis $\mathcal H_{0, 5 }$ of the Tukey-type contrast matrix with $\delta =$ 1.5 and with sample size unbalanced medium} 
\end{table}
\begin{table}[ht]
\centering
%
\caption{Rejection rates for hypothesis $\mathcal H_{0, 6 }$ of the Tukey-type contrast matrix with $\delta =$ 1.5 and with sample size unbalanced medium} 
\end{table}
\begin{table}[ht]
\centering
%
\caption{Rejection rates for hypothesis $\mathcal H_{0, 3 }$ of the Tukey-type contrast matrix with $\delta =$ 1.5 and with sample size unbalanced small} 
\end{table}
\begin{table}[ht]
\centering
%
\caption{Rejection rates for hypothesis $\mathcal H_{0, 5 }$ of the Tukey-type contrast matrix with $\delta =$ 1.5 and with sample size unbalanced small} 
\end{table}
\begin{table}[ht]
\centering
%
\caption{Rejection rates for hypothesis $\mathcal H_{0, 6 }$ of the Tukey-type contrast matrix with $\delta =$ 1.5 and with sample size unbalanced small} 
\end{table}
\begin{table}[ht]
\centering
%
\caption{Rejection rates for hypothesis $\mathcal H_{0, 1 }$ of the Grand-mean-type contrast matrix with $\delta =$ 1.5 and with sample size balanced large} 
\end{table}
\begin{table}[ht]
\centering
%
\caption{Rejection rates for hypothesis $\mathcal H_{0, 2 }$ of the Grand-mean-type contrast matrix with $\delta =$ 1.5 and with sample size balanced large} 
\end{table}
\begin{table}[ht]
\centering
%
\caption{Rejection rates for hypothesis $\mathcal H_{0, 3 }$ of the Grand-mean-type contrast matrix with $\delta =$ 1.5 and with sample size balanced large} 
\end{table}
\begin{table}[ht]
\centering
%
\caption{Rejection rates for hypothesis $\mathcal H_{0, 1 }$ of the Grand-mean-type contrast matrix with $\delta =$ 1.5 and with sample size balanced medium} 
\end{table}
\begin{table}[ht]
\centering
%
\caption{Rejection rates for hypothesis $\mathcal H_{0, 2 }$ of the Grand-mean-type contrast matrix with $\delta =$ 1.5 and with sample size balanced medium} 
\end{table}
\begin{table}[ht]
\centering
%
\caption{Rejection rates for hypothesis $\mathcal H_{0, 3 }$ of the Grand-mean-type contrast matrix with $\delta =$ 1.5 and with sample size balanced medium} 
\end{table}
\begin{table}[ht]
\centering
%
\caption{Rejection rates for hypothesis $\mathcal H_{0, 1 }$ of the Grand-mean-type contrast matrix with $\delta =$ 1.5 and with sample size balanced small} 
\end{table}
\begin{table}[ht]
\centering
%
\caption{Rejection rates for hypothesis $\mathcal H_{0, 2 }$ of the Grand-mean-type contrast matrix with $\delta =$ 1.5 and with sample size balanced small} 
\end{table}
\begin{table}[ht]
\centering
%
\caption{Rejection rates for hypothesis $\mathcal H_{0, 3 }$ of the Grand-mean-type contrast matrix with $\delta =$ 1.5 and with sample size balanced small} 
\end{table}
\begin{table}[ht]
\centering
%
\caption{Rejection rates for hypothesis $\mathcal H_{0, 1 }$ of the Grand-mean-type contrast matrix with $\delta =$ 1.5 and with sample size unbalanced large} 
\end{table}
\begin{table}[ht]
\centering
%
\caption{Rejection rates for hypothesis $\mathcal H_{0, 2 }$ of the Grand-mean-type contrast matrix with $\delta =$ 1.5 and with sample size unbalanced large} 
\end{table}
\begin{table}[ht]
\centering
%
\caption{Rejection rates for hypothesis $\mathcal H_{0, 3 }$ of the Grand-mean-type contrast matrix with $\delta =$ 1.5 and with sample size unbalanced large} 
\end{table}
\begin{table}[ht]
\centering
%
\caption{Rejection rates for hypothesis $\mathcal H_{0, 1 }$ of the Grand-mean-type contrast matrix with $\delta =$ 1.5 and with sample size unbalanced medium} 
\end{table}
\begin{table}[ht]
\centering
%
\caption{Rejection rates for hypothesis $\mathcal H_{0, 2 }$ of the Grand-mean-type contrast matrix with $\delta =$ 1.5 and with sample size unbalanced medium} 
\end{table}
\begin{table}[ht]
\centering
%
\caption{Rejection rates for hypothesis $\mathcal H_{0, 3 }$ of the Grand-mean-type contrast matrix with $\delta =$ 1.5 and with sample size unbalanced medium} 
\end{table}
\begin{table}[ht]
\centering
%
\caption{Rejection rates for hypothesis $\mathcal H_{0, 1 }$ of the Grand-mean-type contrast matrix with $\delta =$ 1.5 and with sample size unbalanced small} 
\end{table}
\begin{table}[ht]
\centering
%
\caption{Rejection rates for hypothesis $\mathcal H_{0, 4 }$ of the Grand-mean-type contrast matrix with $\delta =$ 1.5 and with sample size unbalanced small} 
\label{tab:MultPower2}
\end{table}

\FloatBarrier
\subsection{Figures}




\begin{figure}[H]
    \centering
    \includegraphics[width=0.75\textwidth]{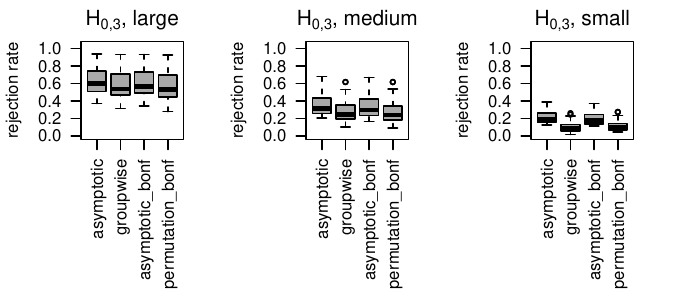}
    \caption{Rejection rates of the false local hypothesis over all settings under the alternative hypothesis for the Dunnett-type contrast matrix.}
    \label{fig:Dunnett_H1_asymptotic}
\end{figure}
\begin{figure}[H]
    \centering
    \includegraphics[width=0.75\textwidth]{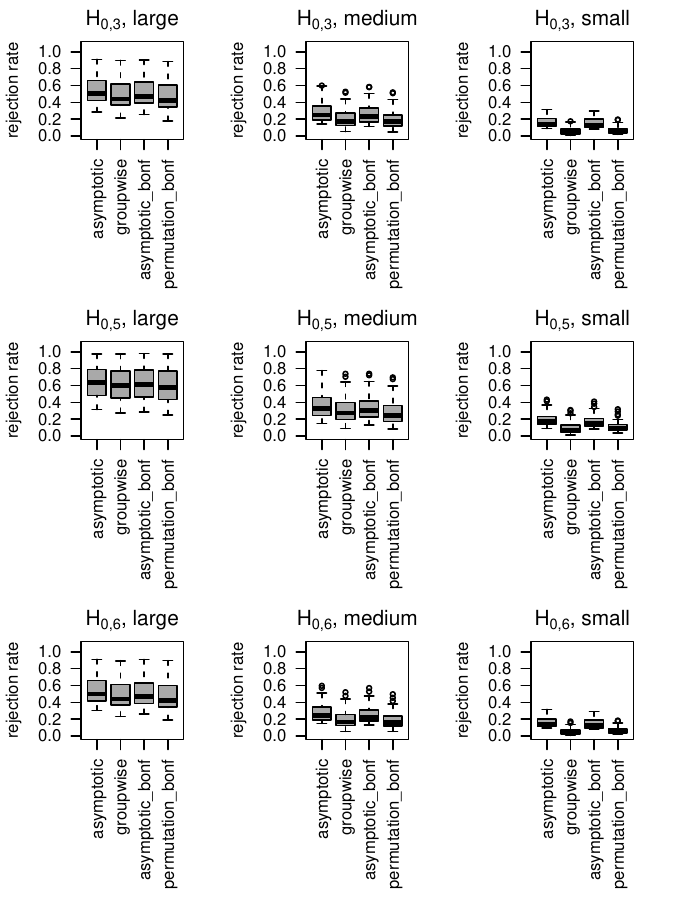}
    \caption{Rejection rates of all false local hypotheses over all settings under the alternative hypothesis for the Tukey-type contrast matrix.}
    \label{fig:Tukey_H1_asymptotic}
\end{figure}
\begin{figure}[H]
    \centering
    \includegraphics[width=0.75\textwidth]{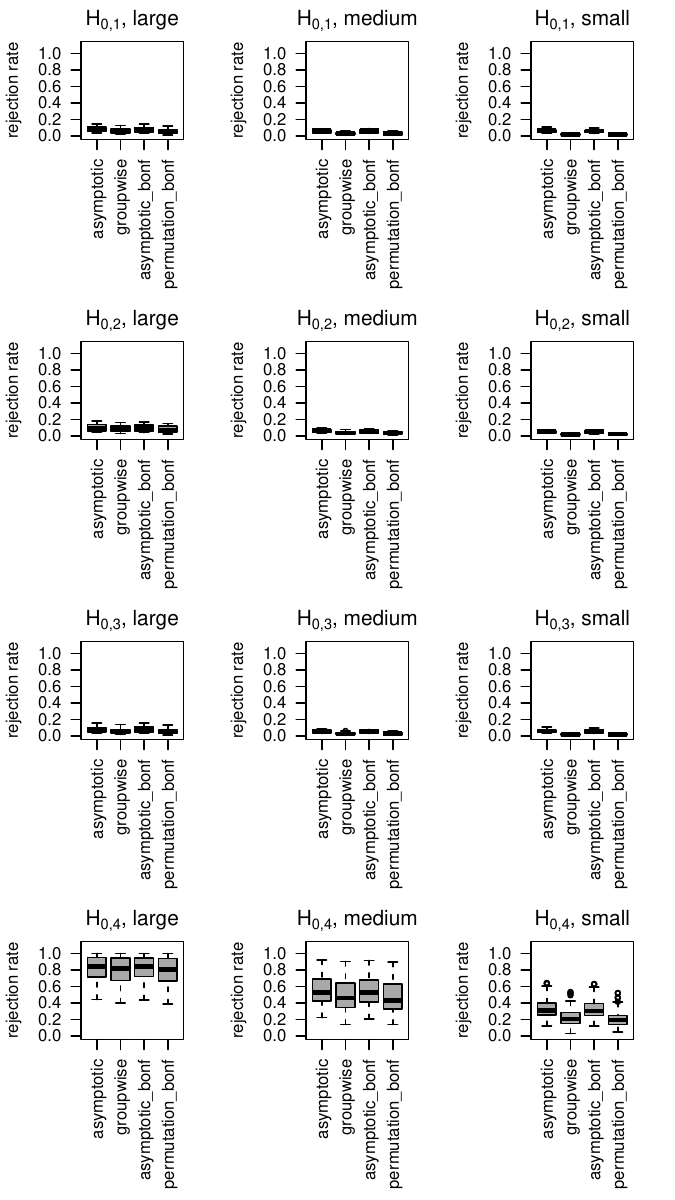}
    \caption{Rejection rates of the false local hypothesis over all settings under the alternative hypothesis for the Grand-mean-type contrast matrix.}
    \label{fig:GrandMean_H1_asymptotic}
\end{figure}

\end{document}